%% file: SNRs_nonthermal_v3.1.tex
\documentclass{aastex63}
\usepackage{nicefrac}
\usepackage{ulem}
\input{mypreset.tex}
\shorttitle{Acceleration Efficiency in Young SNRs}
\shortauthors{Tsuji et al.}
\begin{document}

\title{Systematic Study of Acceleration Efficiency in Young Supernova Remnants with Nonthermal X-ray Observations}

\correspondingauthor{Naomi Tsuji}
%\email{n.tsuji@rikkyo.ac.jp}
\email{naomi.tsuji@riken.jp}

\author[0000-0002-0786-7307]{Naomi Tsuji}
\affiliation{Interdisciplinary Theoretical \& Mathematical Science Program (iTHEMS), RIKEN, 2-1 Hirosawa, Saitama 351-0198, Japan}
\affiliation{Department of Physics, Rikkyo University, 3-34-1 Nishi Ikebukuro, Toshima-ku, Tokyo 171-8501, Japan}

\author{Yasunobu Uchiyama}
\affiliation{Department of Physics, Rikkyo University, 3-34-1 Nishi Ikebukuro, Toshima-ku, Tokyo 171-8501, Japan}
\affiliation{Department of Artificial Intelligence and Science, Rikkyo University, 3-34-1 Nishi Ikebukuro, Toshima-ku, Tokyo 171-8501, Japan}

\author{Dmitry Khangulyan}
\affiliation{Department of Physics, Rikkyo University, 3-34-1 Nishi Ikebukuro, Toshima-ku, Tokyo 171-8501, Japan}
%\collaboration{(AAS Journals Data Scientists collaboration)}

\author{Felix Aharonian}
\affiliation{Dublin Institute for Advanced Studies, 31 Fitzwilliam Place, Dublin 2, Ireland}
\affiliation{Max-Planck-Institut f\"ur Kernphysik, P.O. Box 103980, D 69029 Heidelberg, Germany}
%\affiliation{National Academy of Sciences of the Republic of Armenia, Marshall Baghramian Avenue, 24, 0019 Yerevan, Republic of Armenia}
%\affiliation{Gran Sasso Science Institute, Viale Francesco Crispi 7, 67100, L'Aquila, Italy}




\begin{abstract}
\input{SNRs_nonthermal_abstract.tex}

\end{abstract}



\section{Introduction} \label{sec:intro}
%%%%%%%%%%%%%%%%%%%%%%%%%%%%%%%%%%%%%%%%%%%%%%%%%%%%%%%%%%%%
%%%%%%%%%%%%%%%%%%%%%%%%%%%%%%%%%%%%%%%%%%%%%%%%%%%%%%%%%%%%

%% GCRs
\Acp{snr} are believed to be prominent accelerators of the galactic \acp{cr} with energies up to a few PeV. Under the
conditions typical for shocks in \acp{snr}, \ac{dsa} should be capable of boosting protons to the PeV energies. The rate
of \acp{sn} in the Galaxy and their typical energy output are high enough to explain the entire flux of PeV \ac{cr}
particles at the Earth. It favors the scenario where the dominant fraction of \acp{cr} below the knee is produced in
Galactic \acp{snr}.  However, this scenario contains some problems, among which the absence of any empirical evidence
for existence of PeV particles in \acp{snr} is the most critical.  Nearly all \acp{snr} detected in gamma-ray show
power-law energy spectra with cutoffs at energies of at most tens of TeV. Thus, it is questionable whether \acp{snr} are
indeed capable of producing PeV particles (i.e., whether they are ``PeVatron''). From the theoretical point of view, the
maximum attainable energy for particles accelerated by \ac{dsa} in \acp{snr} is not well constrained because of some
unknown physical parameters such as a diffusion coefficient of the non-thermal particle and magnetic field turbulence.

%% DSA, acc. efficiency
In \ac{dsa}, which is widely accepted as the main acceleration mechanism in \acp{snr}, particles gradually gain energy by crossing the shock wave forward and backward.  A particle changes its
direction by being scattered by magnetic field.  Assuming Bohm diffusion, one
can characterize the diffusion coefficient using mean free path that is represented by
the product of the particle gyroradius and energy independent factor $\eta$ (the so-called Bohm factor).  In the case
of $\eta=1$ (also known as the Bohm limit), particle mean free path takes the minimum value, and the particles are
accelerated most efficiently.  The diffusion coefficient (or the $\eta$ parameter) is strongly related to the spectrum
of the turbulent magnetic field that scatters particles.  Theoretical studies and numerical simulations propose that the
turbulence is generated by the instability resulting from the pressure gradient derived from CR streaming (e.g.,
\cite{Bell1978a,Caprioli2014a}).  However, many facets of the mechanism of turbulent production are poorly
understood. % in the process of turbulent production.

%% synchrotron X-ray
Synchrotron radiation is a good channel to trace electron acceleration at shocks in \acp{snr}. 
The synchrotron X-ray, in particular, is emitted by multi-TeV electrons.  
Therefore, an observation of synchrotron radiation in the X-ray
domain is a great means for diagnosing the nature of the acceleration and of the maximum energy attainable in \acp{snr}. 
In young \acp{snr} where the shock speed is high, the conditions for particle acceleration are especially favorable. 
Thus, young \acp{snr}, with ages less than a few thousand years, are expected to radiate synchrotron X-rays.
The detection of the synchrotron emission has been reported in a few tens of galactic \acp{snr} to date.
%G1.9$+$0.3, Cassiopeia A, Kepler's \ac{snr}, Tycho's \ac{snr}, G330.2$+$1.0 (G330), SN~1006, \rxj, RCW 86, Vela Jr., and HESS J1731$-$347 (HESS J1731).
%and others\check.

%%  eta in \acp{snr}
One can indirectly estimate $\eta$ from measuring the shock speed and cutoff energy in the synchrotron X-ray spectrum (e.g., \cite{srcut_a,ZA07}).
\rxj, an TeV emitting \ac{snr}, is recognized as an accelerator operating at the most efficient rate (i.e., the acceleration proceeds in a regime close to the Bohm limit of $\eta=1$ \citep{Tanaka2008,Tsuji2019}).
Since \nustar\ was launched in 2012, it has provided us with spatially resolved hard X-ray spectra, thus enabling precise measurements of the cutoff energy.
Although young \acp{snr} are believed to be efficient accelerators, 
some recent observations with \nustar\ have in fact revealed that the acceleration efficiency $\eta$ varies significantly and depends on the acceleration site and/or the object.
Based on \nustar\ observations, it has been determined that $\eta$ is $\sim$1 in the forward shock and 3--8 in the reverse shock or reflection shock in Cassiopeia A \citep{Sato2018}. 
In addition, $\eta$ turned out to be $\sim 20$ in the youngest galactic \ac{snr}, G1.9$+$0.3 \citep{Aharonian2017}.
Toward a unified understanding of the particle acceleration in young \acp{snr}, gaps between G1.9$+$0.3 ($\sim$190 year), Cassiopeia A ($\sim$330 year), and \rxj\ ($\sim$1600 year) must be filled.

%% toc
In this paper, we present a systematic analysis and measurement of the cutoff energy in the synchrotron radiation spectrum of young 11 \acp{snr}:
G1.9$+$0.3, Cassiopeia A, Kepler's \ac{snr} (hereafter Kepler), Tycho's \ac{snr} (Tycho), G330.2$+$1.0, SN~1006, \rxj, RCW 86, Vela Jr., HESS J1731$-$347, and SN~1987A.
Most of them are historical, and thus their ages are well constrained.
The dataset of observations analyzed here and the data reduction are presented in \secref{sec:obs}.
\secref{sec:analysis} presents imaging and spectral analyses. The results are presented in \secref{sec:results}.
Our analysis for the first time reveals a variety of instances of particle acceleration in the remnants and clarifies a systematic tendency of acceleration efficiency.
Discussion and interpretation are given in \secref{sec:discussion}, and conclusion are provided in \secref{sec:conclusions}.

%\clearpage
\section{Observations and data reduction} \label{sec:obs}
%%%%%%%%%%%%%%%%%%%%%%%%%%%%%%%%%%%%%%%%%%%%%%%%%%%%%%%%%%%%
%%%%%%%%%%%%%%%%%%%%%%%%%%%%%%%%%%%%%%%%%%%%%%%%%%%%%%%%%%%%
We systematically studied eleven young \acp{snr} using X-ray observations.
To reveal a tendency of acceleration efficiency in the young \acp{snr}, they should be strong synchrotron emitters, and their physical parameters, such as age, distance, and shock speed, should be well constrained.
The following 11 \acp{snr} were selected based on the aforementioned criteria: % for our purpose :
G1.9$+$0.3, Cassiopeia A, Kepler, Tycho, G330.2$+$1.0, SN~1006, \rxj, RCW 86, Vela Jr., and HESS J1731$-$347 in our galaxy, and SN~1987A in the \ac{lmc}.
The other candidates that are dominated by synchrotron radiation in the X-ray energy band (e.g., G32.45$+$0.1 and HESS J1640$-$465) were not included because there are large uncertainties regarding their ages and distances. %  G310.6,
The known parameters of the 11 \acp{snr} are listed in \tabref{tab:dataset_SNRs}.

%%%%%%%%%%%% table

\input{make_FigTab/Bohm_dataset_SNRs.tex}
We utilized archival X-ray observations using \chandra\ and \nustar.
%We also make use of archive data with \xmm\ for sources with poor statistics.
The dataset used in the analysis is summarized in \tabref{tab:Bohm_dataset_chandra} and \tabref{tab:Bohm_dataset_nustar} in \secref{sec:appendix_dataset}.

%% data reduction
We reduced all the observational data using the following software and calibration.
The \chandra\ data were processed using CALDB version 4.8.3 in CIAO version 4.11.
The \nustar\ data were calibrated and screened using {\tt nupipeline} of \nustar\ {\it Data Analysis Software} (NuSTARDAS version 1.4.1 with CALDB version 20180814) included in HEAsoft version 6.19. 
To screen the \nustar\ data, we used the strictest mode (SAAMODE = STRICT and TENTACLE = YES cut).
The effective observational time reduced by these processes is shown in \tabref{tab:Bohm_dataset_chandra} and \tabref{tab:Bohm_dataset_nustar} (\secref{sec:appendix_dataset}).
%CALDB version of \xmm\ \check

%%%%%%%%%%%% table
%\input{make_FigTab/Bohm_dataset_Chandra_thispaper.tex}
%\input{make_FigTab/Bohm_dataset_NuSTAR_thispaper.tex}

%\clearpage
\section{Analysis} \label{sec:analysis}
%%%%%%%%%%%%%%%%%%%%%%%%%%%%%%%%%%%%%%%%%%%%%%%%%%%%%%%%%%%%
%%%%%%%%%%%%%%%%%%%%%%%%%%%%%%%%%%%%%%%%%%%%%%%%%%%%%%%%%%%%

\subsection{Image}
%%%%%%%%%%%%%%%%%%%%%%%%%%%%%%%%%%%%

%% images
%The flux images are shown in \figref{fig:Bohm_image_1}, \figref{fig:Bohm_image_2}, \figref{fig:Bohm_image_3}, and
\figref{fig:Bohm_image_all} illustrates flux images taken with \chandra.
The energy band is set to be 0.5--7 keV for synchrotron dominated \acp{snr} (i.e., G1.9$+$0.3, G330.2$+$1.0, SN~1006, and \hessj1731) and RCW 86, while is set to be 4--6 keV for \acp{snr} which contain thermal emissions (i.e., Cassiopeia A, Kepler, and Tycho).
All the epochs were combined by using {\tt merge\_obs}, the exposure was corrected, and the background was not subtracted.
%We used {\tt merge\_obs} to produce the \chandra\ images.
%For the \nustar\ images, we created a count map using XSELECT\footnote{https://heasarc.gsfc.nasa.gov/ftools/xselect/} and the exposure map using {\tt nuexpomap} with no vignetting correction (``vignflag=no'' set). Finally, we divided the count map by the exposure map to generate the flux image using XIMAGE\footnote{https://heasarc.gsfc.nasa.gov/docs/xanadu/ximage/ximage.html}.
%In \figref{fig:Bohm_image_1}, \figref{fig:Bohm_image_2}, and \figref{fig:Bohm_image_3}, we present three-color (red, green, and blue (RGB)) images with \chandra. For \acp{snr} with strong thermal emissions (i.e., Cassiopeia A, Kepler, and Tycho), the RGB bands were set to be 1.7--2 keV (line emissions from Si), 6.4--7 keV (Fe), and 4--6 keV (continuum), respectively. For the other \acp{snr}, which are relatively dominated by  nonthermal radiation, the RGB bands were 0.5--1.2 keV, 1.2--2 keV, and 2--7 keV, respectively. These RGB images helped us to avoid contamination by thermal line emissions and to extract selectively the spectrum from the synchrotron-dominated (nonthermal) region in which we were interested. 
Note that we show the images of the entire remnants taken with \xmm\ and \suzaku\ for \rxj\ and Vela Jr., respectively.
%These images were kindly provided by \cite{Acero2009} for \rxj\ and \cite{Fukuyama} for Vela Jr.
We omitted the image of SN~1987A since it is treated as a point-like source in this paper, although it shows spatial extent (e.g., \cite{McCray2016}).

%% regions
%\mitya{the next sentence is a repeating, is that needed?} The spectrum of synchrotron radiation is a powerful means of exploring accelerated electrons.  
Some \acp{snr} have strong thermal line emissions in addition to the synchrotron component, making it difficult to extract the pure synchrotron spectrum.
For \acp{snr} that have thermal components, the 4--6 keV channel can be dominated by nonthermal radiation, while the other X-ray band can be contaminated by thermal line emissions.
The rims or the filament-like structures, which are likely located in the outermost regions, appear bright in the 4--6 keV energy range (\figref{fig:Bohm_image_all}).
These regions are expected to contain the synchrotron emission from electrons  accelerated at the forward shock, and are thus the best targets for our analysis to study particle acceleration in \acp{snr}. 
We defined regions to extract the spectra along these outer rims (\figref{fig:Bohm_image_all}).
In order to investigate the relation between the shock velocity and synchrotron radiation, we also defined subregions along the regions where the proper motions were measured in the previous works. %e.g., Kepler \citep{Katsuda2008_kepler} and RCW 86 \citep{Yamaguchi2016}.
Five \acp{snr} were picked up for this purpose: proper motions along the entire rims were already measured in G1.9$+$0.3\ \citep{Borkowski2017}, Cassiopeia A \citep{Patnaude2009}, Kepler \citep{Katsuda2008_kepler}, Tycho \citep{Williams2013}, and SN~1006 \citep{Winkler2014}.
It should be noted that the superb angular resolution of \chandra\ enabled us to extract spectra from relatively small (a few arcsec scale) subregions. %, but the spectrum with \nustar\ was integrated over more extended regions.
%The regions used for spectral analyses using both \chandra\ and \nustar\ are indicated by thick-solid lines, whereas those with only \chandra\ are indicated by thin-solid lines.
%The regions for background extraction are indicated by dashed lines.

%\setcounter{figure}{-1} 

\input{make_FigTab/figure_images.tex}
\subsection{Spectrum}
\label{sec:Bohm_spectrum}
%%%%%%%%%%%%%%%%%%%%%%%%%%%%%%%%%%%%
%%%%%%%%%%%%%%%%%%%%%%%%%%%%%%%%%%%%

%% extract spectrum
To extract spectra, we used {\tt specextract} and {\tt nuproducts} for \chandra\ and \nustar\ data, respectively.
The source regions are shown in \figref{fig:Bohm_image_all}, while the background spectra are extracted from nearby regions of the source otherwise mentioned.
The ``extended=yes'' set was applied to  all the \nustar\ spectra except for SN~1987A, which is spatially compatible with a point-like source.
%The source and background regions are shown in \figref{fig:Bohm_image_1}, \figref{fig:Bohm_image_2}, and \figref{fig:Bohm_image_3}.
The spectra of different epochs and different detector modules were combined using {\tt addascaspec}.
%% proper motion effect
Although most of the \acp{snr} examined in this paper showed expanding motions, we could safely combine the spectra of the different epochs because the extracted regions were substantially larger than the shifts due to  their proper motions.
%% time evolution of Chandra response
It also should be noted that summing up the spectra of the different epochs is not affected by time evolution and variations across the \ac{fov} of the response files.
We checked this by simultaneously fitting the spectra of the different epochs, showing the consistent result with when fitting the combined spectrum.
The spectra obtained from the representative regions of each \ac{snr} are shown in \figref{fig:Bohm_spectra}.

%% nustar stray light
%(usage of \nustar\ \check\ --$>$ G1.9$+$0.3, Cas A SE and NE,SN~1006, \rxj, Vela Jr., and SN~1987A, while only \chandra\ for Kepler)
%\check\ Re-analyzing observations with \nustar\ using nuskybgd. The results, especially the spatially resolved spectral analysis in the smaller regions, do not change. This paragraph would be revised after the reanalysis. \check\ \\
In many cases \nustar\ spectra contain uncertainties related to stray light. 
We therefore used \nustar\ data only for sources which it can be reliably dealt with: namely, for G1.9$+$0.3, Cassiopeia A (the SE and NE regions), SN~1006 (the subregions 2--4 in NE and 11--13 in SW), \rxj, Vela Jr., and SN~1987A. 
The background spectrum with \nustar\ here does include uncertainty of non-uniform distribution due to stray light and instrumental components \citep{Wik2014}, except for \rxj\ which the background was adequently subtracted as reported in \cite{Tsuji2019}. 
As G1.9$+$0.3 and SN~1987A have small angular sizes, the background can be extracted from the surrounding region of the source, and the non-uniform distribution of the background should not noticeably affect the results.
For largely extended sources, such as SN~1006, \rxj, and Vela Jr., a careful treatment with the non-uniform background is necessary.
%Particularly for SN~1006, \rxj, and Vela Jr., which were spatially extended across nearly the entire \ac{fov}, a careful treatment with the non-uniform background is necessary. %\footnote{The first results of \nustar\ observations of Kelper (PI: L. Lopez) and Vela Jr. (PI: F. Acero) have yet to be published. A detailed analysis would be performed therein.}.
See \cite{Grefenstette2014}, \cite{Li2018} and \cite{Tsuji2019} for the detailed studies of the background spectra in \nustar\ observations of Cassiopeia A, SN~1006, and \rxj, respectively.
In the case of the \nustar\ spectrum of Vela Jr., we simply checked that changing normalization of the present background by $\pm$10\% and 20\% resulted in differences in the spectral parameters within 7\% and 15\%, respectively. %does not have any effect on the results.

%%%    figures 
\input{make_FigTab/Bohm_spectra.tex} % figure of spectra

\subsection{Model}
\label{sec:Bohm_model}
%%%%%%%%%%%%%%%%%%%%%%%%%%%%%%%%%%%%
%%%%%%%%%%%%%%%%%%%%%%%%%%%%%%%%%%%%

%% model fitting
We applied the model of synchrotron radiation from cooling-limited electrons, as proposed in \cite{ZA07} (hereafter the ZA07 model), given by %same as \eqref{eq:ZA07_model_electron} and \eqref{eq:ZA07_model_synch} in \secref{sec:nustar_rxj1713}), 
\if0
\begin{eqnarray}
% --- Electron spectrum
%\frac{dN_e}{dE} &\propto & E^{-3} \Bigg[ 1+0.66\left( \frac{E}{E_0} \right)^{\frac{5}{4}} \Bigg]^{\frac{9}{5}}\exp \Bigg[ -\left(\frac{E}{E_0}\right)^2 \Bigg] \label{eq:ZA07_model_electron} , \\ 
\frac{dN_e}{dE} &\propto & \left(\frac{E}{E_0}\right)^{-3} 
\Bigg[ 
\left\{ 1+0.523\left( \frac{E}{E_0} \right)^{\frac{9}{4}} \right\}^{2}
-  0.0636 \left(\frac{E}{E_0}\right)^{2} \left\{ 1+1.7\left( \frac{E}{E_0} \right)^3 \right\}^{\frac{5}{6}}
\Bigg] 
\exp \Bigg[ -\left(\frac{E}{E_0}\right)^2 \Bigg] \label{eq:ZA07_model_electron2} 
\end{eqnarray}
for electrons, and
\fi
\begin{eqnarray}
% --- Synch spectrum
%\frac{dN}{d\varepsilon} &\propto & \varepsilon^{-2} \Bigg[ 1+0.46\left( \frac{\varepsilon}{\varepsilon_0} \right)^{0.6} \Bigg]^{2.29}\exp \Bigg[ -\left(\frac{\varepsilon}{\varepsilon_0}\right)^{\frac{1}{2}}\Bigg] .\label{eq:ZA07_model_synch} 
\frac{dN_X}{d\varepsilon} &\propto & \left(\frac{\varepsilon}{\varepsilon_0}\right)^{-2} \Bigg[ 1+0.38\left( \frac{\varepsilon}{\varepsilon_0} \right)^{\nicefrac{1}{2}} \Bigg]^{\nicefrac{11}{4}}\exp \Bigg[ -\left(\frac{\varepsilon}{\varepsilon_0}\right)^{\nicefrac{1}{2}}\Bigg] , \label{eq:ZA07_model_synch2} 
\end{eqnarray}
with $\varepsilon_0$ being a cutoff energy parameter.
% --- kappa
In this model, we adopted the $\kappa=\sqrt{1/11}$ case in which $\kappa$ is the ratio of the upstream magnetic field to the downstream magnetic field, $\kappa = B_{\rm up} / B_{\rm down}$.
This corresponds to an enhancement of random isotropic magnetic field due to the standard shock compression ratio of $\sigma=4$.

The model is described with an absorbed ZA07, where the interstellar absorption is considered  by the TBabs model in XSEPC. % for the synchrotron-dominated \acp{snr}, 
%% energy band
The energy band of the \chandra\ spectra is set to be 0.5--7 keV unless otherwise mentioned.
For the \nustar\ spectra, the energy band is set to be 3--20 keV for SN~1006, \rxj, and Vela Jr.; 3--40 keV for G1.9$+$0.3\ and SN~1987A; and 3--50 keV for Cassiopeia A.
We performed spectral fitting of the broadband X-ray observations (i.e., \chandra\ $+$ \nustar\ joint fitting) if \nustar\ data are available. %, in relatively larger area with archival \nustar\ data.
%\textcolor{red}{
Although the combination of the \chandra\ and \nustar\ spectra gives us the more precise measurement of $\varepsilon_0$, the fitting result with solely the \chandra\ data would be less affected by systematic errors. We confirmed this by measuring the $\varepsilon_0$ parameter separately using only the \chandra\ and \nustar\ spectra, showing the consistent results with those obtained by the joint fit.
%}
%Since for G330.2$+$1.0 and HESS J1731$-$347 there are no available observations with \nustar\ and the \chandra\ spectra do not have enough statistics, we use the spectra with \xmm\ in addition to \chandra\ in order to enforce the quality of the spectra.
%The column density was fixed to values in the literature for some \acp{snr}.
The fitting results are presented in \figref{fig:Bohm_spectra} and \tabref{tab:Bohm_result_all_in_one}.
Spectral fitting was performed using XSPEC version 12.9.0.

\subsubsection{Synchrotron dominated spectra} \label{sec:synch_spectra}
%%%%%%%%%%%%%%%%%%%%%%%%%%%%%%%%%%%%%%%
%\sout{The synchrotron (nonthermal) dominated spectra are targets of our interest in particle acceleration in \acp{snr}.}
X-ray spectra can constrain the process of particle acceleration in \acp{snr} where the synchrotron emission dominates.
Such completely featureless and nonthermal spectra were obtained from G1.9$+$0.3, G330.2$+$1.0, \rxj, Vela Jr., and \hessj1731\footnote{Note that these synchrotron dominated remnants also have thermal components, particularly in the central regions.}. 
They are well reproduced by the absorbed ZA07 model.

%% thermal \acp{snr}
On the other hand, some spectra contain non-negligible thermal emission in Cassiopeia A, Kepler, Tycho, SN~1006, RCW 86, and SN~1987A.
% (1) ZA07 + thermal model
For these \acp{snr}, we added a thermal model, namely, TBabs $\times$ (ZA07 + thermal plasma).
The thermal model is described by VNEI for Cassiopeia A, Kepler, and Tycho, and by Vpshock in SN~1006 and RCW 86.
We added two-component thermal plasma, Vpshock and Vequil, in SN~1987A \citep[see, e.g,][]{Frank2016}.
In the former three remnants, a plasma temperature ($kT$) and an ionization parameter ($nt$) are fixed to the values which are derived from the reference regions in \figref{fig:Bohm_image_all}.
Thermal parameters are fixed to the values in the literature for SN~1006 \citep{Miceli2009}, RCW 86 \citep{Tsubone2017}, and SN~1987A \citep{Frank2016}.
Additional line emissions described by Gaussian are added when necessary.
The parameters of the thermal component are summarized in \tabref{tab:Bohm_result_tab_thermal} (\secref{sec:appendix_thermal}).
Although the spectra are well fitted with this model, X-rays in the higher energy range might originate from thermal Bremsstrahlung and/or a hot plasma instead of synchrotron radiation.
Therefore, we defined {\it synchrotron dominated} spectra as those which 1.) were featureless (i.e., line emissions are week), and 2.) the normalization ratio of the synchrotron model to the thermal model was relatively large.
The threshold normalization ratio is $\sim$10--20, but we cannot adopt a common value for all \acp{snr} because the thermal components differ.
This selection resulted in that 5 regions (subregions 4--8) in Kepler,  11 regions (subregions 3, 4, 6--9, 11--13, 17, and 20) in Tycho, and 12 region (subregions 1--6 and 10--15) in SN~1006 were synchrotron dominated, and the others are referred to as {\it thermal dominated}.
We presumed that all the spectra in Cassiopeia A, RCW 86, and SN~1987A defined in this paper belonged to the synchrotron dominated group.
It should be noted that the origin of a hard X-ray component in SN~1987A is unknown yet.

% (2) ZA07 model in >2.5 keV
For those \acp{snr} containing strong thermal components, we also fitted with only the ZA07 model, but using the spectra above 2.5 keV where the thermal components can be suppressed to some extent. 
We excluded the 6.4--6.8 keV channel if the spectrum has a sign of iron line emission.
The cutoff energy parameter was roughly consistent between these two methods, although the one including the thermal model yielded slightly higher cutoff energy parameter.
It should be noted that the thermal-dominated spectra defined above cannot be fitted only with the synchrotron model, resulting in large reduced chi-squared values of $\chi_\nu^2 \geq 2$.
We focus on the results of the synchrotron dominated spectra in this paper and leave, to the future publication, discussion of the relation between the thermal properties and the synchrotron radiation.

\subsubsection{Acceleration efficiency}
%%%%%%%%%%%%%%%%%%%%%%%%%%%%%%%%%%%%%%%%%%%%%%%%%%%%%%%%%%%%
%% eta: equation
In the present framework, the cutoff energy parameter $\varepsilon_0$ is a key factor characterizing shock acceleration in \acp{snr} as it is determined by the balance between acceleration and synchrotron cooling.
\cite{ZA07} derived a relation between the cutoff energy parameter and the shock velocity,
\begin{eqnarray}
%\varepsilon_0 = 0.93 \left(\frac{v_{\rm sh}}{3900 ~{\rm \kms}}\right)^2 \eta^{-1} ~{\rm keV} ,    % ZA07, case1
\varepsilon_0 = 1.6 \left(\frac{v_{\rm sh}}{4000 ~{\rm km}~{\rm s}^{-1} }\right)^2 \eta^{-1} ~{\rm keV} .    % ZA07, case2
\label{eq:ZA07_e02}
\end{eqnarray}
\eqref{eq:ZA07_e02} yields a relation determining the Bohm factor: 
\begin{eqnarray}
\eta  = 1.6 \left(\frac{v_{\rm sh}}{4000 ~{\rm km}~{\rm s}^{-1} }\right)^2 \left( \frac{\varepsilon_0}{\rm keV} \right)^{-1} .    % ZA07, case2
\label{eq:ZA07_eta}
\end{eqnarray}
Combining the measured cutoff energy parameter and the known shock speed, we can estimate the value of $\eta$ using \eqref{eq:ZA07_eta}.
\tabref{tab:Bohm_result_all_in_one} lists the obtained $\eta$ parameter as well as the shock velocity in each region of \ac{snr} analyzed in \secref{sec:Bohm_spectrum}.

%%%%%%%%%%%%%%%%%%%%%%%%%%%%%%%%%%%%
%%%%%%%%%%%%%%%%%%%%%%%%%%%%%%%%%%%%
%\clearpage
\subsubsection{Validity of cooling-limited model} %\label{sec:Bohm_result_tau}

We need to test the validity of the models (\eqref{eq:ZA07_model_synch2} and \eqref{eq:ZA07_e02}) which are derived based on the assumption of the cooling-limited case, since acceleration in young \acp{snr} can be limited by an age instead.
%Our models (\eqref{eq:ZA07_model_electron2}; \eqref{eq:ZA07_model_synch2}; \eqref{eq:ZA07_e02}) were based on the assumption of the cooling-limited case.
\if0
The characteristic timescale of electrons to gain energy via \ac{dsa} is generally given by:
\begin{eqnarray}
\tau_{\rm acc} &\sim & \frac{D}{v_{\rm sh} ^2}  ,
\label{eq:Bohm_tau_acc}
\end{eqnarray}
where the diffusion coefficient of $D$ is now assumed to be the product of $\eta$ and the gyroradius.
The characteristic timescale of electrons at the energy around the cutoff energy parameter ($E \approx E_0$) to suffer the synchrotron cooling is described by:
\begin{eqnarray}
%% ZA, eq. 23:  tau
\tau_{\rm synch} &=& \frac{E_0}{b(E_0)} \\
     &=& 53  \left(\frac{v_{\rm sh}}{3000 ~{\rm km}~{\rm s}^{-1} }\right)^{-1}  \left(\frac{B}{100 ~\mu{\rm G} }\right)^{-\frac{3}{2}}   \eta^{\frac{1}{2}}  \ {\rm yr} ,
     \label{eq:Bohm_tau_synch}
\end{eqnarray}
which is the same as (23) in \cite{ZA07} for $\kappa = \sqrt{11}^{-1}$ and $\gamma_s = 4$.
Here, $b(E)$ denotes the energy loss rate of synchrotron radiation: $b(E) = 4q^4 B^2 E^2 /9m^4c^8$.
\fi
In the synchrotron-cooling-limited case, the acceleration timescale ($\tau_{\rm acc}$) is comparable to the synchrotron-cooling timescale ($\tau_{\rm synch}$; defined as, e.g., Equation (23) in \cite{ZA07} for $\kappa = \sqrt{11}^{-1}$) and is shorter than the characteristic dynamical timescale of the source, such as the age of the accelerator ($\tau_{\rm age}$).
Therefore, to fulfill our assumption, the condition that follows should be satisfied for the energy of synchrotron-emitting electrons:
\begin{eqnarray}
\tau_{\rm acc} \approx \tau_{\rm synch} \leq \tau_{\rm age} .
\end{eqnarray}
%% this work
%The characteristic timescales, by calculating \eqref{eq:Bohm_tau_acc} and \eqref{eq:Bohm_tau_synch}, are presented in \figref{fig:Bohm_result_tau}.
We can place a lower limit value on the magnetic field by imposing the condition of $\tau_{\rm synch} \leq \tau_{\rm age}$,
\begin{eqnarray}
B \geq B_{\rm low} = 12 \left(\frac{t_{\rm age}}{1~{\rm kyr}}\right)^{-\frac{2}{3}} \left(\frac{u_1}{4000~\kms}\right)^{-\frac{2}{3}} \eta^{\frac{1}{3}} \  \uG .
\label{eq:Blow}
\end{eqnarray}
%For example, with the parameter set of \rxj\ NW (i.e., $u_1=3900~ \kms$, $\eta=1.4$, and $\tau_{\rm age}=1600$ yr), $B$ should be greater than 10 \uG.
The lower limit of $B$ calculated for each region is listed in \tabref{tab:Bohm_result_all_in_one}.
$B_{\rm low}$ was estimated to be 10--20 \uG\ in the galactic \acp{snr} older than G330.2$+$1.0 with the age of at most a thousand years.
This can be reasonably achieved given the magnetic field strength in the interstellar medium of $\sim 4$ \uG\ and the standard shock compression of 4. 
For the younger \acp{snr}, we obtained $B_{\rm low}$ of 30--40 \uG, which is still acceptable when considering magnetic field amplification.
Indeed, it has been reported that the magnetic field strength is enhanced up to $\sim$100 \uG, inferred from the filament width \citep[see, e.g.,][]{Bamba2005_filament,Berezhko2006}.
%\textcolor{red}{
An exceptionally high value of $B_{\rm low} =170~ \uG$ was required for SN~1987A.
It is worth noting that, for gamma-ray emitting \acp{snr}, $B_{\rm low}$ derived from \eqref{eq:Blow} is compatible with or smaller than that obtained by modelling the \ac{sed} with the leptonic (\ac{ic}) models in the literature. %except for G1.9$+$0.3.
%An exceptionally high value of $B_{\rm low} =170~ \uG$ for SN~1987A might be challenging, taking into account the extremely young age of the remnant.
%}
%We conclude that the synchrotron-cooling-limited condition is reasonable for all of the galactic \acp{snr} analyzed in this thesis, and the case of SN~1987A might be challenging.

%%%     table
%\clearpage
%\input{make_FigTab/Bohm_result_all_fitting.tex} % table of spectral fitting
\input{make_FigTab/Bohm_result_all_in_one.tex}
\section{Results} \label{sec:results}

\subsection{Relation between shock speed and cutoff energy}   \label{sec:vshe0}
%%%%%%%%%%%%%%%%%%%%%%%%%%%%%%%%%%%%%%%%%%%%%%%%%%%%%%%%%%%%

%% vsh-e0 diagram
In this section we study the possible correlation between the cutoff
energy from our analysis in \secref{sec:Bohm_model} and the shock speed taken from the literature (see references in Tables~\ref{tab:dataset_SNRs} and \ref{tab:Bohm_result_all_in_one}). 
Our results show that there is a significant variance of both the cutoff energy parameter and the shock velocity in the considered SNR (see in
\figref{fig:vshe0}). 
Below we discuss the result for each SNR, and in \secref{sec:Bohm_result_all} we consider the source averaged acceleration efficiencies. 
We also show in \figref{fig:vshe0} (top left panel) a compilation of the largest-$\varepsilon_0$ region of each SNR, which is highlighted by the thick line in \figref{fig:Bohm_image_all} and nearly corresponds to the maximum shock-speed region. 
In this maximum-$\varepsilon_0$ region, the acceleration is the most efficient, and the result is discussed in \secref{sec:Bohm_result_all} and \secref{sec:evolution_eta}.
\if0
%\figref{fig:vshe0} illustrates the relation between the cutoff energy parameter obtained in \secref{sec:Bohm_model} and the shock speed taken from the literature (see references in Tables~\ref{tab:dataset_SNRs} and \ref{tab:Bohm_result_all_in_one}).
The top left panel of \figref{fig:vshe0} represents the results of the largest-$\varepsilon_0$ region of each \ac{snr}, which is highlighted by the thick line in \figref{fig:Bohm_image_all} and nearly corresponds to the maximum shock-speed region.
%{\bf
In this maximum-$\varepsilon_0$ region, the acceleration is the most efficient, and the result is discussed in \secref{sec:Bohm_result_all} and \secref{sec:evolution_eta}.
Apparently, varieties of both the cutoff energy parameter and the shock velocity exist in the top left panel of \figref{fig:vshe0}.
Because only the largest-$\varepsilon_0$ region cannot be representative of the entire remnant, we investigate the $v_{\rm sh}$--$\varepsilon_0$ relation of each SNR in the following and the averaged acceleration efficiency in \secref{sec:Bohm_result_all}.
\fi
%We also confirmed variation in the cutoff energy parameter and/or shock velocity inside the remnants, as shown in each $\varepsilon_0$-$v_{\rm sh}$ diagram of G1.9$+$0.3, Cassiopeia A, Kepler, Tycho, and SN~1006.

In \figref{fig:vshe0}, we also show the obtained $v_{\rm sh}$--$\varepsilon_0$ relation of G1.9$+$0.3, Cassiopeia A, Kepler, Tycho, and SN~1006.
%The proper motions of the subregions were already measured in these \acp{snr}.
For Kepler, Tycho, and SN~1006, we separately showed the thermal dominated and nonthermal (synchrotron) dominated spectra (see \secref{sec:synch_spectra}).
As mentioned in \secref{sec:synch_spectra}, in the thermal dominated spectrum the cutoff energy might not be appropriate because the spectrum in the higher energy range, fitted by the ZA07 model, may not be of synchrotron radiation origin, but Bremsstrahlung or hotter thermal plasma.
Furthermore, the proper motion in the thermal-dominated regions might represent the speed of ejecta components rather than blast waves.
Hence, our objective to examine acceleration efficiency cannot be achieved in these thermal regions (i.e., both cutoff energy and shock speed would not be robust in terms of particle acceleration). 
There exist variations of the shock speed and cutoff energy across the remnants in \figref{fig:vshe0}, and the summary will be given in \secref{sec:discussion_vshe0}.
Note that the spatially resolved $v_{\rm sh}$--$\varepsilon_0$ plots can be obtained only in these five \acp{snr}, since the shock velocity was not measured across the remnant in the other \acp{snr}\footnote{The proper motion was measured to be different from region to region in G330.2$+$1.0, as reported in \cite{Williams2018}, but due to its faint radiation we could not obtain the spatially resolved cutoff energy. 
In RCW 86, proper motion measurement with X-ray observations was conducted in the northeast rims \citep{Yamaguchi2016}, while proper motion with optical observations was measured in the other regions. However X-ray and optical measurements were different by a factor of about 2. We made use of the shock speed measured with X-rays since accelerated electrons are traced by the synchrotron X-ray emission.}.
In addition to them, the $v_{\rm sh}$--$\varepsilon_0$ relation in the north west rim of \rxj\ was already provided in details in \cite{Tsuji2019}.
We make a brief description of the nature of particle acceleration of individual \acp{snr} in the following and present a systematical tendency of particle acceleration of young \acp{snr} in \secref{sec:Bohm_result_all}.

%% vsh-e0 diagram (resolved)
%%%%%%%%%%%%%%%%%%%%%%%%%%%%%%%%%%%%%%%
%%%    figures: Vsh-e0
\input{make_FigTab/vshe0_diagram.tex}

%%% individual interpretation

\input{SNRs_nonthermal_individual_vshe0.tex}

%\clearpage
\subsection{Systematic tendency of $\eta$} \label{sec:Bohm_result_all}
%%%%%%%%%%%%%%%%%%%%%%%%%%%%%%%%%%%%%%%%%%%%%%%%%%%%%%%%%%%%

We obtained the $\eta$ values that are indicative of acceleration efficiency (Bohm factor) in the 11 individual \acp{snr}.
This subsection presents the systematic trend of $\eta$ when all these \acp{snr} are considered together.
Because five \acp{snr} showed significantly different properties of particle acceleration depending on the sites (\secref{sec:vshe0}), we investigated the systematical $\eta$ values estimated from the selected regions based on the largest-$\varepsilon_0$ regions in \secref{sec:Bohm_result_brightest}.
We also present the results using the $\eta$ values averaged over the remnants.
\secref{sec:Bohm_result_type} presents the result of different supernova explosion types.

%%%%%%%%%%%%%%%%%%%%%%%%%%%%%%%%%%%%
\subsubsection{Largest-$\varepsilon_0$ (maximum-$v_{\rm sh}$) region} 
\label{sec:Bohm_result_brightest}

%%%%%%%%%%%%%%
%%%    figures: eta 
%%%%%%%%%%%%%%
\begin{figure*}
\begin{center}
\plottwo{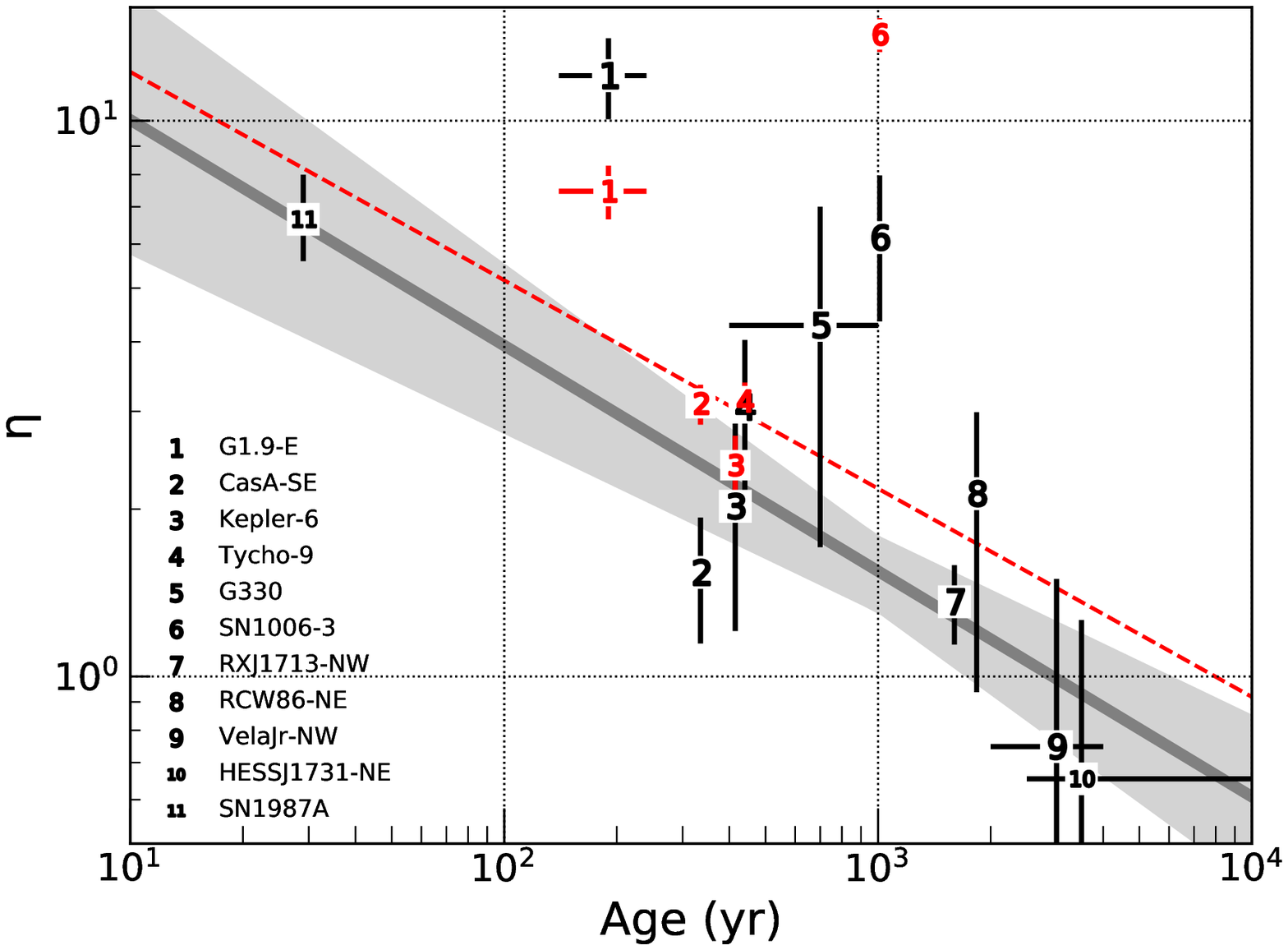}{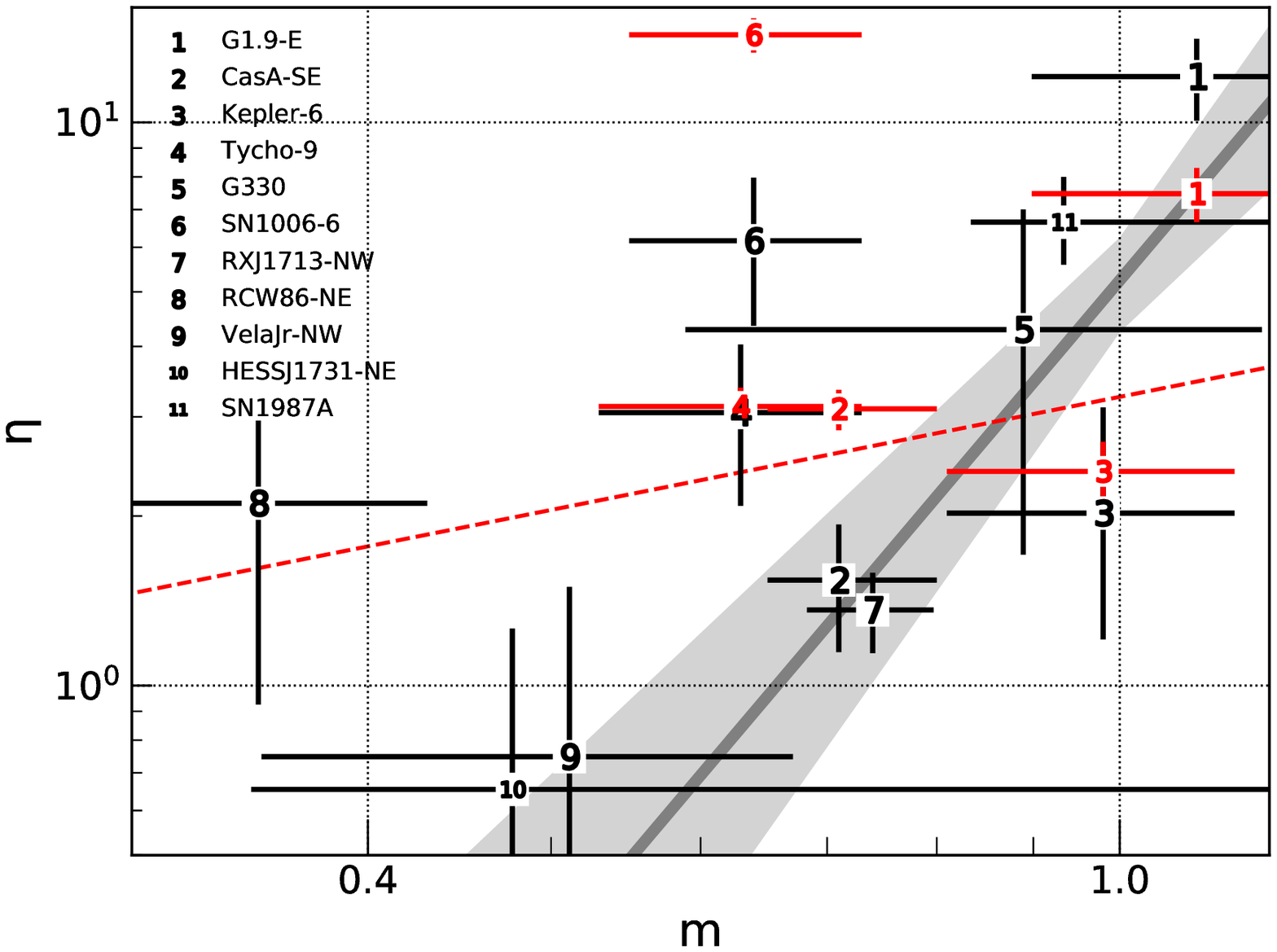}
\caption{
$\eta$ as a function of the age (left) and expansion parameter $m$ (right).
The $\eta$ parameters averaged over the synchrotron dominated regions are shown in red, and the model fitted with the averaged values is shown with the red dashed line.
}
\label{fig:Bohm_result_brighest}
\end{center}
\end{figure*}

%% eta vs age
%The cutoff energy parameter in nonthermal spectrum may contain important information of particle acceleration in the \ac{snr}.
We investigated acceleration efficiency ($\eta$) obtained in the region with the largest $\varepsilon_0$ of each \ac{snr}.
Note that the largest-$\varepsilon_0$ region nearly corresponded to the fastest-$v_{\rm sh}$ region, except for SN~1006, which showed a slightly inverse correlation between $\varepsilon_0$ and $v_{\rm sh}$.
\figref{fig:Bohm_result_brighest} (left) illustrates the relation between the estimated $\eta$ value and the age of each \ac{snr}.
We confirmed a tendency in which $\eta$ decreased as the age of \ac{snr} increased.
In \acp{snr} younger than a few 100 years, the acceleration efficiency substantially deviated from the Bohm limit (i.e., $\eta > 1$).
In the later evolutional stage older than a few 1,000 years, the acceleration proceeded at the most efficient rate (i.e., Bohm limit with $\eta \approx 1$).

%% eta vs age; fitting
To quantify the relation between $\eta$ and $t_{\rm age}$, we fit the observed diagram using an empirical equation, 
\begin{eqnarray}
\eta = C_{\rm age} \left( \frac{t_{\rm age}}{1\ {\rm kyr}} \right) ^{-\delta_{\rm age}} ,
\label{eq:Bohm_eta_age}
\end{eqnarray}
where $C_{\rm age}$ and $\delta_{\rm age}$ indicate a constant value of $\eta$ at the age of 1 kyr and a slope of the age--$\eta$ relation, respectively. 
%(change the result \check)
With chi-squared fitting, we obtained $C_{\rm age}=1.5\pm 0.2$ and $\delta_{\rm age} = 0.41 \pm 0.08$. % with $\chi^2$ of 35 and \ac{dof} of 10.
A correlation coefficient of the logarithm-scale plots in \figref{fig:Bohm_result_brighest} (left) was approximately $-$0.70 with a significance of 2.4$\sigma$.
The best-fit model is indicated by the grey line, and the 90\% uncertainty is indicated by the light grey region in \figref{fig:Bohm_result_brighest}.
The best-fit parameters are summarized in \tabref{tab:eta}.
When we considered the averaged $\eta$ instead of the largest-$\varepsilon_0$ region for G1.9$+$0.3, Cassiopeia A, Kelper, Tycho, and SN~1006, $C_{\rm age}$ and $\delta_{\rm age}$ were $2.2 \pm 0.2$ and $0.37 \pm 0.06$, respectively, and a correlation coefficient was $-0.65$ with a significance of 2.2$\sigma$.
\eqref{eq:Bohm_eta_age} nicely reproduces the observation, although G1.9$+$0.3 and SN~1006 significantly deviate from the best-fit model.
The deviations in the two \acp{snr} might arise from the fact that they are limited by some different factors, such as magnetic field or age, instead of cooling. %, as mentioned in the summary of the previous section.
If we omitted the results of G1.9$+$0.3 and SN~1006, the correlation coefficient was obtained to be $-$0.81 with 2.7$\sigma$ ($-$0.89 with 3.0$\sigma$ using the averaged $\eta$).

%% eta vs m
\figref{fig:Bohm_result_brighest} (right) indicates the observed relation between the $\eta$ value and an expansion parameter $m$, which is an alternative parameter indicating the evolutional phase of \acp{snr}.
Note that $m$ refers to the evolutional stage at which the \ac{snr} currently resides, and it depends not only on the \ac{snr} age, but also on other physical parameters
%\sout{whereas the age depends on the physical parameters surrounding the supernova explosion} 
(e.g., ambient density, progenitor mass, and total energy of \ac{sne}). 
The radius of the \ac{snr}, $R$, and $m$ are described respectively with
\begin{eqnarray}
R \propto t^m ,
\end{eqnarray}
and
\begin{eqnarray}
m \equiv \frac{v_{\rm obs}}{ R/t_{\rm age}} ,
\end{eqnarray}
where $v_{\rm obs}$ is the observed expansion velocity.
$m=1$ indicates the earliest evolutional stage (also known as the free expansion phase or the ejecta dominated (ED) phase).
$m=0.4$ corresponds to the Sedov stage at which the \ac{snr} continues to expand self-similarly.
As the \acp{snr} evolve from $m=1$ to 0.4, $\eta$ decreases to reach $\sim$1 in \figref{fig:Bohm_result_brighest} (right).
We also fit the observed $m$--$\eta$ diagram using an experimental model given by
\begin{eqnarray}
\eta = C_m m^{\delta_m} .
\label{eq:mEta}
\end{eqnarray}
The best-fit values were $C_{\rm m} = 5.2 \pm 1.0$ and $\delta_{\rm m} = 4.0 \pm 0.8$. % with $\chi^2$ of 39 and \ac{dof} of 10.
The correlation coefficient of the logarithm plots in \figref{fig:Bohm_result_brighest} (right) was approximately 0.60 with a significance of 1.9$\sigma$.
Using the average value of the $\eta$ parameter, we obtained $C_{\rm m} = 3.3 \pm 0.4$ and $\delta_{\rm m} = 0.67 \pm 0.34$ and a correlation coefficient of 0.52 with a significance of 1.6$\sigma$.

%\textcolor{red}{
Here we summarize the results of 6 SNRs (G1.9$+$0.3, Cassiopeia A (the SE and NE regions), SN~1006 (the NE and SW regions), \rxj, Vela Jr., and SN~1987A) which were analyzed with the available \nustar\ data.
Although the analysis using only the \chandra\ is reliable to some extent, 
the \nustar\ observation in the higher X-ray domain enables us to derive the more robust result.
When we reduce to those 6 SNRs, the best-fit parameters are $(C_{\rm age},\ \delta_{\rm age}) = (1.5\pm 0.3, 0.41\pm 0.09)$ with $\chi^2$ of 16 (\ac{dof} of 3) 
and $(C_m,\ \delta_m) = (5.3\pm 0.7, 7.7\pm 1.2)$ with $\chi^2$ of 6.1,
which are roughly consistent with those derived by using all SNRs.
%}

%\input{make_FigTab/Bohm_result_brightest.tex}

%\clearpage
%%%%%%%%%%%%%%%%%%%%%%%%%%%%%%%%%%%%
\subsubsection{Supernova explosion type} \label{sec:Bohm_result_type}

In this subsection, we present the $\eta$ values from two types of \ac{sne}:
Type Ia SN which is driven by a thermonuclear explosion of a white dwarf star and Type II which is a core-collapse explosion.
In general, the ambient interstellar medium of a Type Ia \ac{snr} is rarefied,
whereas the circumstellar medium surrounding a core-collapse \ac{snr} is clumpy and complex due to stellar wind from the massive progenitor star.
%Thus one might expect that the pre-existing complexity of Type II makes it  shorter to produce magnetic field turbulence, resulting in smaller $\eta$ values.

\begin{figure*}
\begin{center}
\plottwo{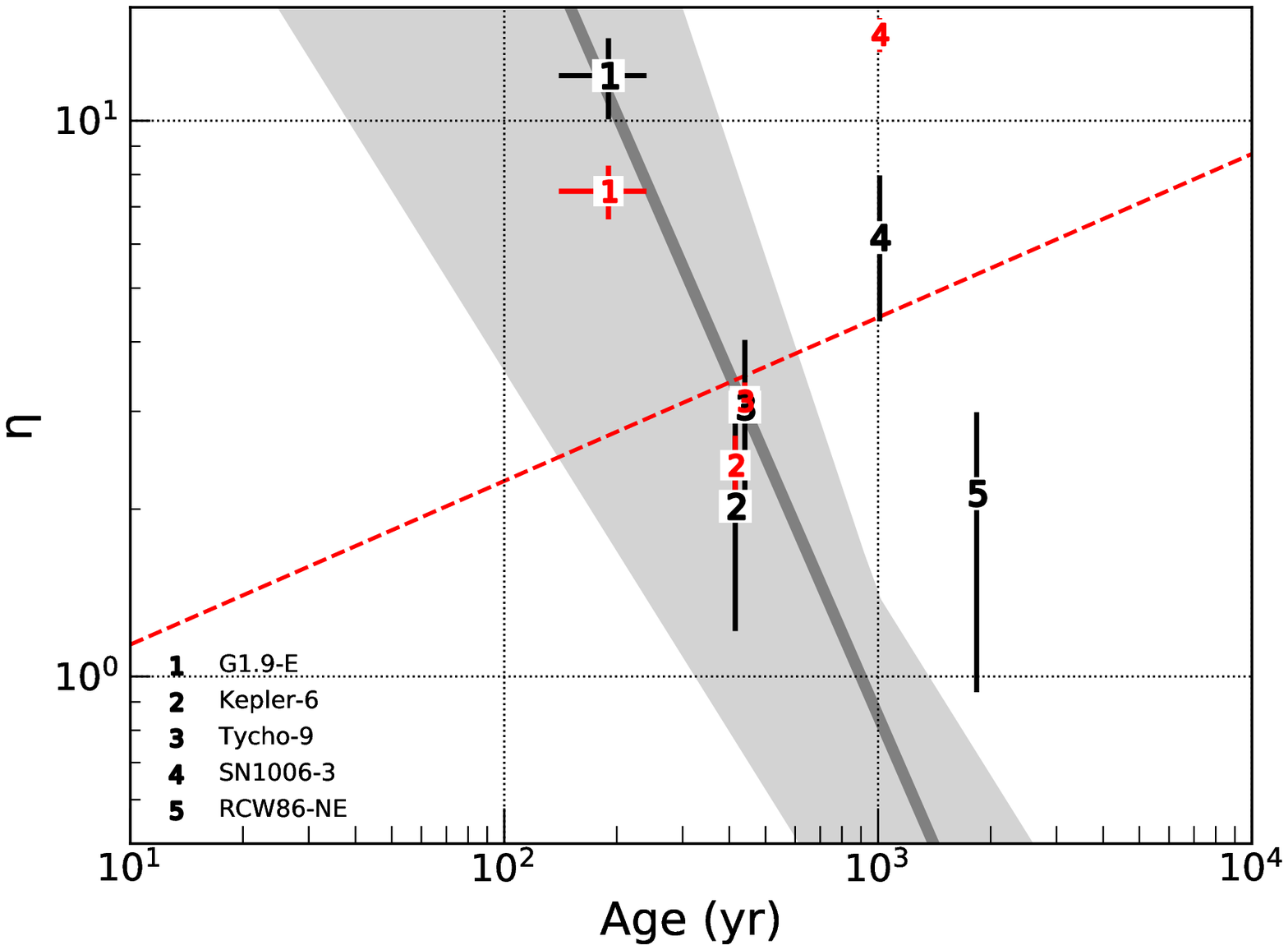}{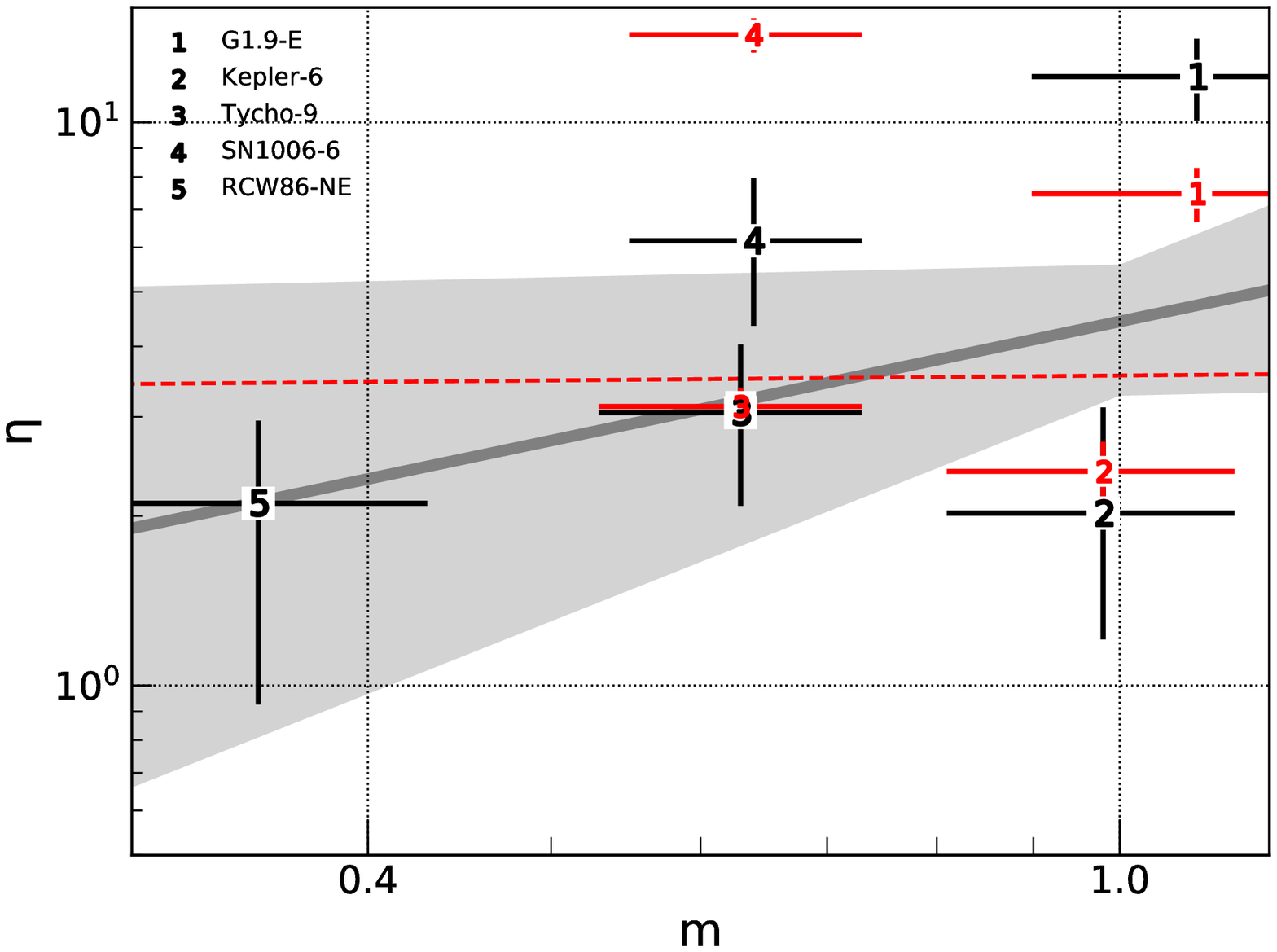}
\plottwo{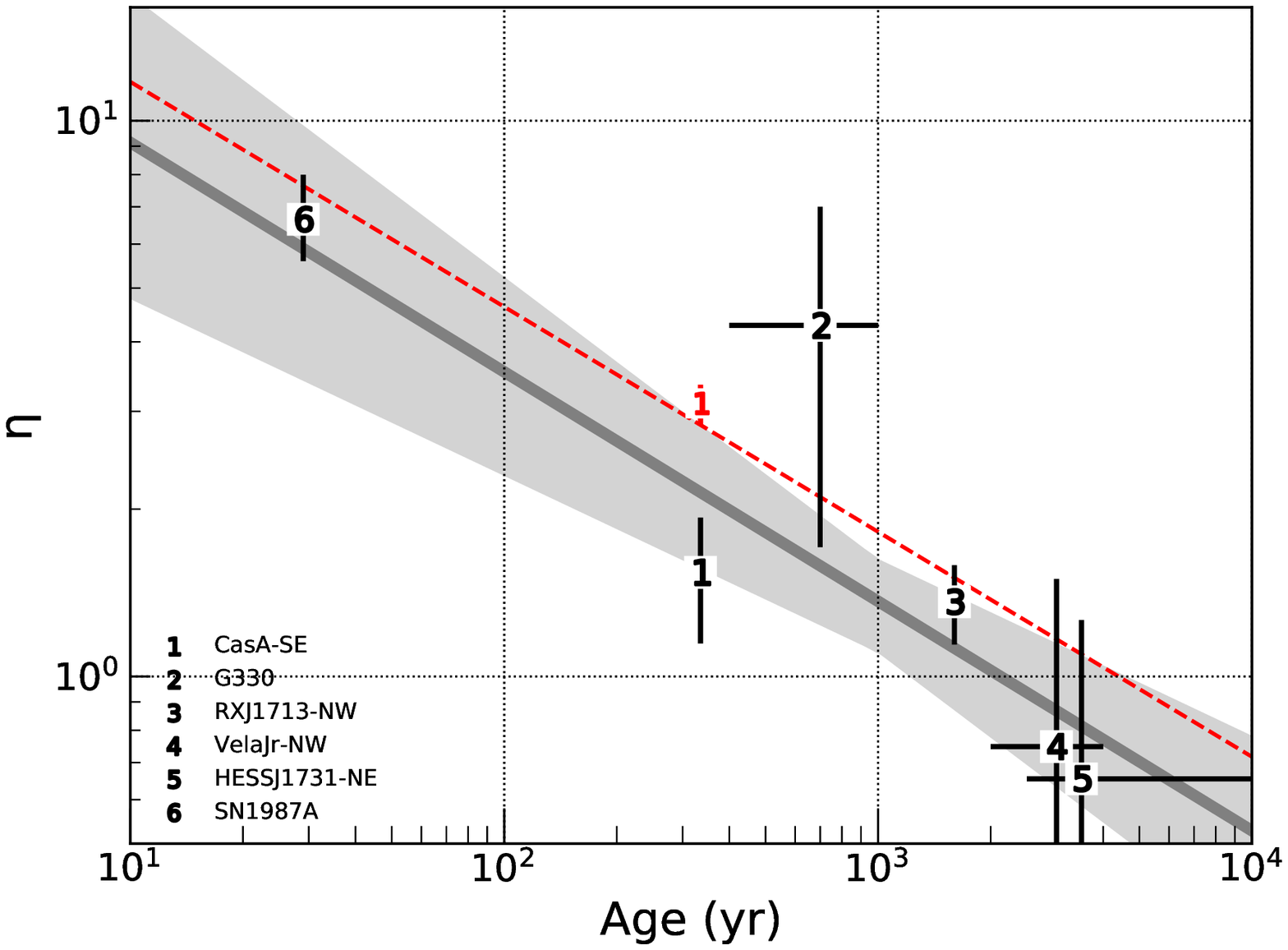}{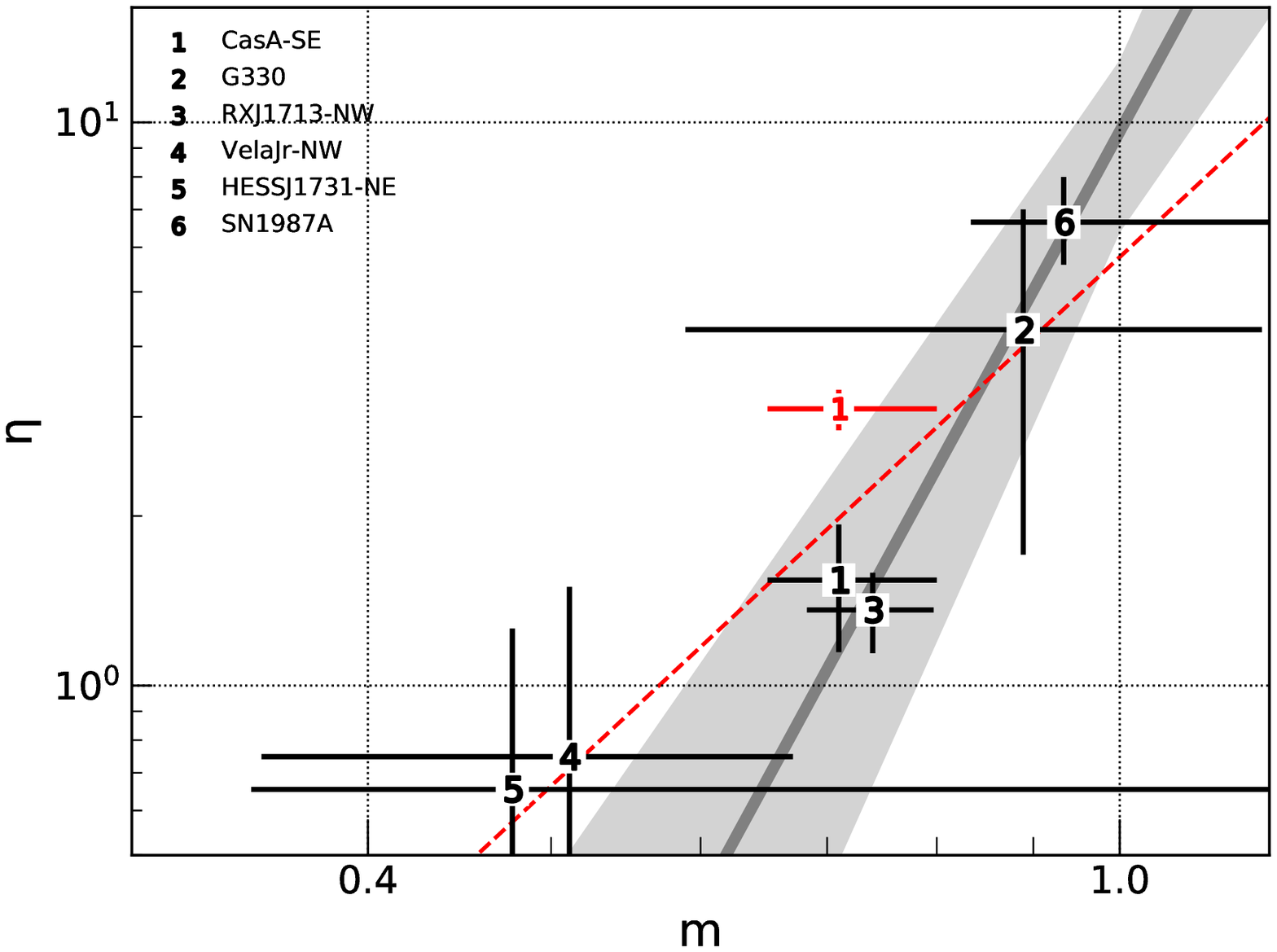}
\caption{
Same as \figref{fig:Bohm_result_brighest} for Type I (top) and Type II (bottom).
}
\label{fig:Bohm_result_type}
\end{center}
\end{figure*}

In this paper, G1.9$+$0.3, Kepler, Tycho, SN~1006, and RCW 86 are classified as Type Ia,
and Cassiopeia A, G330.2$+$1.0, \rxj, Vela Jr., HESS J1731$-$347, and SN~1987A are classified as core-collapse \acp{snr}.
We group \acp{snr} with \acp{cco} as Type II.
It should be noted that this might not be a precise classification.
\figref{fig:Bohm_result_type} shows age--$\eta$ and $m$--$\eta$ diagrams of Type Ia (top) and Type II (bottom).
The best-fit parameters are $(C_{\rm age},\ \delta_{\rm age}) = (0.85\pm 0.55, 1.5\pm 0.5)$ for Type Ia 
and $(C_{\rm age},\ \delta_{\rm age}) = (1.4\pm 0.3,\ 0.41\pm 0.09)$ for Type II.
%The core-collapse \acp{snr} have a flatter slope and a higher $\eta$ value at 1 kyr.
If we took the averaged $\eta$, $(C_{\rm age},\ \delta_{\rm age}) = (4.4\pm 0.6,\ -0.29\pm 0.17)$ for Type Ia and $(C_{\rm age},\ \delta_{\rm age}) = (1.8\pm 0.2, 0.40\pm 0.07)$ for Type II.
%For Type Ia \acp{snr}, over 10,000 years are required for $\eta$ to reach unity, whereas $\sim$3000 years are required for Type II \acp{snr}. 
As seen in \figref{fig:Bohm_result_type} and \tabref{tab:eta}, the observed $\eta$ value of Type Ia SNRs does not have a significant correlation with the age nor the expansion parameter, while Type II SNRs show the relatively strong correlation.
Overall, $\eta$ of core-collapse \acp{snr} appears smaller than that of Type Ia \acp{snr} at $t_{\rm age} \lesssim 10000$ years.
We should be careful regarding the small number of data samples and the fact that Type II \acp{snr} presented here consist of relatively older \acp{snr}.

%%%%%%%%%%%%%%
%%%%%%%%%%%%%%

%% eta-* table
%%%%%%%%%%%%%%
\input{make_FigTab/EvolvingEta_table.tex}

%\clearpage
\section{Discussion} \label{sec:discussion}
%%%%%%%%%%%%%%%%%%%%%%%%%%%%%%%%%%%%%%%%%%%%%%%%%%%%%%%%%%%%
%%%%%%%%%%%%%%%%%%%%%%%%%%%%%%%%%%%%%%%%%%%%%%%%%%%%%%%%%%%%

\subsection{Variety in $v_{\rm sh}$--$\varepsilon_0$ } \label{sec:discussion_vshe0}
%%%%%%%%%%%%%%%%%%%%%%%%%%%%%%%%%%%%%%%%%%%%%%%%%%%%%%%%%%%%

%%%%%%%%%%%%%%%%%%%%%%%%%%%%%%%%%%%%
%\subsubsection*{Summary}

\begin{figure}
\begin{center}
%%    >>> run anaZA07synch.py 6
%\includegraphics[height=6cm]{\pathwork/Vshe0/Vshe0_case2_summary_20191127.eps}
\plotone{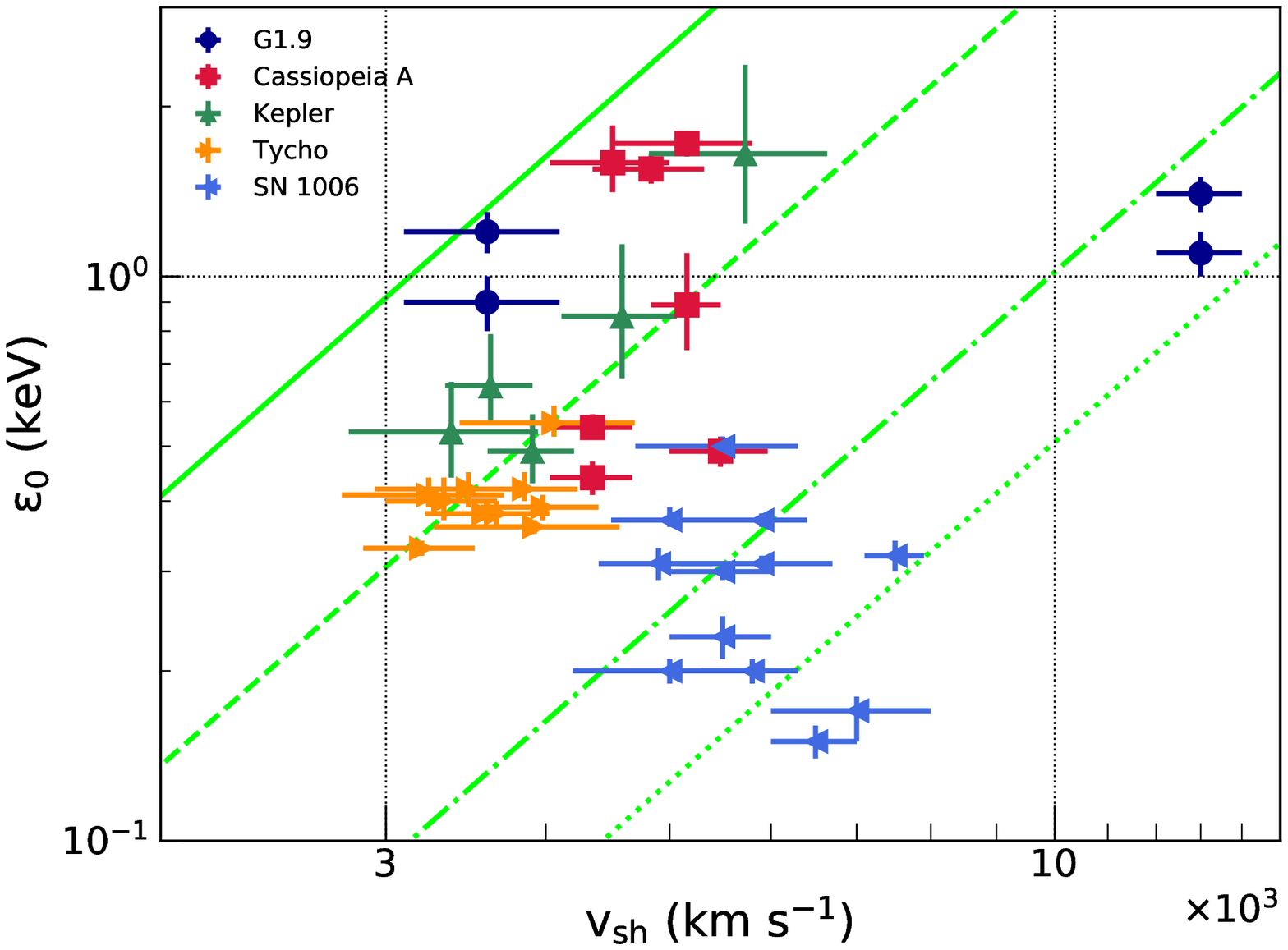}
\caption{
$v_{\rm sh}$--$\varepsilon_0$ scatter plots of the synchrotron-dominated regions in G1.9$+$0.3, Cassiopeia A, Kepler, Tycho (expect for {\it knot g}), and SN~1006. 
Green solid, dashed, dash-dotted, and dotted lines indicate $\eta$ of 1, 3, 10, and 20, respectively.
%SN~1006 does not include the subregions defined in \cite{Li2018}.  The results of the synchrotron dominated regions are presented for Kepler and Tycho. To be updated\check
}
\label{fig:Bohm_result_summary}
\end{center}
\end{figure}

First, we summarize the properties of particle acceleration in the individual sources. Significant variations across the remnants were confirmed in G1.9$+$0.3, Cassiopeia A, Kepler, Tycho, and SN~1006  \citep[see also][for the NW of \rxj]{Tsuji2019}. 
Spatially resolved studies could not be conducted in the other \acp{snr} due to lacking the observational data. 
\figref{fig:Bohm_result_summary} shows the $v_{\rm sh}$--$\varepsilon_0$ scatter plots of the former five \acp{snr}, which showed significant varieties,  indicating four types of $v_{\rm sh}$--$\varepsilon_0$ dependencies.
In G1.9$+$0.3, the cutoff energy parameters showed small variations, but the shock speeds were different from region to region.
This was opposite to the case of Cassiopeia A, which the cutoff energy parameters were variable, whereas the shock speeds were nearly constant.
Only Kepler and Tycho (expect for {\it knot g}) presented the $v_{\rm sh}$--$\varepsilon_0$ scatter plots nicely described with the theoretical prediction of $\varepsilon_0 \propto v_{\rm sh}^2$. 
Finally, SN~1006 showed a slightly inverse trend, in which the cutoff energy parameters tended to be smaller in the higher shock-speed regions.

These features were interpreted as follows.
Particle acceleration in Kepler and Tycho exhibited the theoretically predicted $v_{\rm sh}$--$\varepsilon_0$ relation. Therefore, the standard framework was applicable in these two \acp{snr}.
However, in the other \acp{snr} in which the observed $v_{\rm sh}$--$\varepsilon_0$ plots were not described by the theoretical curve, particle acceleration may have been strongly affected by the surrounding environment.
In the case of SN~1006, particle acceleration was probably determined by the ambient magnetic field, because the resulting Bohm factors were correlated with the shock obliquities  \citep[see \figref{fig:Bohm_result_SN1006} and also][]{Miceli2009}.
For G1.9$+$0.3, the different shock speeds may have been caused by differences in the ambient density, resulting in variable Bohm factors across the remnant.
It is more likely that G1.9$+$0.3, at the age of $\sim$190 years, is too young to accelerate sufficiently energetic particles, and our framework of the cooling-limited electron is not appropriate. Instead, the particle acceleration in G1.9$+$0.3 might be limited by the age.
%As already reported in \cite{Tsuji2019}, the NW rim of \rxj\ showed the maximum rate ($\eta \approx 1$) of acceleration in the forward-shock regions and the invalid values of $\eta < 1$ in the slower-shock regions. The latter case may imply that the X-ray emission was not attributed to the acceleration site but rather to the enhancement of the magnetic field.
Finally, in the case of Cassiopeia A, the acceleration efficiency might be related to thermal emission \citep{Tsuji_PhD}. The investigation of the nonthermal and thermal components will be discussed in the separated paper.

%%%%%%%%%%%%%%

%%% individual interpretation
%\input{SNRs_nonthermal_individual_vshe0.tex}

%\clearpage
\subsection{Evolution of $\eta$}  \label{sec:evolution_eta}
%%%%%%%%%%%%%%%%%%%%%%%%%%%%%%%%%%%%%%%%%%%%%%%%%%%%%%%%%%%%

%% interpretation 
The evolution of acceleration efficiency, as previously mentioned and shown in \figref{fig:Bohm_result_brighest}, was observationally revealed for the first time in this study.
%In the following, we discuss how to interpret the observational results.
The Bohm factor $\eta$ characterizes the diffusion coefficient and is explicitly related to the magnetic turbulence.
$\eta$ can also be described as:
\begin{eqnarray}
\eta \equiv \left( \frac{B_0}{\delta B} \right)^2  ,
\end{eqnarray}
where $B_0$ and $\delta B$ are the initial background magnetic field and the turbulent magnetic field, respectively.
A higher value of $\eta$ (i.e., a larger diffusion coefficient) implies that insufficient turbulent magnetic field to scatter the particles exists.
A lower value of $\eta$ (i.e., a smaller diffusion coefficient) implies that the magnetic field is sufficiently turbulent.
The observed age--$\eta$ plots may suggest generation of the turbulent magnetic field is related to the evolution of \acp{snr}:
\figref{fig:Bohm_result_brighest} implies the turbulence of magnetic field growing with time.
Here we address some open issues, such as a quantitative explanation and a physical meaning of the growth index of $-\delta$.

%% eta -- Mach 
The acceleration efficiency depends on a Mach number of the shock wave, as studied in \cite{Caprioli2014a,Caprioli2014b,Caprioli2014c} using numerical simulations.
\cite{Caprioli2014c} derived the relation (Eq. (16) therein), 
\begin{eqnarray}
\eta \propto {\cal M}^{-1/2}     ,
\label{eq:Bohm_eta_M}
\end{eqnarray}
where ${\cal M}$ is the Mach number of the \ac{snr} shock.
However, the Mach number decreases with the evolution of \acp{snr} because the shock is decelerated, and $\eta$ increases following \eqref{eq:Bohm_eta_M}.
Thus, the theoretical ${\cal M}$--$\eta$ relation predicts the opposite tendency of the observation.

%% eta -- obliquity 
The shock obliquity ($\theta_{\rm Bn}$) also plays a critical role in particle acceleration in \acp{snr} \citep{Petruk2011,Caprioli2014a}.
Whether the acceleration occurs at parallel ($\theta_{\rm Bn} \sim 0$\degr) or perpendicular ($\theta_{\rm Bn} \sim 90$\degr) shocks remains controversial.
Some numerical calculations have suggested a quasi-parallel shock produces particle injection more efficiently and generates a more amplified magnetic field \citep[e.g.,][]{Caprioli2014a}.
They found that acceleration efficiency ($\varepsilon_{\rm cr}$)\footnote{$\varepsilon_{\rm cr}$ is defined as the fraction of the postshock energy density of particles with energies greater than a certain threshold energy. This is different from our definition of acceleration efficiency with the Bohm factor, $\eta$.} drastically drops off for $\theta_{\rm Bn} \geq$45\degr.
SN~1006 is the only target to investigate the dependence on shock obliquity, as we have known the direction of the field.
Although one should be cautious about the different definitions of acceleration efficiency, our result also indicates inefficiency (larger $\eta$) above $\sim$50\degr\ for shock obliquity in SN~1006, as shown in \figref{fig:Bohm_result_SN1006}.
The observed gradual decrease of acceleration efficiency for increasing $\theta_{\rm Bn}$ might be consistent with the simulation result with a higher Mach number, for example ${\cal M}=50$.
From a morphological point of view, G1.9$+$0.3 and Vela Jr. resemble bilateral structures similar to SN~1006, but shock obliquities are unknown.
If the initial ambient field is aligned, the acceleration efficiency could depend on the shock obliquity in the earlier phase, as confirmed in SN~1006. 
In the later evolution stage or in the preexisting randomly turbulent field, the shock obliquity is averaged over the outer rim, resulting in efficient acceleration.

%% non-linear effect
Another possible explanation for the mismatch between the measured and theory predicted values of $\eta$ is a nonlinear modification of the shock profile.
In the test-particle limit, the shock front is presumed to be a sharp jump.
Considering the backreation of accelerated particles, the jump profile is modified, and this may have a considerable impact on the acceleration regime %\sout{the configuration of acceleration becomes different}  
\cite[see, for example,][for a review of \acp{snr} and references therein]{Reynolds2008_SNRreview}. 
For fast shock waves in young \acp{snr}, this nonlinear effect is not important.
However, for slow shocks in older \acp{snr}, the nonlinear effect becomes non-negligible, which might change the acceleration efficiency.
%Add any references or comments\check.

%% selection effect
It should be noted that the observed age--$\eta$ plot is somewhat biased because of a selection problem in systematical analyses.
There might exist \acp{snr} with larger Bohm factors at the older ages (a few 1,000 years), but their synchrotron emission cannot be detected due to the inefficient acceleration.
At that point, the age--$\eta$ plot here is considered as the lower limit of $\eta$ as a function of age.
%\sout{Our analysis determined that younger \acp{snr} (of a few 100 years) exhibiting  efficient acceleration with $\eta\sim 1$ are hardly present.} 
Our analysis disfavours efficient acceleration with $\eta\sim 1$ in  young \acp{snr} (of a few 100 years).% except for the effective acceleration at the lower density region in Cassiopeia A.

\if0
%% age (m) -Vsh and e0
(Need this part\check) \figref{fig:Bohm_result_brighest_2} shows the cutoff energy parameters and shock velocities taken from the same regions in \figref{fig:Bohm_result_brighest} as a function of age (and $m$).
Although the decreasing $v_{\rm sh}$ (shock deceleration) is naturally expected, the $\varepsilon_0$ does not show any correlation with time.
Moreover, the shock speed varies by a factor of $\sim$6 at most, which should generate a difference in $\varepsilon_0$ by $\sim$36 if the $\eta$ parameter is constant.
The observed cutoff energy parameters range from 0.3 to 2 keV, corresponding to the difference of a factor of $\sim$7.
This inconsistency is also supportive of the time dependence on $\eta$, which is determined by the balance between $\varepsilon_0$ and $v_{\rm sh}$.
\fi

%%%%%%%%%%%%%%%%%%%%%%%%%%%%%%%%%%%%
\subsubsection{Maximum attainable energy  }%\sout{: Are \acp{snr} PeVatrons?}  }
\label{sec:Bohm_result_pevatron}

%The maximum energy attainable in \acp{snr} was presented in \secref{sec:review_acc}.
Evolving acceleration efficiency, as confirmed in \figref{fig:Bohm_result_brighest}, would modify a picture of the maximum attainable energy of accelerated particles in \acp{snr}.
Because the energy loss timescale for protons is significantly longer than the age of \ac{snr} at the earlier evolutional phase, the maximum energy of the proton is expected to be limited by its age.
With the governing equation of $\tau_{\rm acc} = \tau_{\rm age} $,
the maximum energy of the accelerated proton is derived as
\begin{eqnarray}
E_{\rm max,age} = \frac{3}{20} \frac{q}{c} t v_{\rm sh}^2 B \eta^{-1}  .
\label{eq:Bohm_pevatron_Emax0}
\end{eqnarray}
We simply assume that the shock velocity is constant in the ED phase and is given by a self-similar solution in the Sedov-Taylor (ST) phase:
\begin{eqnarray}
v_{\rm sh}(t) \propto 
  \left\{
  \begin{array}{ll}
        t^0 &  {\rm (ED)}  \\
        t^{-3/5} &  {\rm (ST)} .
  \end{array}
  \right.
  \label{eq:Bohm_pevatron_vsh}
\end{eqnarray}
Note that sophisticated analytical solutions to express a smooth connection between the ED and ST phases have been well studied \citep[e.g.,][]{TM99,Tang2017}.
Since the evolution of the magnetic field is not well understood, we presume that $B$ is dependent on the evolutional age as
\begin{eqnarray}
 B(t) \propto t^{-\mu} .
\end{eqnarray}
Our analysis (see in \figref{fig:Bohm_result_brighest}) suggests that the $\eta$ value also depends on the time evolution of the \ac{snr}. Thus, we apply 
\begin{eqnarray}
\eta (t) \propto t^{-\delta} .  \label{eq:Bohm_eta_age2}
\end{eqnarray}
We assume that this empirical relation holds until $\eta$ becomes unity with $\delta$ being $\sim 0.4$ (\eqref{eq:Bohm_eta_age}).
Substituting into \eqref{eq:Bohm_pevatron_Emax0}, we obtain time dependence on the maximum energy of the proton:
\begin{eqnarray}
E_{\rm max,age} \propto 
  \left\{
  \begin{array}{ll}
        t^{1 -\mu +\delta}   & {\rm (ED)}\\
        t^{-1/5 -\mu +\delta}  & {\rm (ST)}    
  \end{array}
  \right. .
  \label{eq:Bohm_pevatron_Emax}
\end{eqnarray}
\eqref{eq:Bohm_pevatron_Emax} insists that $E_{\rm max,age}$ can be greater than expected before because of the newly added term of $\delta$.
Our result, $\delta \sim 0.4$, suggests that $E_{\rm max,age}$ increases as $t^{0.2}$ even in the ST stage on the assumption of $\mu=0$.

\if0
%% case of Tycho 
%% !!! correction: Ep = 44 *(2.3 /kpc)^2 TeV
Let us demonstrate the prediction of Equation~(\ref{eq:Bohm_pevatron_Emax}) for the case of Tycho \ac{snr}. %(could be Cas A with recent VERITAS paper with the averaged $\eta$ of 3\check).
The GeV gamma-ray spectrum from this \ac{snr} seems to show a \pizero\ bump that is a characteristic feature of the hadronic scenario. The TeV gamma-ray spectrum shows exponential cutoff at $\sim$2 TeV, which roughly translates into the proton spectrum cutoff energy of \(44\)~TeV \citep{VERITAS2017_tycho}.
Tycho is obviously not a PeVatron in the current stage with its age of 440 years.
Our measurements showed that the averaged $\eta$ of Tycho is about 3.
If we apply $\delta = 0.4$ and $\mu=0$ for \eqref{eq:Bohm_pevatron_Emax}, the maximum energy of the proton becomes 1.5 PeV at $\sim$5000 years when the $\eta$ value reaches 1, assuming it is still in the ED phase.
This maximum energy is approximately three times greater than that in the same later stage with $\delta=0$, which has been sometimes assumed.
The time evolution of $\eta$ suggests that \acp{snr} may accelerate particles up to the PeV range.
\fi

%% case of Cas A 
Let us demonstrate the prediction of \eqref{eq:Bohm_pevatron_Emax} for the case of Cassiopeia A. %(could be Cas A with recent VERITAS paper with the averaged $\eta$ of 3\check).
The GeV gamma-ray spectrum from this \ac{snr} seems to show a \pizero\ bump that is a characteristic feature of the hadronic scenario. 
The gamma-ray spectrum shows exponential cutoff at $\sim$2 TeV, which roughly translates into the proton spectrum cutoff energy of \(20\)~TeV \citep{VERITAS2020_CasA}.
Cassiopeia A is obviously not a PeVatron in the current stage with its age of $\sim$300 years.
Our measurements showed that in Cassiopeia A $\eta$ varies from 1 to 6 depending on the site, and the averaged $\eta$ is about 3.
We adopt the averaged value here since the gamma-ray observation could not spatially resolve the exact location of the gamma-ray emission.
%If we apply $\delta = 0.4$ and $\mu=0$ for \eqref{eq:Bohm_pevatron_Emax}, the maximum energy of the proton becomes 1 PeV at $\sim$5000 years when the $\eta$ value reaches 1, assuming it is still in the ED phase. This maximum energy is approximately three times greater than that in the same later stage with $\delta=0$, which has been sometimes assumed.
%\textcolor{red}{
If we apply $\delta = 0.4$ and $\mu=0$ for the ED (ST) phase of \eqref{eq:Bohm_pevatron_Emax}, the maximum energy of the proton becomes $\sim$400 TeV ($\sim$30 TeV) at $\sim$3000 years when the $\eta$ value reaches 1 according to \figref{fig:Bohm_result_brighest} and \eqref{eq:Bohm_eta_age}. 
This maximum energy is approximately twice as large as that in the same later stage with $\delta=0$, which has been sometimes assumed.
The time evolution of $\eta$ may suggest that there would be a possibility for \acp{snr} to become more efficient accelerators of multi-TeV particles at the late stage of $\gtrsim$1 kyr. This can be consistent with the evolution of the gamma-ray radiation, which the $\sim$kyr \acp{snr} appeared to be strong TeV gamma-ray emitters \citep{Funk2015}.
However, it should be emphasised that the assumption above is the most optimistic, and that the maximum energy would be predominantly dependent on other parameters, such as the shock velocity, the magnetic field, and the condition of escape. 
%}

%% issues; future work
%Defining when exactly an \ac{snr} enters into the ST phase is difficult. \mitya{Is that a correct statement? I think it is rather straightforward to estimate} The transition between the ED and ST stages is most likely smoother than as described in \eqref{eq:Bohm_pevatron_vsh}.
Observational studies of $\eta$ in the later stage of \acp{snr} remain as a future work.
The \acp{snr} we analyzed in this study included those in both ED and ST phases, but most them were relatively young and might be in the very transition stage from ED to ST.
Altogether, they indicate the Bohm factor decreases until it becomes close to unity.
Our work could not fully determine the behavior of $\eta$ in the later phase, particularly the phase that has significantly entered into the ST stage.
It is also interesting but challenging to measure the $\eta$ value of escaping particles from the SNRs.

%%%%%%%%%%%%%%%%%%%%%%%%%%%%%%%%%%%%
\subsubsection{Supernova explosion type}
\label{sec:discussion_Bohm_result_type}

\figref{fig:Bohm_result_type} showed that Type II \acp{snr} tend to have the lower value of $\eta$. %, although the difference in Type I and II \acp{snr} was not significant.
Since a core-collapse (Type II) \ac{snr} has exploded in the surroundings produced by stellar wind of the massive progenitor, the ambient medium is more irregular and complex than that of Type I.
If the obtained evolution of $\eta$ represents the growth of turbulence, the complexity of the circumstellar medium of a Type II \ac{snr} is expected to make the turbulent production faster, which might be observed as smaller $\eta$ value of Type II.
This would be investigated in the future by measurements of the cutoff energy and shock speed with higher accuracy, which can help us reduce the uncertainty of $\eta$ and distinguish the difference between Type I and II \acp{snr}.

%\clearpage
\section{Conclusions} \label{sec:conclusions}
%%%%%%%%%%%%%%%%%%%%%%%%%%%%%%%%%%%%%%%%%%%%%%%%%%%%%%%%%%%%
%%%%%%%%%%%%%%%%%%%%%%%%%%%%%%%%%%%%%%%%%%%%%%%%%%%%%%%%%%%%

We analyzed X-ray observations of 11 young \acp{snr} to determine the cutoff energy parameter in the synchrotron spectrum and constrain the corresponding Bohm factor of each \ac{snr}.
Our model of synchrotron radiation is based on the framework that the accelerated electron is limited by synchrotron cooling and Bohm diffusion.
This assumption is reasonable for \acp{snr} older than 1,000 years if $B \geq  10$--$20~ \uG$ and for \acp{snr} as young as a few 100 years if $B \geq 30$--$40~ \uG$.
We should be cautious about G1.9$+$0.3 (150--190 years) and SN~1987A (30 years) because $B$ should be much larger for the assumption to be valid in these sources, 
and it is questionable for such young \acp{snr} to be capable of amplifying magnetic fields in their short lifetimes.
 
%% individual source
The $v_{\rm sh}$--$\varepsilon_0$ relations obtained for the five individual \acp{snr} are interpreted as the following cases:
\begin{enumerate}
    \setlength{\parskip}{0mm}
    \setlength{\itemsep}{0mm}
    \item The $v_{\rm sh}$--$\varepsilon_0$ scatter plot is nicely represented by the theoretical curve, $\varepsilon_0 \propto v_{\rm sh}^2 \eta^{-1}$, with constant $\eta$ throughout the remnant in the cases of Kepler and Tycho.
    \item Cassiopeia A showed variable cutoff energy parameters but nearly constant shock velocities, resulting in different acceleration efficiency depending on the sites. %The acceleration is affected by a surrounding density in the case of Cassiopeia A. The different $\eta$ values are attributed to the different cutoff energy parameters, which are more correlated with the ambient number density. The kinetic energy of shock is transferred to acceleration and heating in the lower and higher densities, respectively.
    \item The acceleration is affected by a surrounding magnetic-field configuration in the case of SN~1006: the $\eta$ parameters is small near the polar limbs where quasi-parallel shocks are expected to form. %The $\varepsilon_0$, $v_{\rm sh}$, and $\eta$ show an azimuth variation, and we can confirm that the acceleration is more efficient near the polar limbs where quasi-parallel shocks are expected to form.
    \item The different $\eta$ values are attributed to the different shock speeds in the case of the youngest \ac{snr} in our Galaxy, G1.9$+$0.3. %It should be noted that the cooling-limited assumption might not be appropriate for this remnant.
    %\item In the case of the northwest of \rxj, the particle acceleration is required to proceed in the regime close to the Bohm limit near the forward shock. However, we require another scenario to explain the higher cutoff energy than theoretically predicted in the regions with slow speeds, such as the inner edge and the filamentary structure. The parsec-scale amplification of the magnetic field and/or the acceleration at the reflection shock might be the case.
    %\item The spatially resolved $\varepsilon_0$--$v_{\rm sh}$ plots cannot be presented due to limited dataset, in the cases of G330, \rxj, RCW 86, Vela Jr., HESS J1731, and SN~1987A.
\end{enumerate}

%% systematical trend
%With all eleven \acp{snr} together, the systematical tendency of Bohm factor, for the first time, was confirmed.
With all 11 \acp{snr} together, including G330.2$+$1.0, \rxj, RCW 86, Vela Jr., HESS J1731$-$347, and SN~1987A, the systematic tendency of the Bohm factor has been unveiled for the first time.
The $\eta$ in the maximum-$\varepsilon_0$ (or maximum-$v_{\rm sh}$) region of each \ac{snr} depends on the evolutional age as $\eta = 1.5(t_{\rm age} / 1~{\rm kyr}) ^{-0.41}$ or on the expansion parameter as $\eta = 5.2 m^{4.0}$.
This can be related to the turbulent generation: the turbulence becomes more self-generated as particles become more accelerated with time.
Comparing the time dependence on $\eta$ between Types I and II supernova explosions, Type II shows a relatively lower $\eta$ value and a flatter rate of the growth, which might suggest that Type II \acp{snr} exploded in the more complex surroundings of the \ac{csm} and could facilitate the turbulence to grow faster.
%Although the difference between \ac{sne} is not significant, a lower $\eta$ of Type II might be expected because it exploded in the more complex surroundings of the \ac{csm} and could facilitate the turbulence to grow.
%We also present the evolution of $\eta$ at the reverse shock. In addition to the known reverse shock of Cassiopeia A, possible reverse shocks were reported in G1.9$+$0.3 and SN~1987A, showing $\eta$ greater than 1.
%% pevatron
Finally, if we consider the time dependence on $\eta$ as $\eta \propto t^{-\delta}$ with $\delta \approx 0.4$, which has not been expected before, and assume this condition holds until $\eta$ reaches unity even in the Sedov-Taylor phase,
the attainable maximum energy appears greater by the term of $\delta$, possibly in the multi-TeV range.

\acknowledgments

The scientific results reported in this paper are based on observations made by the \chandra\ X-ray Observatory.
%the \chandra\ X-ray Observatory Center, which is operated by the Smithsonian Astrophysical Observatory for and on behalf of the National Aeronautics Space Administration
This work was also made use of data from the \nustar\ mission, a project led by the California Institute of Technology, managed by the Jet Propulsion Laboratory, and funded by the National Aeronautics and Space Administration. 
%This work is supported by the \nustar\ Cycle 1 observation program.
We appreciate the \nustar\ Operations, Software, and Calibration teams for support with the execution and analysis of these observations.
N.T. is supported by the Japan Society for the Promotion of Science (JSPS) KAKENHI grant No. JP17J06025.
%R.K. acknowledges support from the Russian Science Foundation (grant 19-12-00369).
This work was partially supported by JSPS KAKENHI grant Nos. JP18H03722 and JP18H05463.


\vspace{5mm}

%\facilities{HST(STIS), Swift(XRT and UVOT), AAVSO, CTIO:1.3m,CTIO:1.5m,CXO}


%\software{astropy \citep{2013A&A...558A..33A}, Cloudy \citep{2013RMxAA..49..137F}, SExtractor \citep{1996A&AS..117..393B} }
% --- software
\software{NuSTARDAS (v1.4.1), HEAsoft (v6.19), nuskybgd (https://github.com/NuSTAR/nuskybgd; \cite{Wik2014}), XSPEC (v12.9.0, \cite{Arnaud1996}), CIAO (v4.9, \cite{Fruscione2006}) }



\appendix
%%%%%%%%%%%%%%%%%%%%%%%%%%%%%%%%%%%%%%%%%%%%%%%%%%%%%%%%%%%%

\section{Dataset} \label{sec:appendix_dataset}
%%%%%%%%%%%%%%%%%%%%%%%%%%%%%%%%%%%%%%%%%%%%%%%%%%%%%%%%%%%%
The dataset of archival observations of \chandra\ and \nustar\ are listed in \tabref{tab:Bohm_dataset_chandra} and \tabref{tab:Bohm_dataset_nustar}, respectively.

%%%%%%%%%%%% table
\input{make_FigTab/Bohm_dataset_Chandra_thispaper.tex}

\input{make_FigTab/Bohm_dataset_NuSTAR_thispaper.tex}

\section{Thermal parameters} \label{sec:appendix_thermal}
%%%%%%%%%%%%%%%%%%%%%%%%%%%%%%%%%%%%%%%%%%%%%%%%%%%%%%%%%%%%
\tabref{tab:Bohm_result_tab_thermal} presents parameters of the thermal model which is used for fitting the spectra of Cassiopeia A, Kepler, Tycho, SN~1006, RCW 86, and SN~1987A in addition to the synchrotron radiation model (see \tabref{tab:Bohm_result_all_in_one} for the nonthermal parameters).

\input{make_FigTab/Bohm_result_tab_thermal.tex}

\bibliography{zotero.bib}
\bibliographystyle{aasjournal}



\end{document}

%% file: mypreset.tex
\def\figref#1{Figure~\ref{#1}} %\def\figref#1{\hbox{Figure~\ref{#1}}} 
\def\tabref#1{Table~\ref{#1}} %\def\tabref#1{\hbox{Table~\ref{#1}}} 
\def\eqref#1{Equation~\ref{#1}} %\def\eqref#1{\hbox{Eq.~(\ref{#1})}} 
\def\secref#1{Section~\ref{#1}} %\def\secref#1{\hbox{Section~\ref{#1}}} 

\def\check{\textcolor{red}{?????}}

\def\pathwork{/Users/naomi/Dropbox/Documents/Tex_documents_mine/RXJ1713_nustar_ApJ/make_FigTab/fig3_vsh_e0_forSNRs/}

\def\rxj{RX J1713.7$-$3946}

\def\hessj1731{HESS J1731$-$347}

\def\chandra{{\it Chandra}}
\def\nustar{{\it NuSTAR}}
\def\xmm{{\it XMM-Newton}}
\def\suzaku{{\it Suzaku}}

\def\fermi{{\it Fermi}}

\def\dag{\hbox{$^\dagger$}}

\def\uG{\hbox{$\mu \mathrm{G}$}}
\def\flux{\hbox{${\rm erg}~{\rm cm}^{-2}~{\rm s}^{-1}$}}
\def\kms{\hbox{${\rm km}~ {\rm s}^{-1}$}}
\def\cc{\hbox{${\rm cm}^{-3}$}}
\def\columnd{\hbox{${\rm cm}^{-2}$}}  % (/cm2) column density

\def\pizero{\hbox{$\pi^0$}}

\usepackage[single=true]{acro}

\DeclareAcronym{ic}{
  short = IC ,
  long  = inverse Compton ,
  first-style = default }

\DeclareAcronym{nustar}{
  short = {\it NuSTAR} ,
  long  = Nuclear Spectroscopic Telescope Array ,
  first-style = default }

\DeclareAcronym{bi}{
  short = BI ,
  long  = backside illumination ,
  first-style = default }
\DeclareAcronym{fi}{
  short = FI ,
  long  = frontside illumination ,
  first-style = default }
\DeclareAcronym{fov}{
  short = FoV ,
  long  = field of view ,
  first-style = default }
  
\DeclareAcronym{sim}{
  short = SIM ,
  long  = Science Instrument Module ,
  first-style = default }
\DeclareAcronym{hetg}{
  short = HETG ,
  long  = High Energy Transmission Grating ,
  first-style = default }
\DeclareAcronym{letg}{
  short = LETG ,
  long  = Low Energy Transmission Grating ,
  first-style = default }
\DeclareAcronym{hrc}{
  short = HRC ,
  long  = High Resolution Camera ,
  first-style = default }
\DeclareAcronym{acis}{
  short = ACIS ,
  long  = Advanced CCD Imaging Spectrometer ,
  first-style = default }
\DeclareAcronym{hrma}{
  short = HRMA ,
  long  = High Resolution Mirror Assembly,
  first-style = default }

\DeclareAcronym{compton}{
  short = Compton ,
  long  = Compton Gamma Ray Observatory,
  first-style = default }
\DeclareAcronym{hst}{
  short = HST ,
  long  = Hubble Space Telescope ,
  first-style = default }
\DeclareAcronym{iact}{
  short = IACT ,
  long  = Imaging Atmospheric Cherenkov Telescope ,
  first-style = default }
\DeclareAcronym{wcd}{
  short = WCD ,
  long  = Water Cherenkov Detector ,
  first-style = default }

\DeclareAcronym{hawc}{
  short = HAWC ,
  long  = High-Altitude Water Cherenkov ,
  first-style = default }
\DeclareAcronym{cta}{
  short = CTA ,
  long  = Cherenkov Telescope Array ,
  first-style = default }

\DeclareAcronym{em}{
  short = EM ,
  long  = electromagnetic ,
  first-style = default }
\DeclareAcronym{ism}{
  short = ISM ,
  long  = interstellar medium ,
  first-style = default }
\DeclareAcronym{csm}{
  short = CSM ,
  long  = circumstellar medium ,
  first-style = default }
\DeclareAcronym{sne}{
  short = SNe ,
  long  = supernovae , 
  first-style = default }

\DeclareAcronym{iss}{
  short = ISS ,
  long  = International Space Station , 
  first-style = default }

\DeclareAcronym{uhecr}{
  short = UHECRs ,
  long  = ultra high energy cosmic rays , 
  first-style = default }
\DeclareAcronym{ta}{
  short = TA ,
  long  = Telescope Array , 
  first-style = default }
\DeclareAcronym{auger}{
  short = Auger ,
  long  = Pierre Auger Observatory , 
  first-style = default }
\DeclareAcronym{ams}{
  short = AMS ,
  long  = Alpha Magnetic Spectrometer , 
  first-style = default }
\DeclareAcronym{pamela}{
  short = PAMELA ,
  long  = Payload for Antimatter Matter Exploration and Light-nuclei Astrophysics , 
  first-style = default }
   
\DeclareAcronym{cmb}{
  short = CMB ,
  long  = Cosmic Microwave Background , 
  first-style = default }
\DeclareAcronym{sed}{
  short = SED ,
  long  = spectral energy distribution , 
  first-style = default }
\DeclareAcronym{mhd}{
  short = MHD ,
  long  = magnetohydrodynamical ,
  first-style = default }
\DeclareAcronym{dof}{
  short = dof ,
  long  = degree of freedom ,
  first-style = default }
\DeclareAcronym{cco}{
  short = CCO ,
  long  = central compact object ,
  first-style = default
}
\DeclareAcronym{lmc}{
  short = LMC ,
  long  = Large Magellanic Cloud ,
  first-style = default
}
\DeclareAcronym{mpik}{
  short = MPIK ,
  long  = ããã¯ã¹ãã©ã³ã¯æ ¸ç©çç ç©¶æ ,
  first-style = default
}

\DeclareAcronym{hess}{
  short = H.E.S.S. ,
  long  = High Energy Spectroscopic System ,
  first-style = default
}
\DeclareAcronym{snr}{
  short = SNR ,
  long  = supernova remnant ,
  first-style = default
}

\DeclareAcronym{sn}{
  short = SN ,
  short-plural = e ,
  long  = supernova ,
  long-plural  = e ,
  first-style = default
}

\DeclareAcronym{nw}{
  short = NW ,
  long  = northwest ,
  first-style = default
}
\DeclareAcronym{hxc}{
  short = HXC ,
  long  = hard X-ray component ,
  first-style = default
}

\DeclareAcronym{cr}{
  short = CR ,
  long  = cosmic ray ,
  first-style = default }
\DeclareAcronym{psf}{
  short = PSF ,
  long  = point spread function ,
  first-style = default 
}
\DeclareAcronym{hpd}{
  short = HPD ,
  long  = half power diameter ,
  first-style = default 
}
\DeclareAcronym{fwhm}{
  short = FWHM ,
  long  = full width of half maximum ,
  first-style = default 
}

\DeclareAcronym{pic}{
  short = PIC ,
  long  = particle-in-cell ,
  first-style = default 
}
\DeclareAcronym{cxb}{
  short = CXB ,
  long  = Cosmic X-ray Background ,
  first-style = default 
}
\DeclareAcronym{grxe}{
  short = GRXE ,
  long  = Galactic Ridge X-ray Emission ,
  first-style = default 
}
\DeclareAcronym{pa}{
  short = PA ,
  long  = Positional Angle ,
  first-style = default 
}
\DeclareAcronym{dsa}{
  short = DSA ,
  long  = diffusive shock acceleration ,
  first-style = default 
}

\def\pizero{\hbox{$\pi^0$}}

\def\degr{\hbox{$^\circ$}}
\def\arcmin{\hbox{$^\prime$}}
\def\arcsec{\hbox{$^{\prime\prime}$}}
\def\utw{\smash{\rlap{\lower5pt\hbox{$\sim$}}}}
\def\udtw{\smash{\rlap{\lower6pt\hbox{$\approx$}}}}
\def\apjl{ApJ}%
\def\apjs{ApJS}%
\def\ao{Appl.~Opt.}%
\def\aap{A\&A}%

%% file: SNRs_nonthermal_abstract.tex
% %%%%%%%%%%%%%%%%%%%%%%%%%%%%%%%%%%%
%%%%%%%%%%%%%%%%%%%%%%%%%%%%%%%%%%%%

Cutoff energy in a synchrotron radiation spectrum of an suprenova remnant (SNR)  contains a key parameter of ongoing particle acceleration.
We systematically analyze 11 young SNRs, including all historical SNRs, to measure the cutoff energy, thus shedding light on the nature of particle acceleration at the early stage of SNR evolution.
The nonthermal (syncrotron) dominated spectra in filament-like outer rims are selectively extracted and used for spectral fitting because  our model assumes that accelerated electrons are concentrated in the vicinity of the shock front due to synchrotron cooling.
The cutoff energy parameter ($\varepsilon_0$) and shock speed ($v_{\rm sh}$) are related as $ \varepsilon_0 \propto v_{\rm sh}^2 \eta^{-1}$ with a Bohm factor of $\eta$.
Five SNRs provide us with spatially resolved $\varepsilon_0$--$v_{\rm sh}$ plots across the remnants, indicating a variety of particle acceleration.
With all SNRs considered together, the systematic tendency of $\eta$ clarifies a correlation between $\eta$ and an age of $t$ (or an expansion parameter of $m$) as $\eta \propto t^{-0.4}$ ($\eta \propto m^{4}$). 
This might be interpreted  as the magnetic field becomes more turbulent and self-generated, as particles are accelerated at a greater rate with time.
The maximum energy achieved in SNRs can be higher if we consider the newly observed time dependence on $\eta$.

%%%%%%%%%%%%%%%%%%%%%%%%%%%%%%%%%%%%
%%%%%%%%%%%%%%%%%%%%%%%%%%%%%%%%%%%%

%% file: make_FigTab/Bohm_dataset_SNRs.tex
%%     
%%% ------------------------------------------------%%%
\begin{deluxetable*}{cccccc}[ht!]
\tablecaption{Properties of SNRs analyzed in this paper
\label{tab:dataset_SNRs} }
%\tablewidth{700pt}
\tabletypesize{\small}
\tablehead{
\colhead{Name} &  \colhead{Age }&    \colhead{$m$\dag}            & \colhead{Distance}    & \colhead{Shock speed} & \colhead{References} \\
\colhead{}                         & \colhead{(yr)}          &  \colhead{}    & \colhead{(kpc)} & \colhead{(\kms)}   & \colhead{} 
} 
\startdata
%%%%%%%%%%%%%%
G1.9$+$0.3         & 190$\pm$50    & 1.1$\pm$0.2  &  $\sim$8.5  &  3600--13000 & \cite{Borkowski2010,Borkowski2017} \\
Cassiopeia A  (SN 1680)     & 335$\pm$20    & 0.7$\pm$0.1  &  3.4             &  4400--5500 & \cite{Patnaude2009} \\
Kepler (SN 1604)                & 415                  & 0.47--0.82  & 4$\pm$1     &  3400--5700  & \cite{CassamChenai2004_kepler,Katsuda2008_kepler} \\
Tycho (SN 1572)                 & 440                  & 0.6$\pm$0.1  & 2.3             &  3200--4000   & \cite{Williams2013}  \\
G330.2$+$1.0     & 700$\pm$300  & 0.9$\pm$0.3  & $\sim$5     &  3700--9100& \cite{Park2006,Borkowski2018}    \\
SN 1006             & 1010                & 0.5$\pm$0.1  &  1.9$\pm$0.3 &  3000--7200  & \cite{Winkler2014} \\
\rxj\ (SN 393)                      & 1600  $\pm$10 & 0.7$\pm$0.1     & 1             &  800--4000& \cite{Tsuji2016,Acero2017} \\
RCW 86 (SN 185)             & 1835                & 0.3$\pm$0.1  & 2.8                      &  1800--3000& \cite{Rosado1996,Yamaguchi2016} \\
Vela Jr.              & 3000$\pm$1000 & 0.5$\pm$0.2      & 0.5--0.9  &  $\sim$2000  & \cite{Allen2015} \\
HESS J1731$-$347    & 2500--14000 & 0.5$\pm$0.1  & 3.6$\pm$0.4 &  $\sim$2500  & \cite{HESS2011_hessj1731} \\
SN 1987A         & 30                          & 0.9$\pm$0.1   &  51.4  &  $\sim$6700 & \cite{Frank2016}  \\
%%%%%%%%%%%%%%%%
%% end
\enddata
\tablecomments{
%\tnote{\dag} X-ray observation data used for the spectral fitting. C: \chandra. N: \nustar. S: \suzaku. \\ %X: \xmm. 
\dag\ Expansion parameter (i.e., $R \propto t^m$ where $R$ and $t$ are the radius of the SNR and time from the explosion, respectively). 
}
\end{deluxetable*}

%% file: make_FigTab/figure_images.tex
%%%%%%%%%%%%%%%%%%%%%%%%%%%%
%%%%%%%%%%%%%%%%%%%%%%%%%%%%

\begin{figure*}
\gridline{\fig{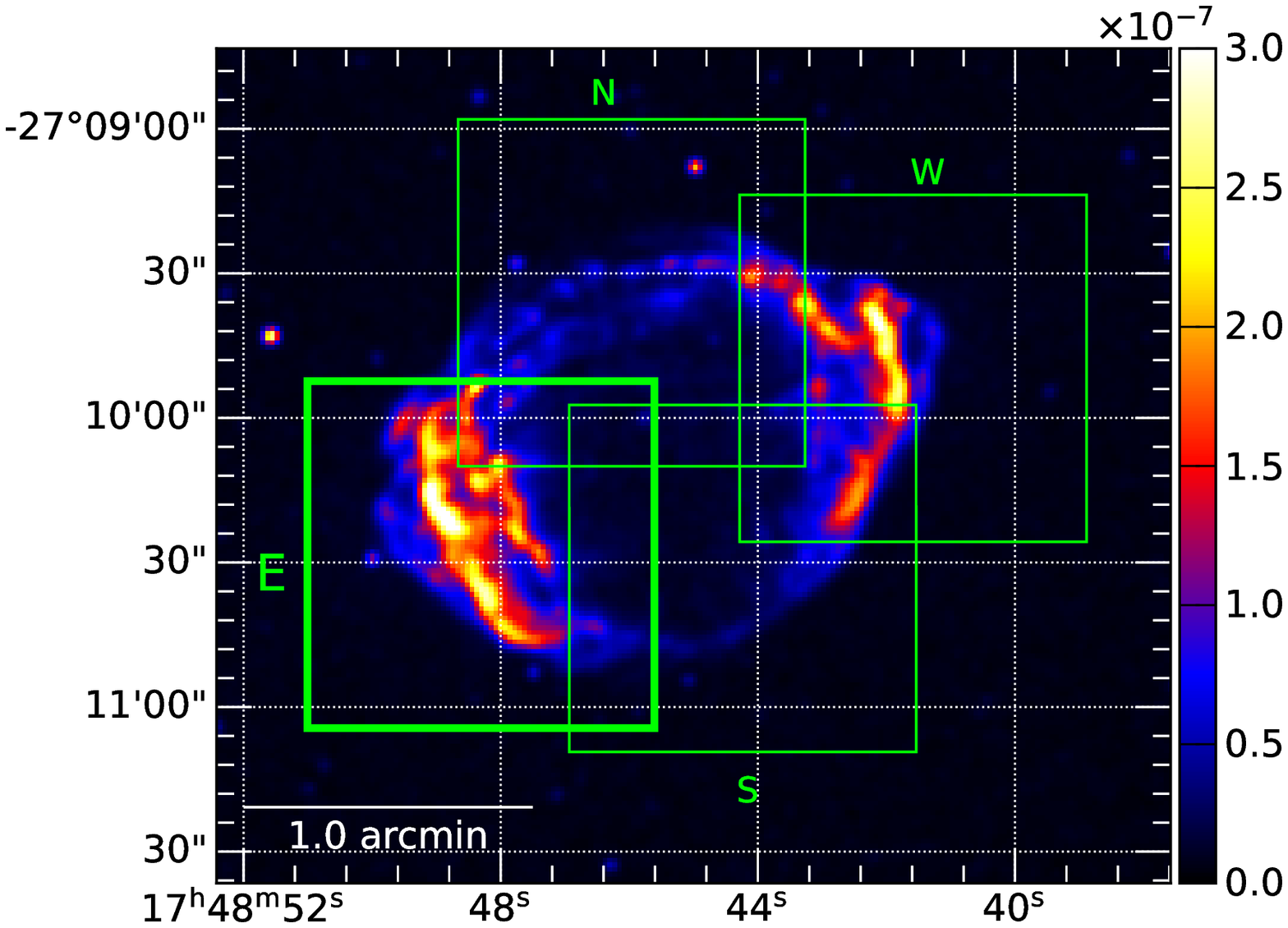}{0.32\textwidth}{G1.9$+$0.3}
          \fig{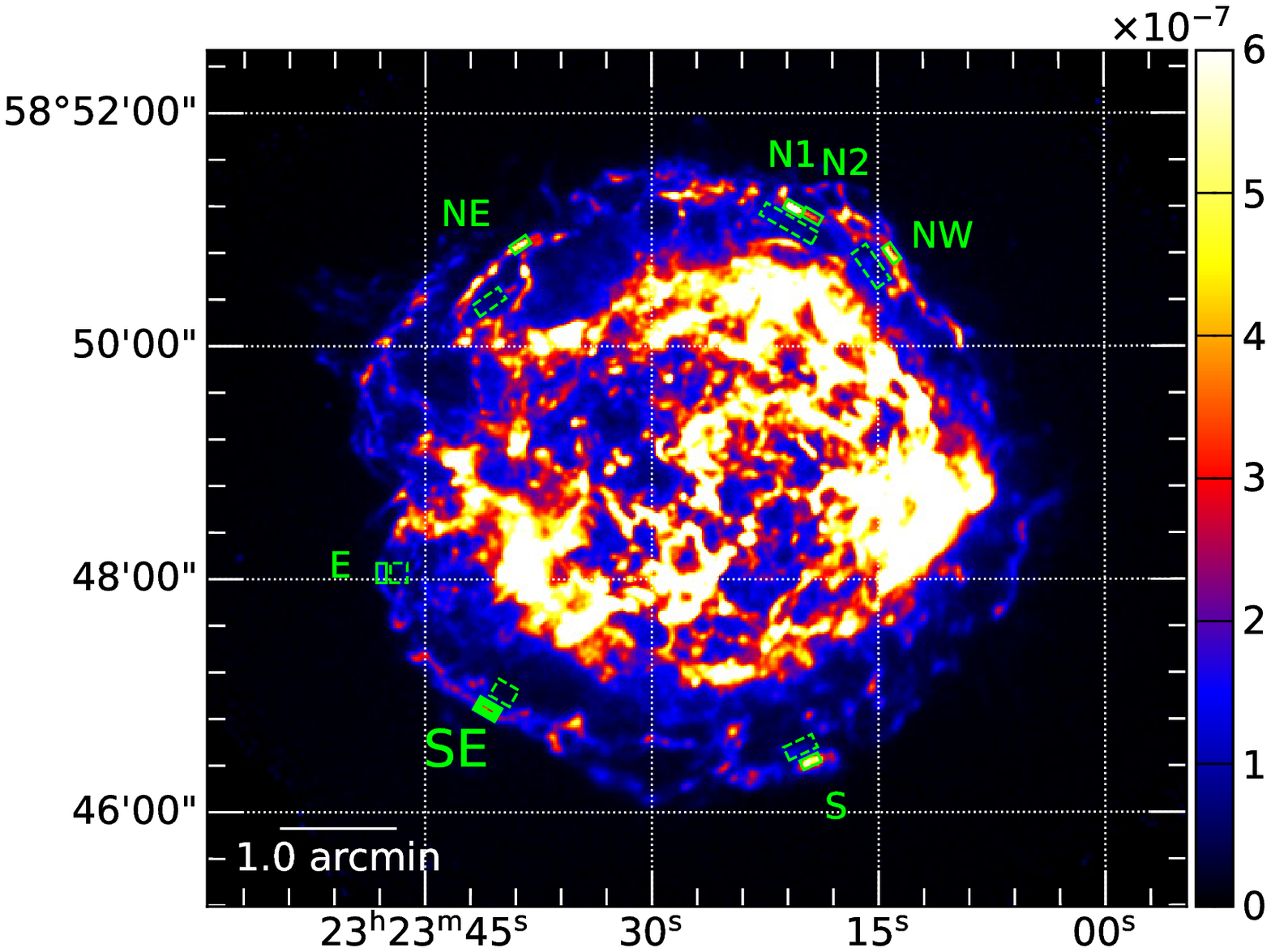}{0.32\textwidth}{Cassiopeia A}
          \fig{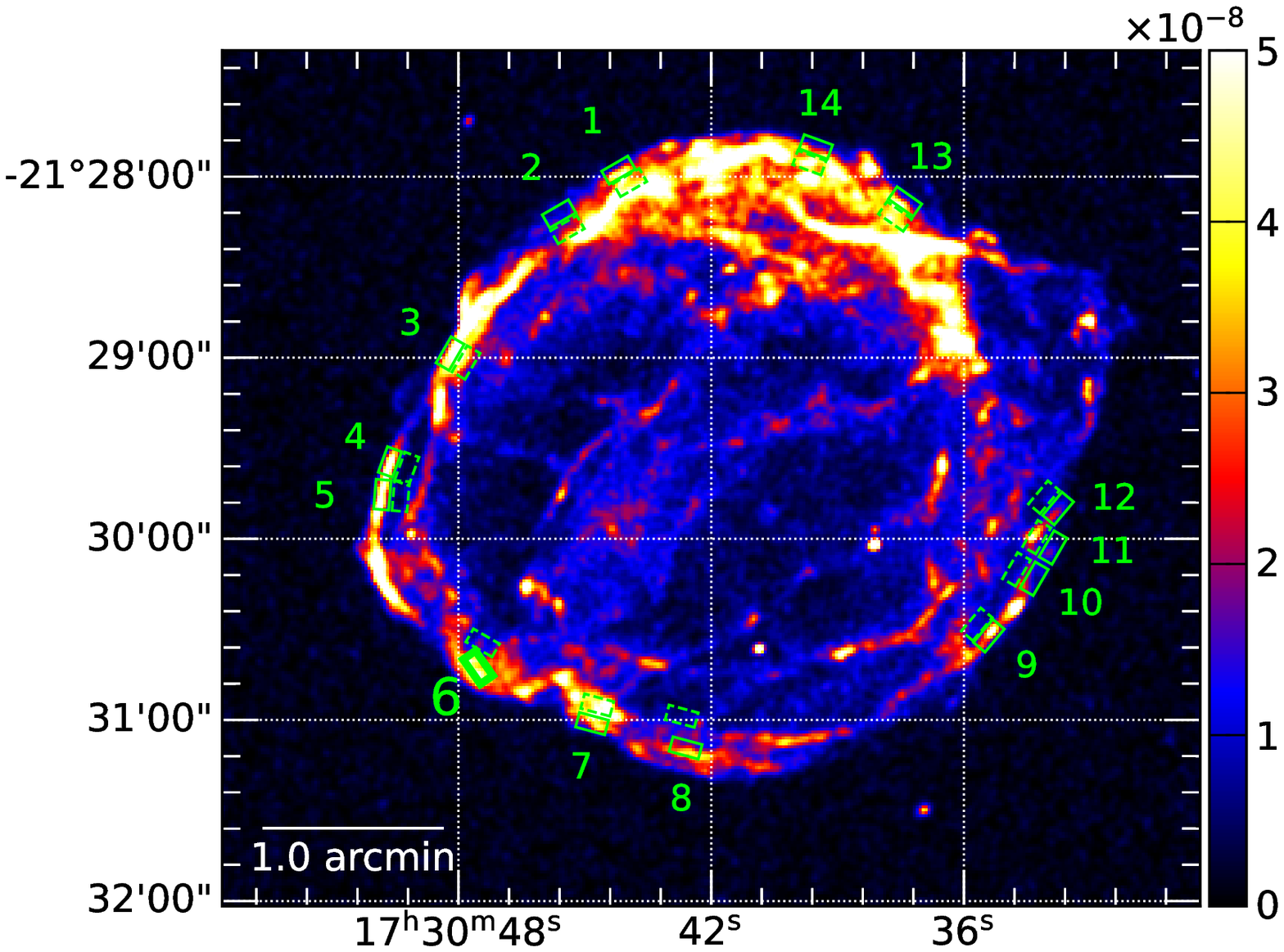}{0.32\textwidth}{Kepler}
          }
\gridline{\fig{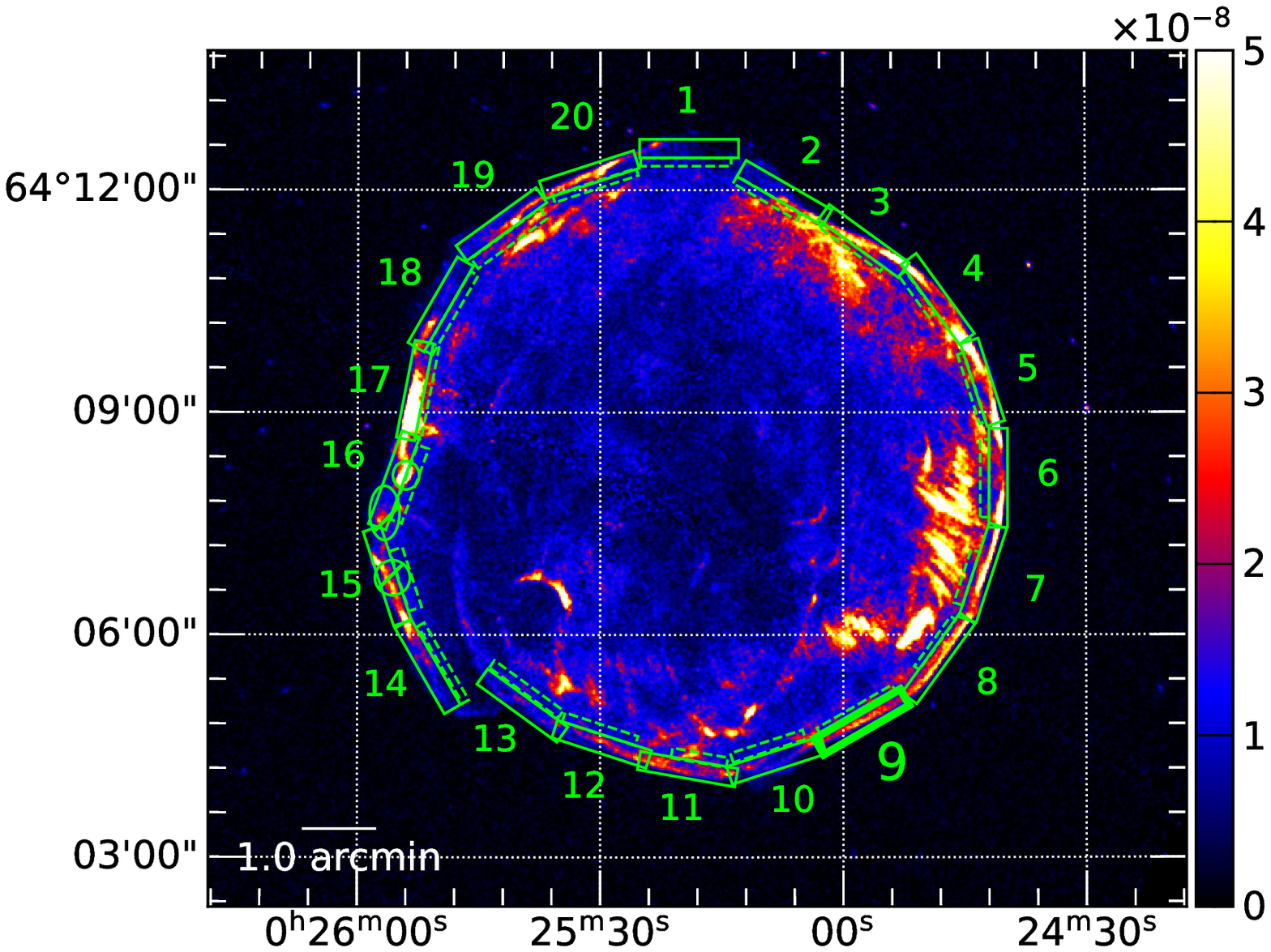}{0.32\textwidth}{Tycho}
          \fig{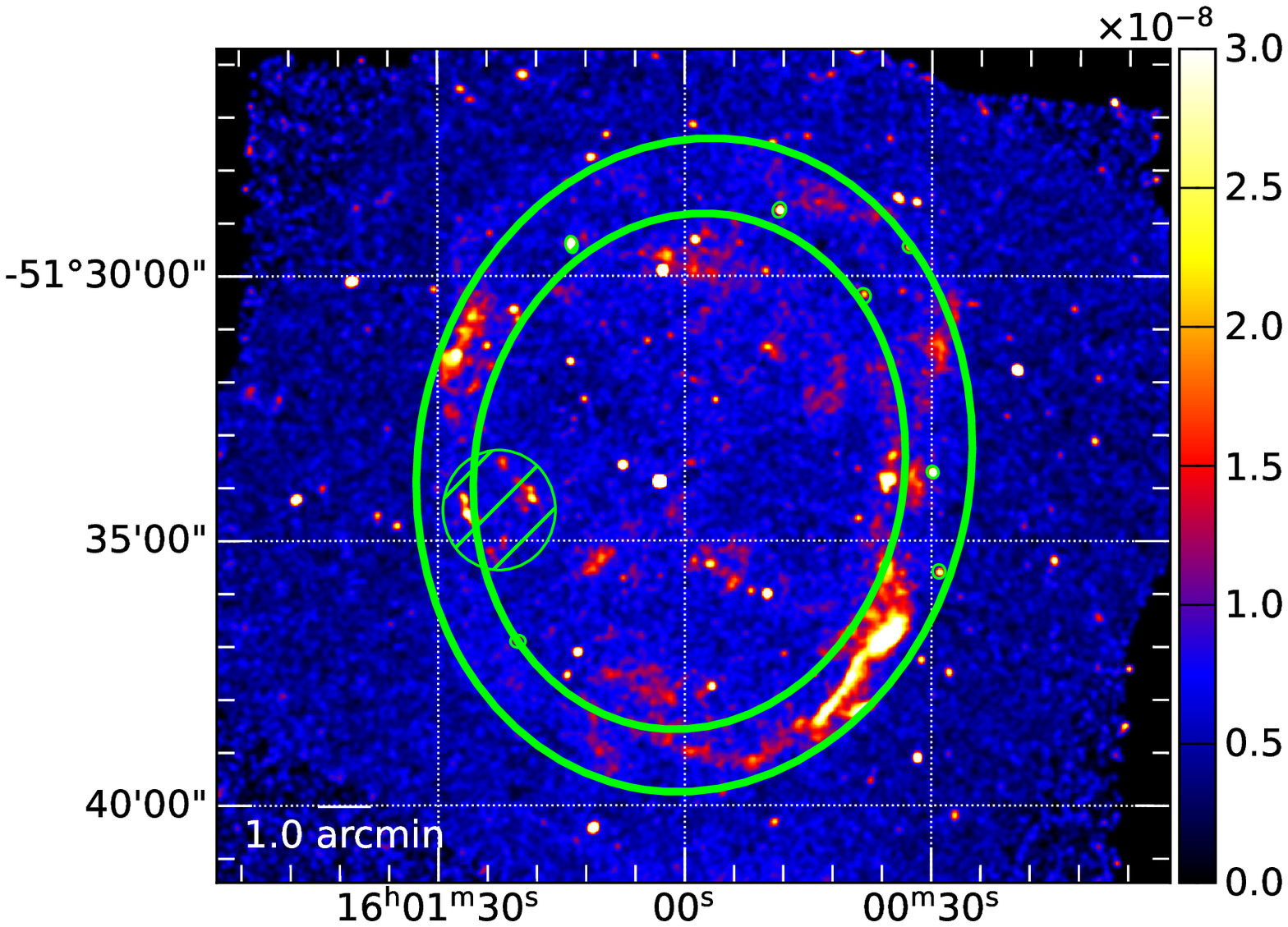}{0.32\textwidth}{G330.2$+$1.0}
          \fig{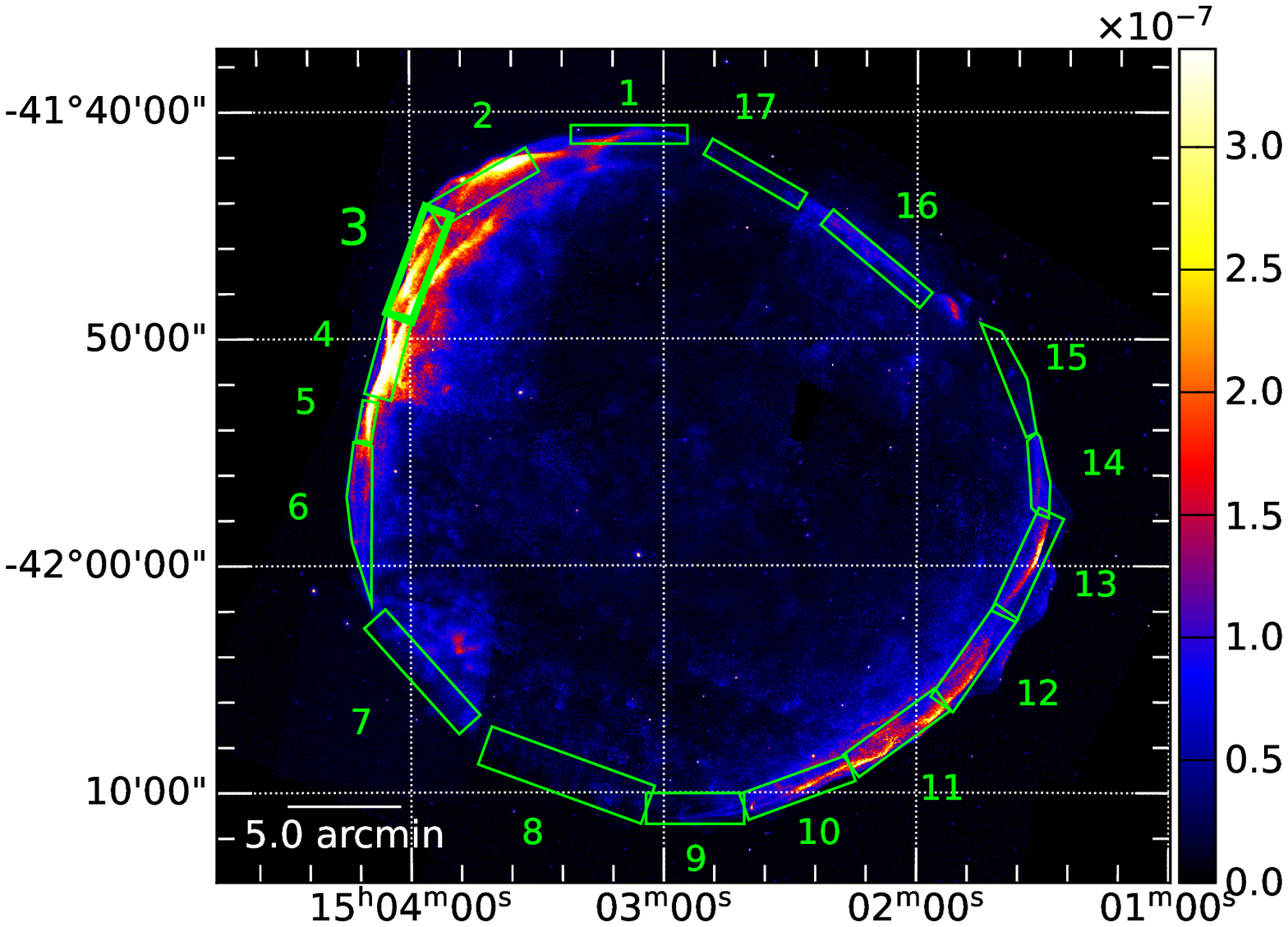}{0.32\textwidth}{SN 1006}
          } 
\gridline{\fig{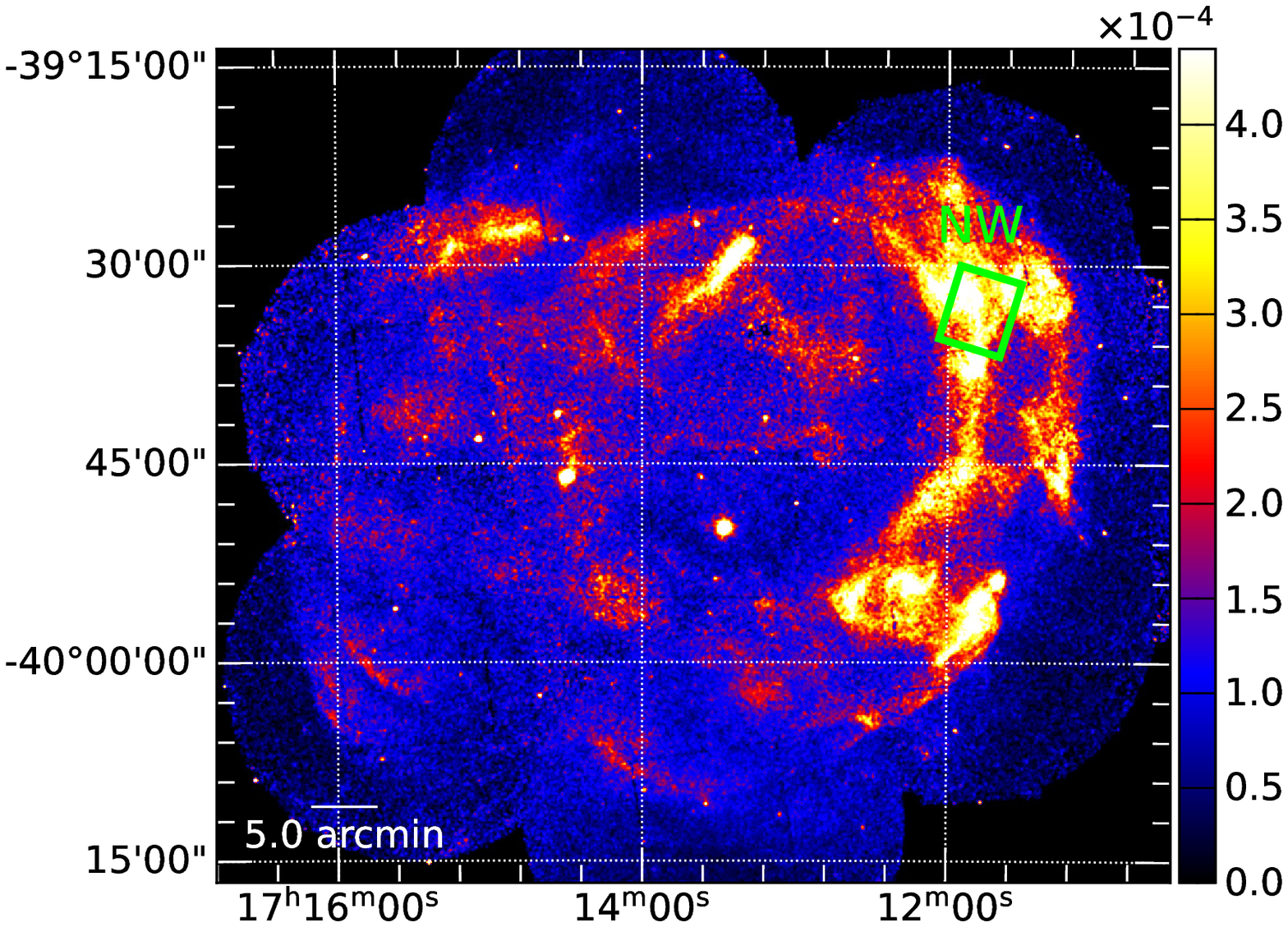}{0.32\textwidth}{\rxj}
          \fig{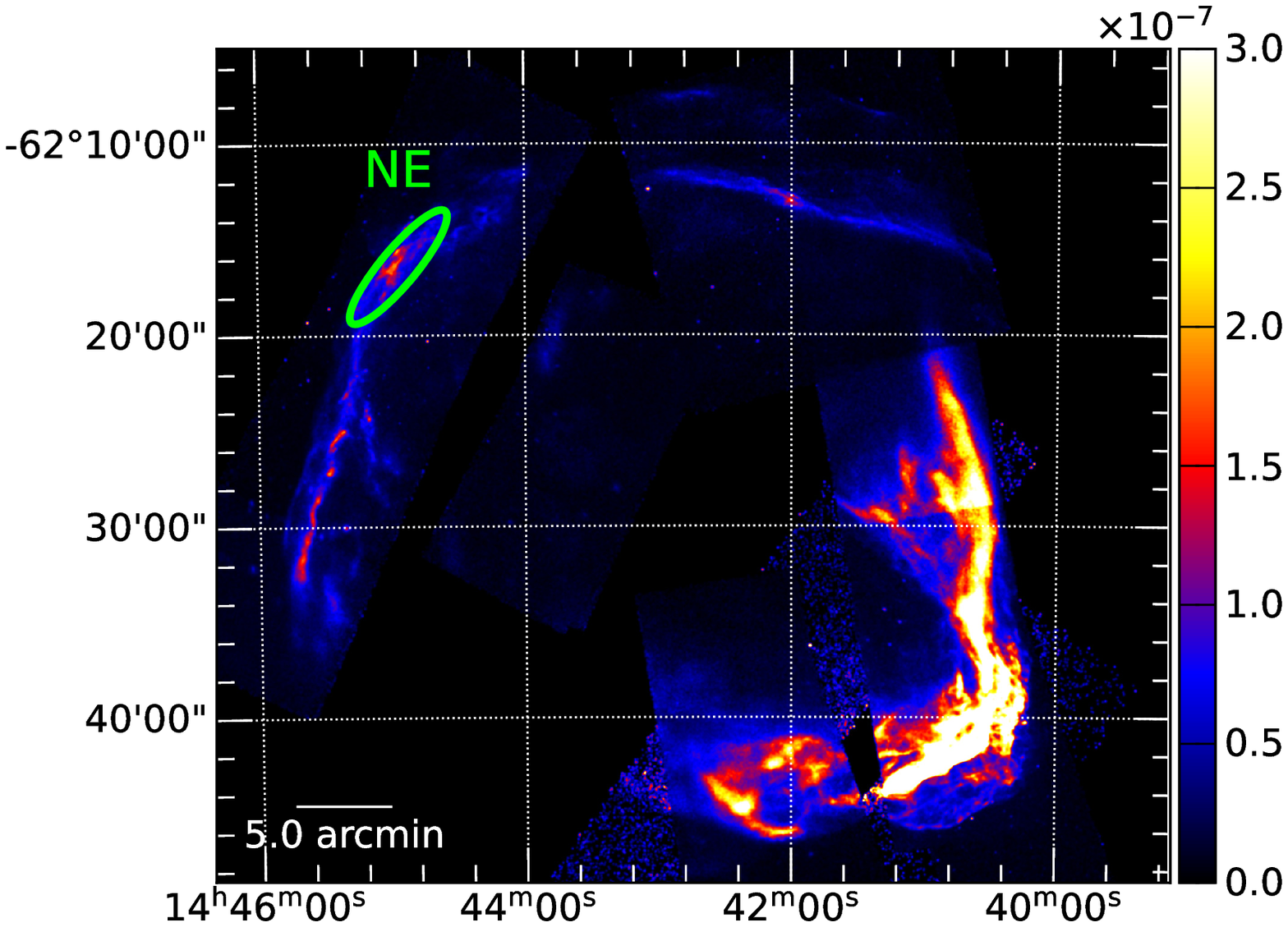}{0.32\textwidth}{RCW 86}
          \fig{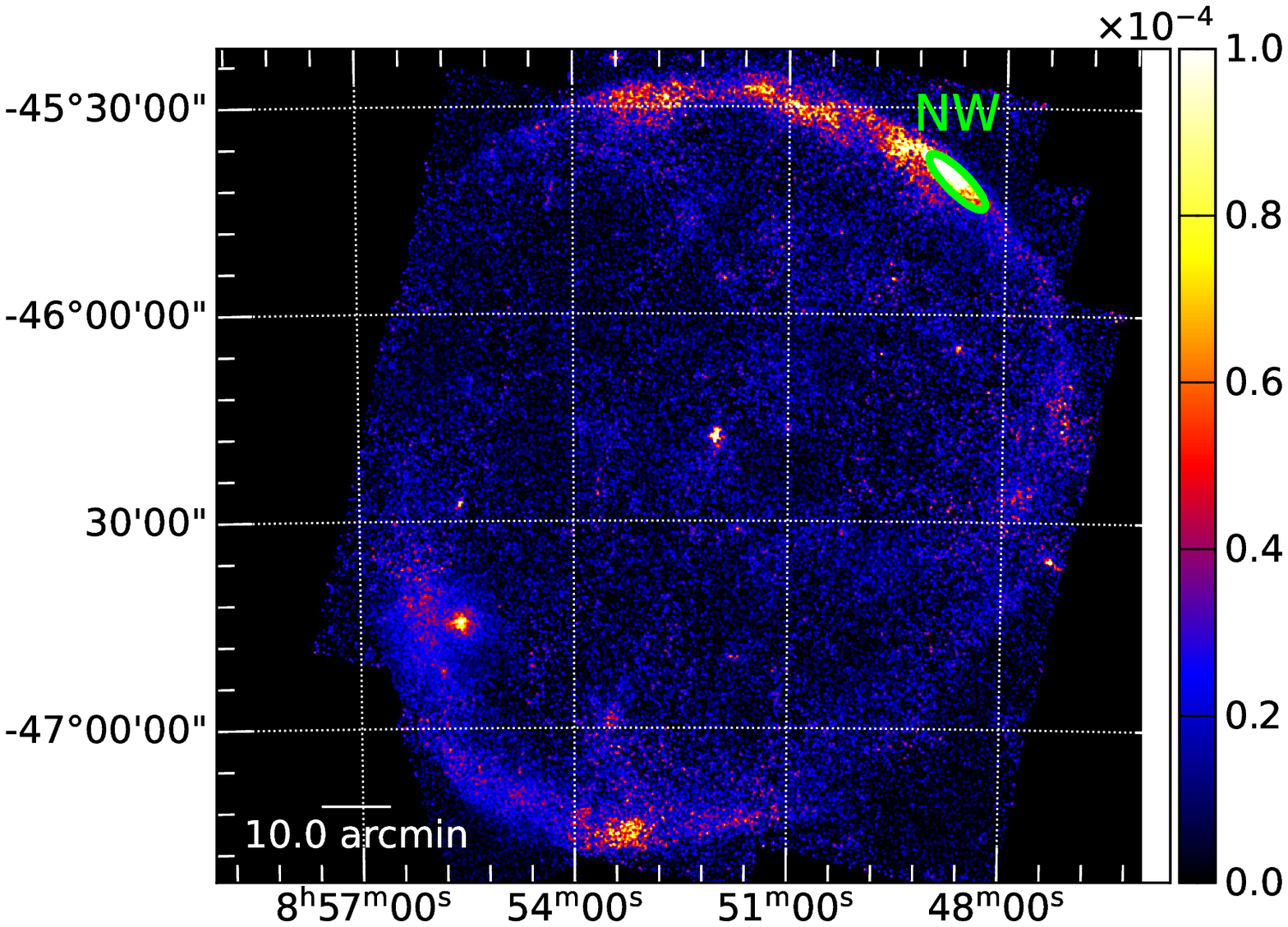}{0.32\textwidth}{Vela Jr.}
          }
\gridline{\fig{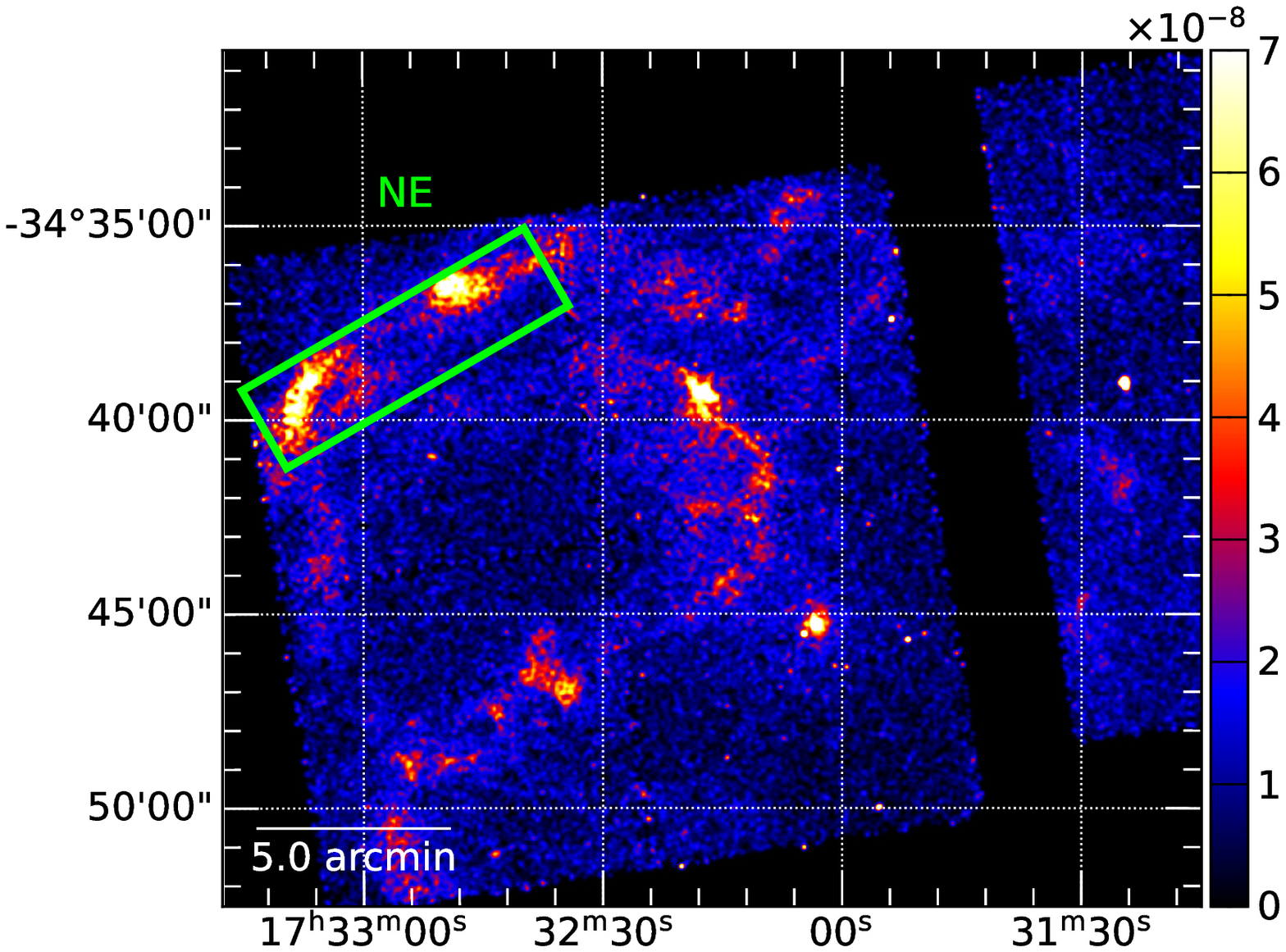}{0.32\textwidth}{HESS J1731$-$347 }
         }                            
%%%%% 
\caption{ 
Flux images in nonthermal dominated energy bands taken with \chandra, except for \rxj\ and Vela Jr. taken with \xmm\ and \suzaku, respectively.
%{\bf 
The color map is in units of photons~cm$^{-2}$~s$^{-1}$. 
1 pixel corresponds to 1\arcsec, except for 5\arcsec\ in \rxj\ and 8\arcsec\ in Vela Jr.
%}
%The color is shown in square root scale in Cassiopeia A for visibility.
The region highlighted by the thick line is used in the analysis in \secref{sec:Bohm_result_all}.
The dashed boxes in Cassiopeia A, Kepler, and Tycho indicate the reference regions used for determination of thermal components (see \secref{sec:Bohm_spectrum}).
%(Remove HESS J1731; check that the label name should be consistent with the following table and the text. Put a length scaler in each panel. \check)
\label{fig:Bohm_image_all} }
\end{figure*}

%% file: make_FigTab/Bohm_spectra.tex
%%%%%%%%%%%%%%%%%%%%%%%%%%%%
%%%%%%%%%%%%%%%%%%%%%%%%%%%%

\begin{figure*}
\gridline{\fig{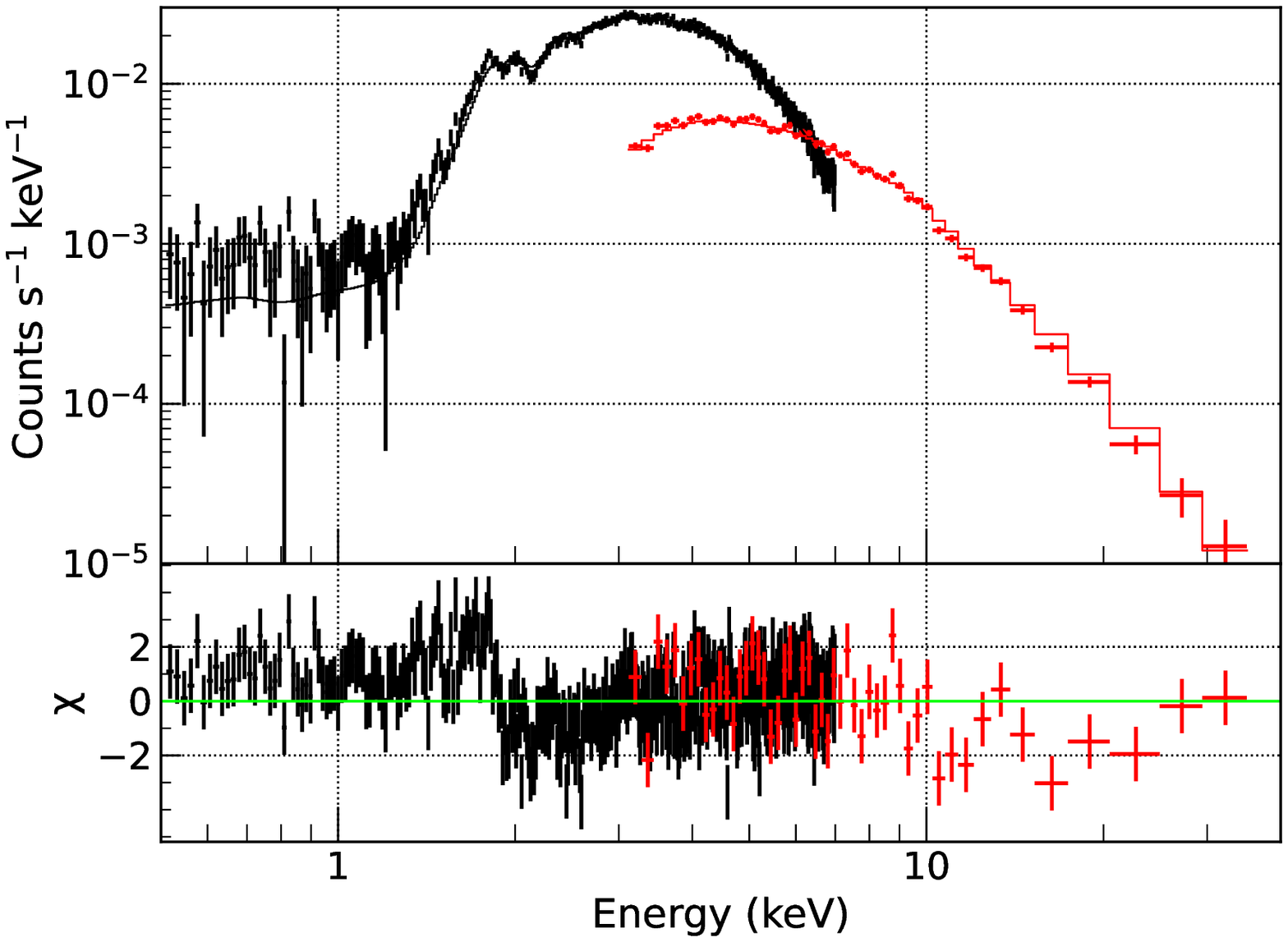}{0.32\textwidth}{G1.9$+$0.3 (E)}
          \fig{make_FigTab/spec_casA_SE2.eps}{0.32\textwidth}{Cassiopeia A (SE)}
          \fig{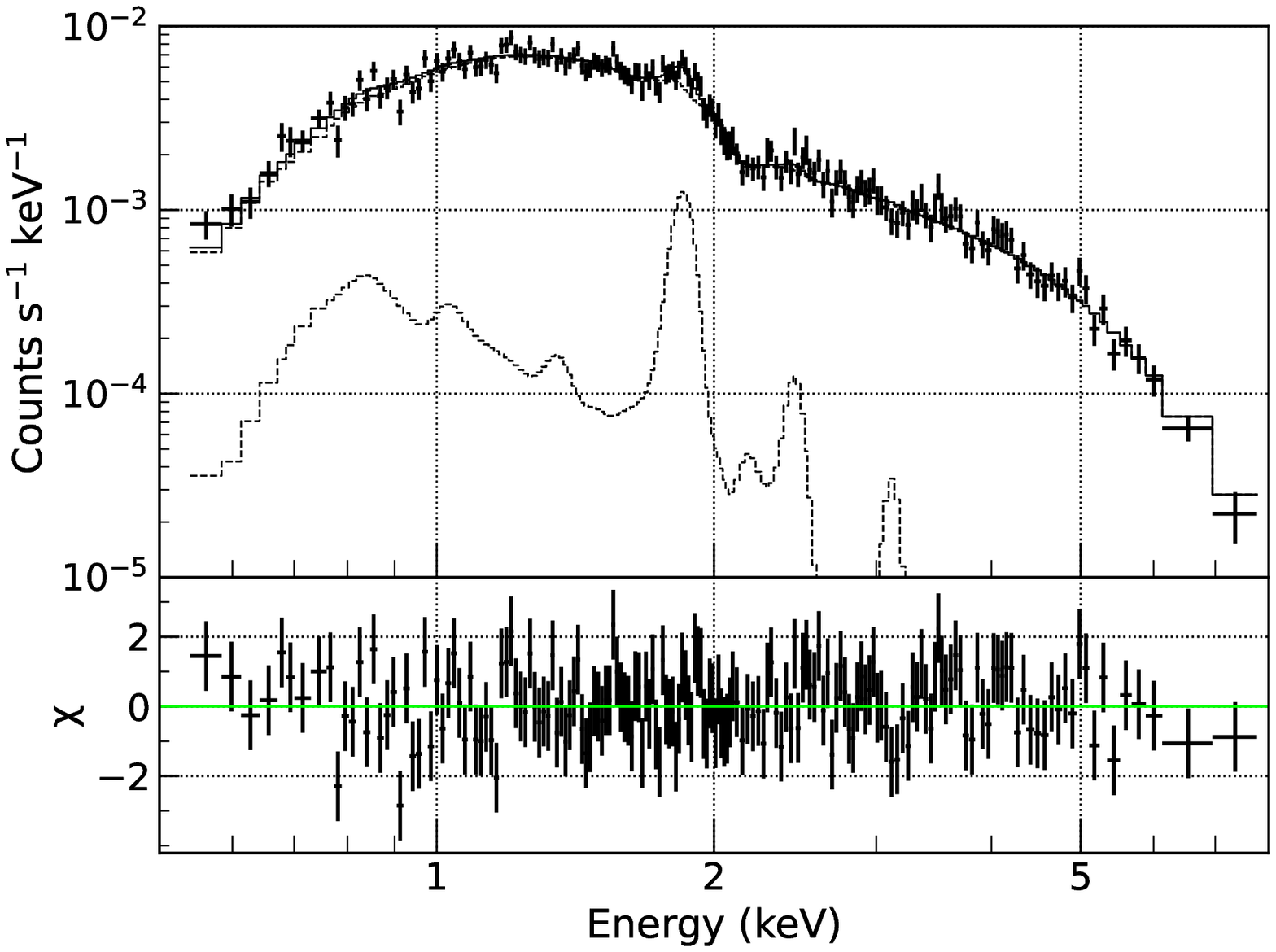}{0.32\textwidth}{Kepler (SE; 6)}
          }
\gridline{\fig{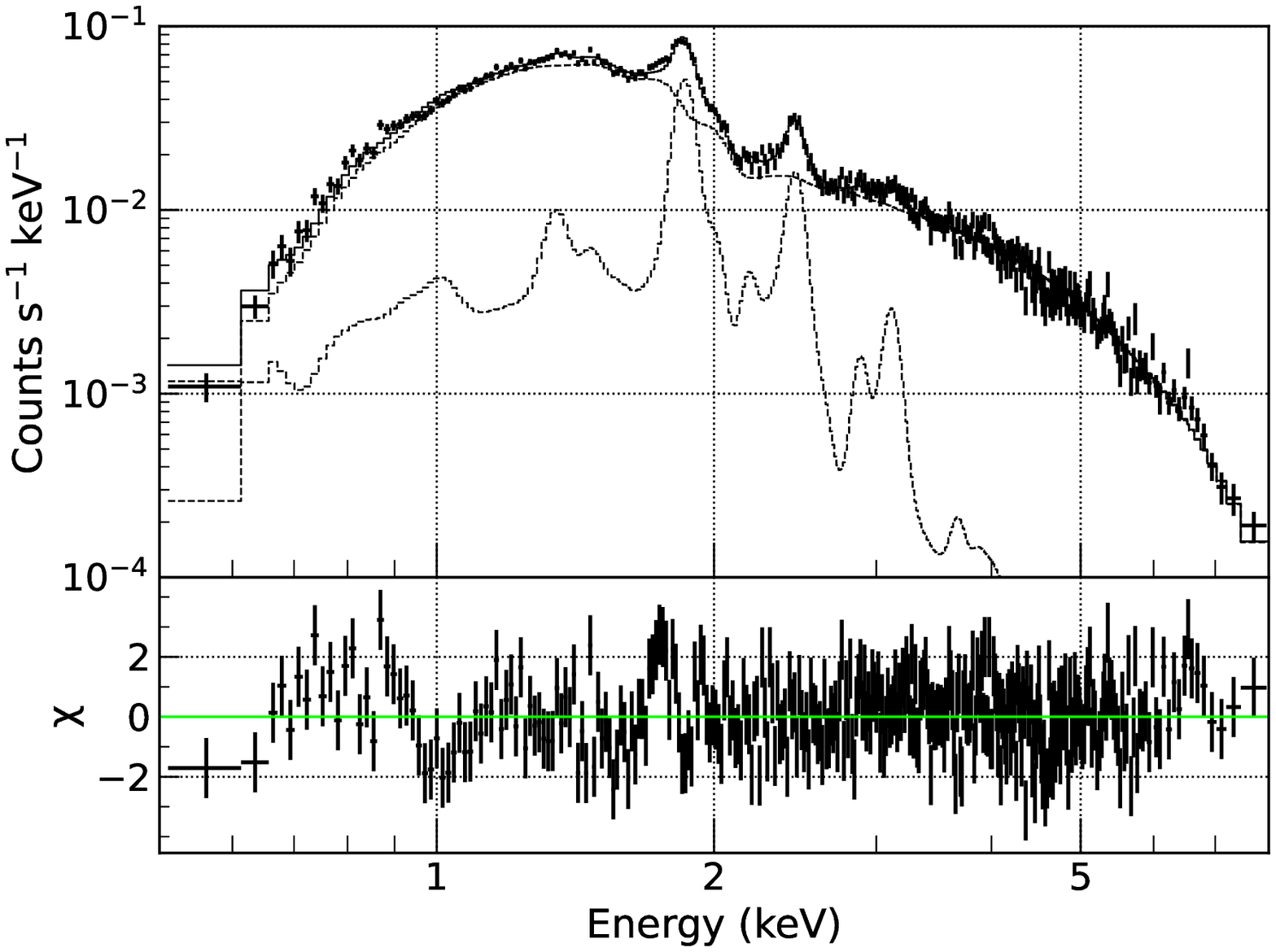}{0.32\textwidth}{Tycho (SW; 9)}
          \fig{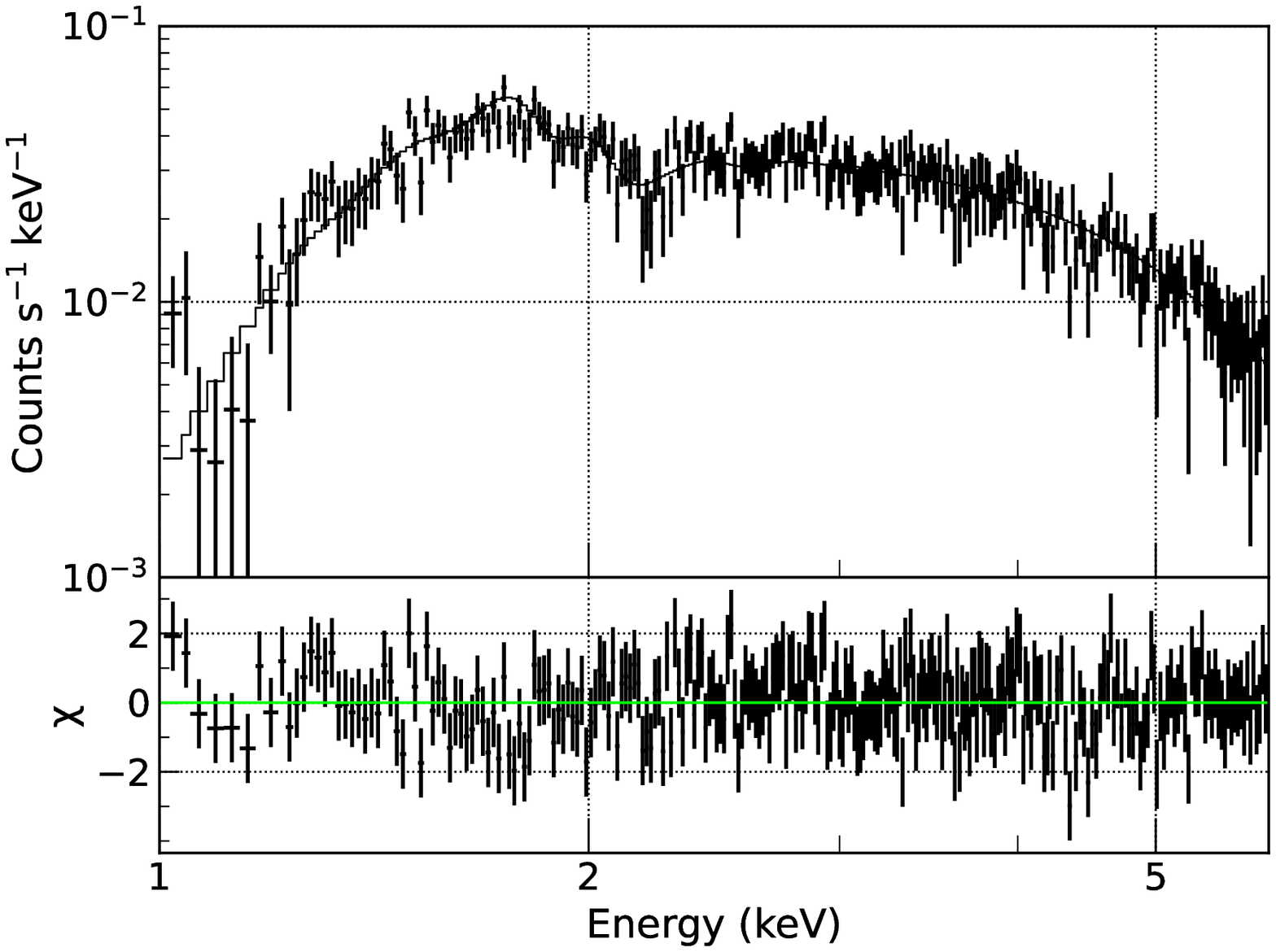}{0.32\textwidth}{G330.2$+$1.0 (whole)}
          \fig{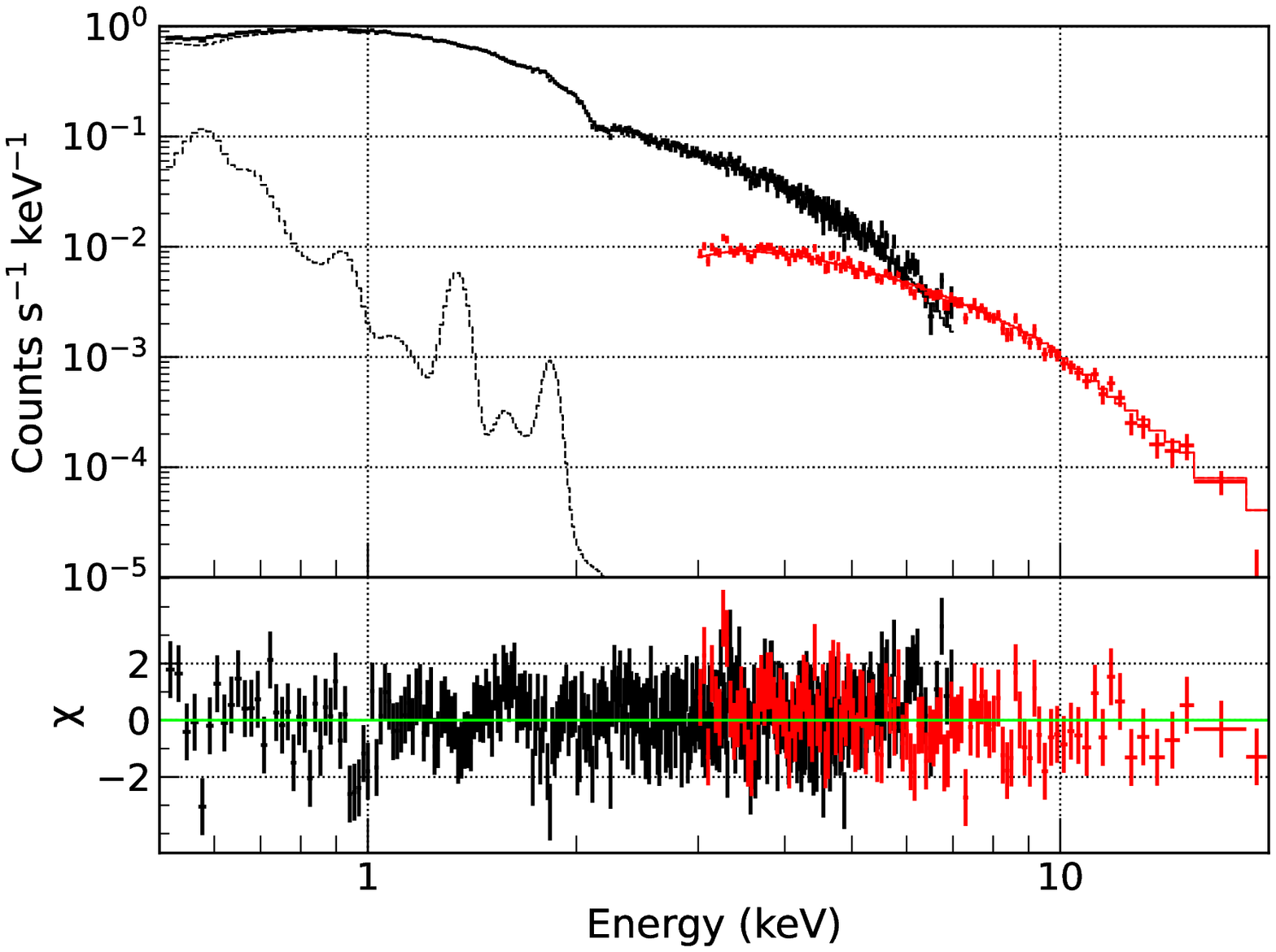}{0.32\textwidth}{SN 1006 (NE; 3)}
          } 
\gridline{\fig{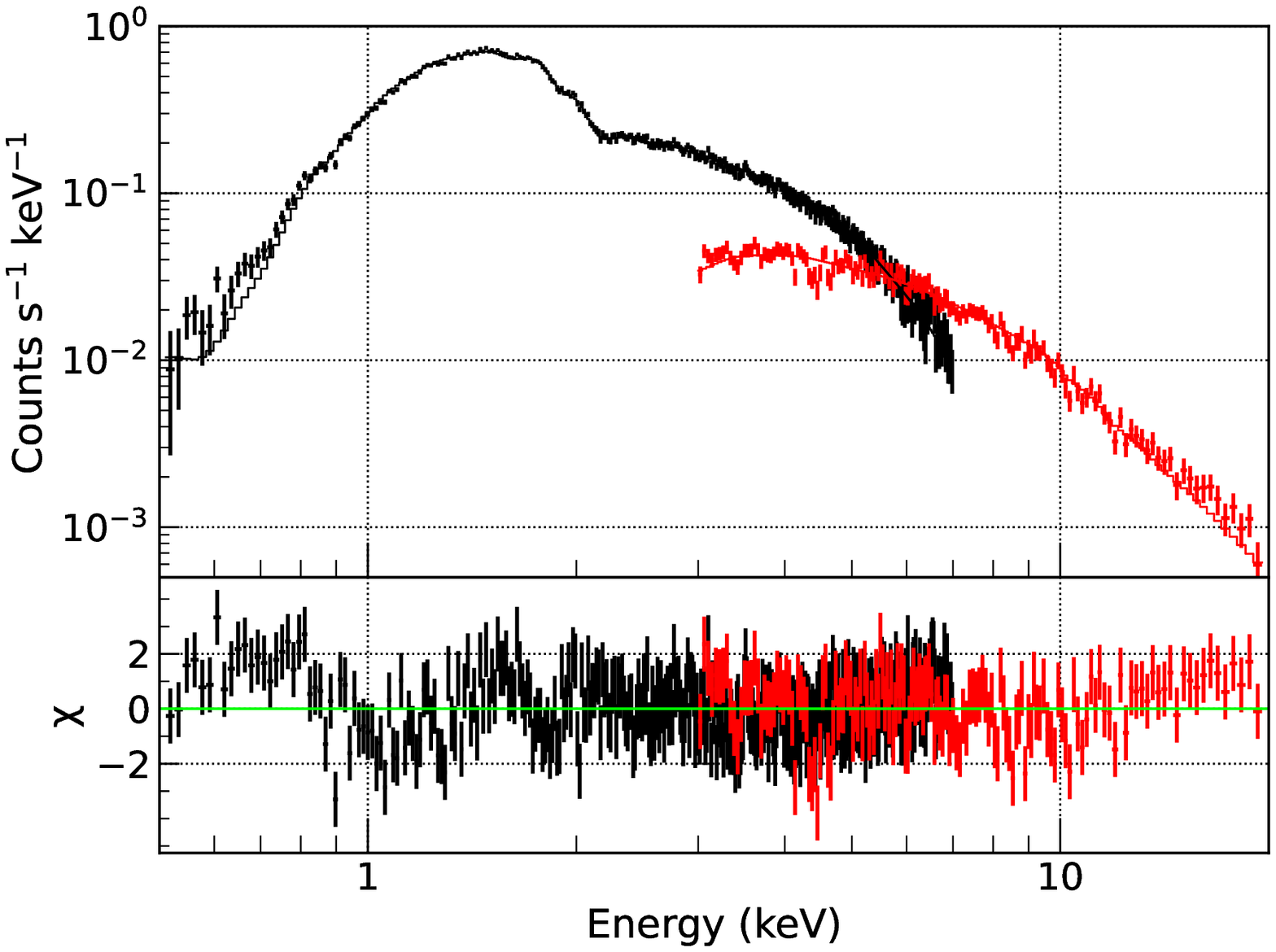}{0.32\textwidth}{\rxj\ (NW)}
          \fig{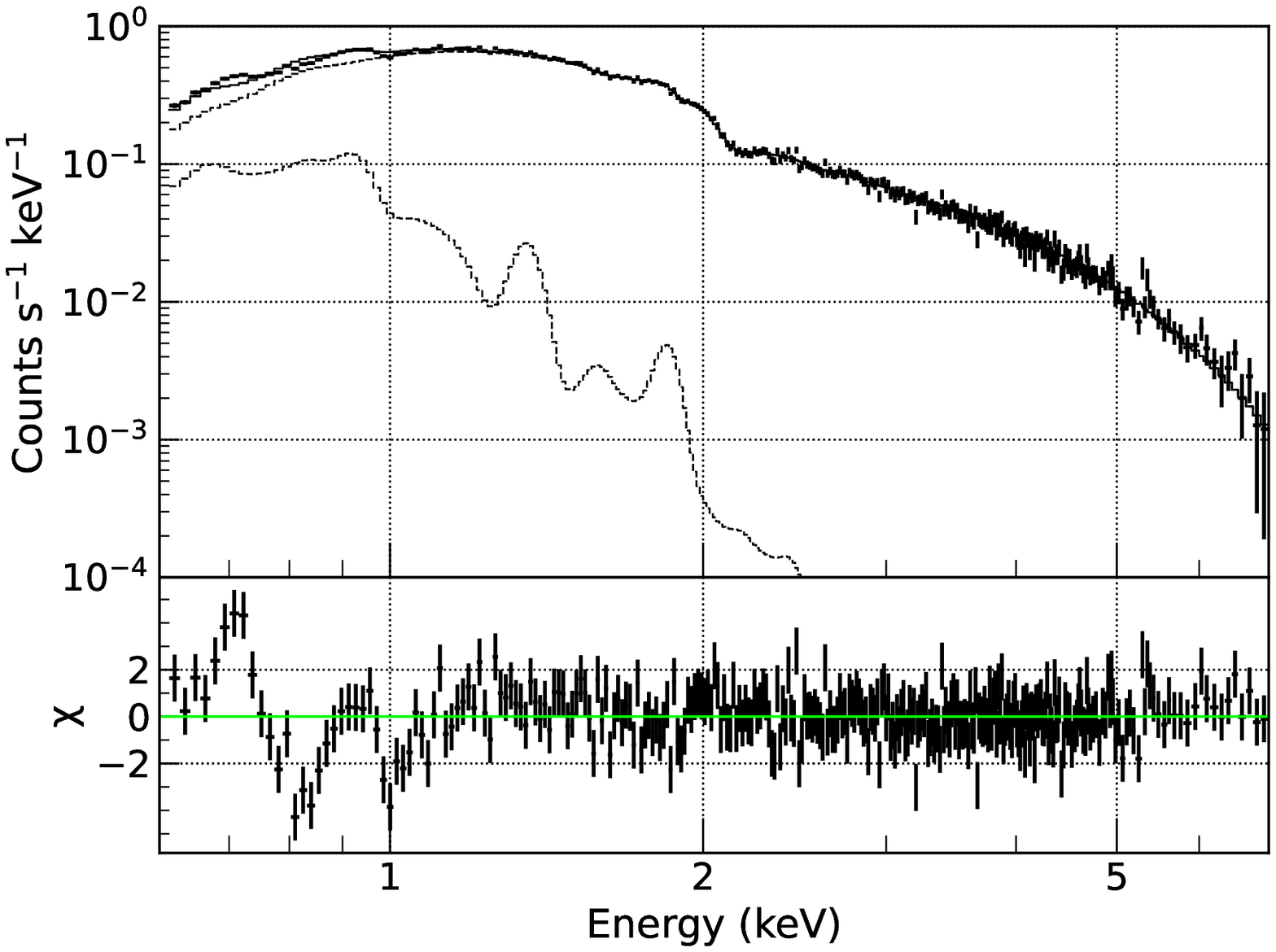}{0.32\textwidth}{RCW 86 (NE)}
          \fig{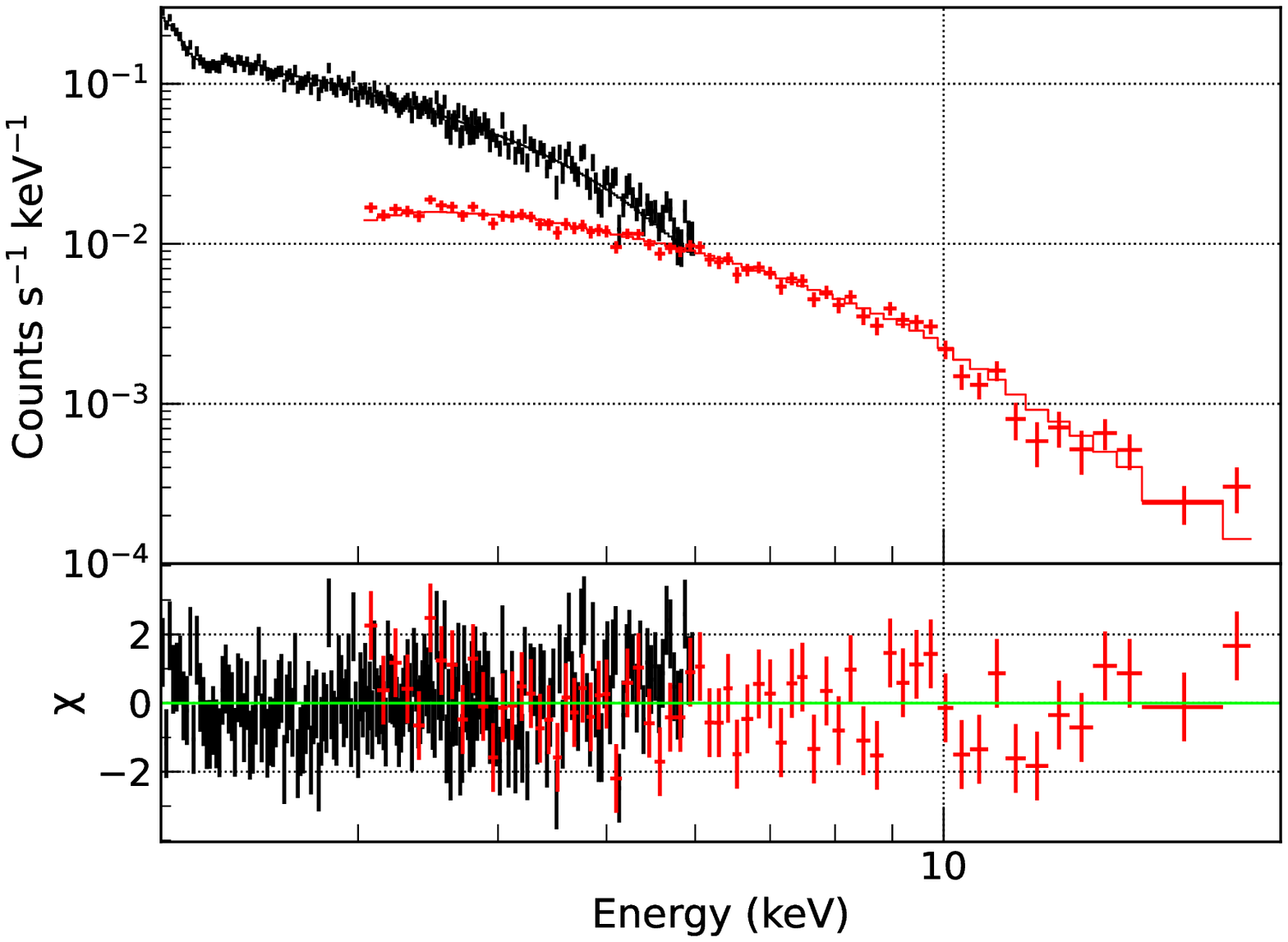}{0.32\textwidth}{Vela Jr. (NW)}
          }
\gridline{\fig{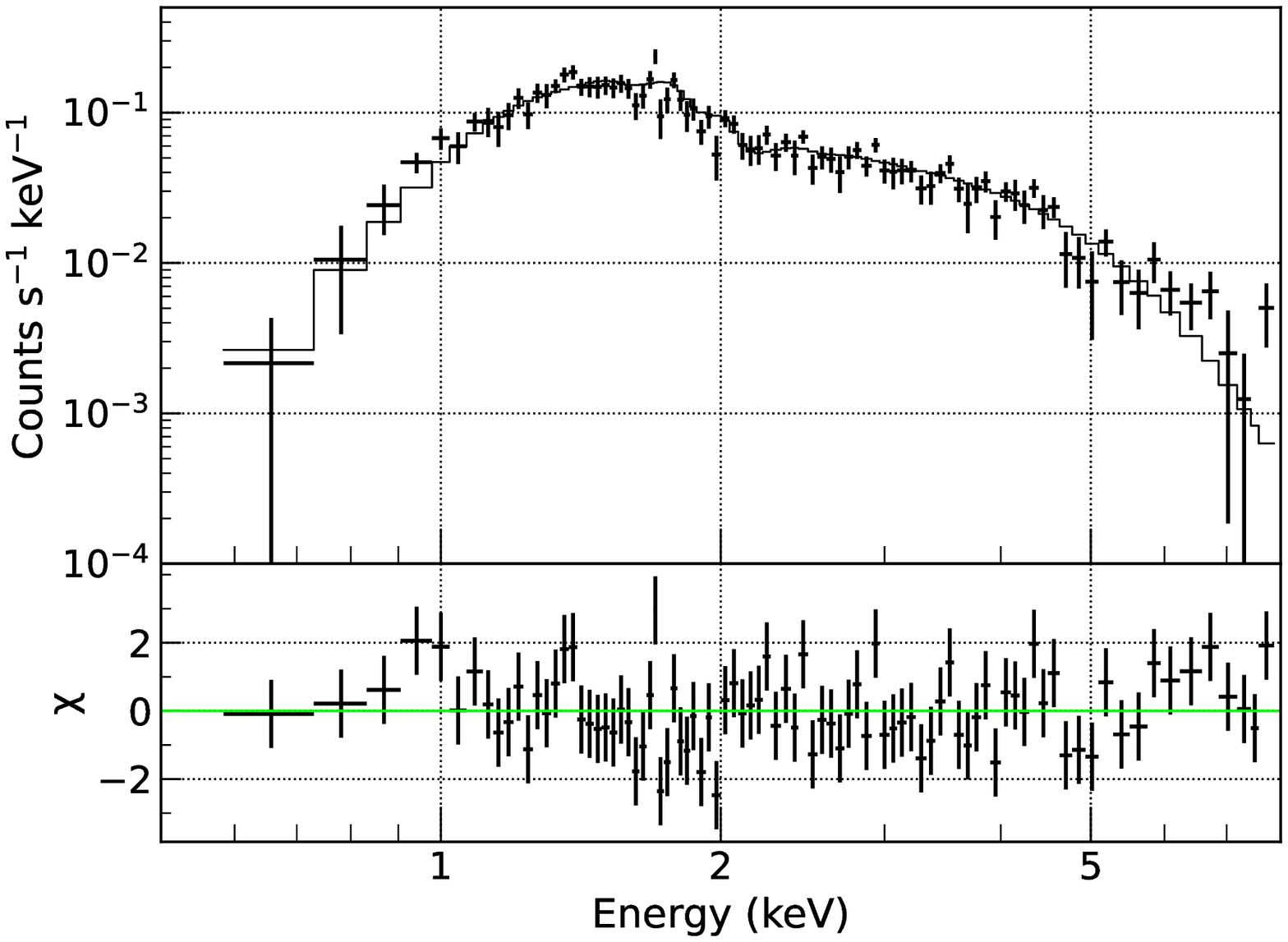}{0.32\textwidth}{HESS J1731$-$347 (NE)}
        \fig{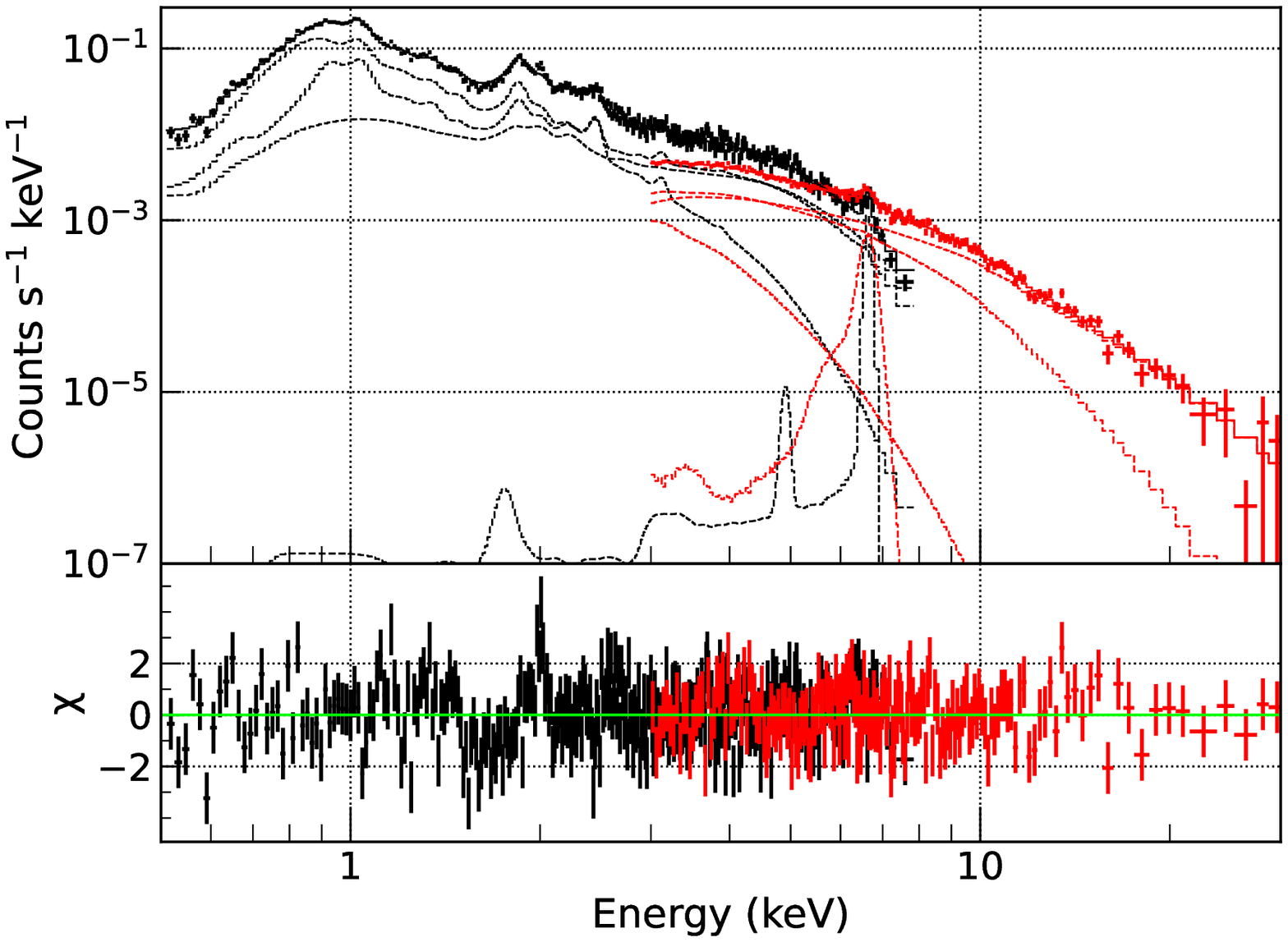}{0.32\textwidth}{SN 1987A (whole)}
         }                            
%%%%% 
\caption{ 
Spectrum of the representative region of each SNR with the best-fit model.
The spectra extracted from the \chandra\ and \nustar\ data are respectively shown in black and red.
%For Cassiopeia A, (Kepler\check) and Tycho, the fitting is performed in higher X-ray energy band (see the text for details), and the spectrum is shown from the softer band.
\label{fig:Bohm_spectra} }
\end{figure*}

%% file: make_FigTab/Bohm_result_all_in_one.tex
%%%%%%%%%%%%%%%%%%%%%%%%%%%%
%%%%%%%%%%%%%%%%%%%%%%%%%%%%

%\begin{longtable}
\startlongtable
\begin{deluxetable*}{ccccccccc}
\tablecaption{Nonthermal parameters and properties of particle acceleration (see also \tabref{tab:Bohm_result_tab_thermal} for the parameters of the thermal model in Cassiopeia A, Kepler, Tycho, SN 1006, RCW 86, and SN 1987A).
\label{tab:Bohm_result_all_in_one}
}
%\tablewidth{700pt}
\tabletypesize{\small}
\tablehead{
\colhead{Region} & 
\colhead{$N_H$}   & 
\colhead{$\varepsilon_0$}   & 
\colhead{Flux $^{\rm (a)}$} & 
\colhead{$\chi^2$ (dof)} & 
\colhead{$v_{\rm sh}$ $^{\rm (b)}$} & 
\colhead{$m$  $^{\rm (c)}$} &  
\colhead{$\eta$} & 
\colhead{$B_{\rm low}$ $^{\rm (d)}$}   
\\ 
\colhead{}  &	\colhead{$\left(10^{22}~{\rm cm}^{-2}\right)$} & \colhead{(keV)} &  \colhead{} &  \colhead{} & 
\colhead{(\kms)} & \colhead{} & \colhead{} & \colhead{(\uG)}
% delete \colhead{$\varepsilon_0$} &
} 
\startdata
%%%%%%%%%%%%%%
%\multicolumn{9}{c}{G1.9$+$0.3}  \\ \cline{2-9}
G1.9$+$0.3  \\ \cline{1-1}
whole    &   6 $\pm$ 0.04   &  1.2 $\pm$ 0.1    &  3.25$^{+0.02} _{-0.04}$ & 902.2 (585) & 13000$\pm$1000 & 1.1$\pm$0.2 & 14.0$\pm$2.8  & 39$\pm$7 \\ 
N    &   5.4 $\pm$ 0.1    &  1.2 $\pm$ 0.1    &  0.99$^{+0.02} _{-0.01}$ & 697.3 (477)  &   3600$\pm$500 & 0.4$\pm$0.2 & 1.1$\pm$0.4  & 39$\pm$9  \\
N-RS$^{\rm (e)}$ & 5.4 $\pm$ 0.1    &  1.2 $\pm$ 0.1    &  0.99$^{+0.02} _{-0.01}$ & 697.3 (477) & 5000$\pm$1000 & 0.4$\pm$0.2 & 2.2$\pm$1.0  & 39$\pm$10  \\ 
E    &   6 $\pm$ 0.1    &  1.4 $\pm$ 0.1    & 2.02$^{+0.03} _{-0.01}$  & 654.2 (492) &   13000$\pm$1000 & 1.1$\pm$0.2 & 12.1$\pm$2.6  & 37$\pm$7  \\ 
S    &   5.7 $\pm$ 0.1    &  0.9 $\pm$ 0.1    & 0.77$^{+0.02} _{-0.01}$  & 573.5 (479) &   3600$\pm$500 & 0.4$\pm$0.2 & 1.5$\pm$0.5  & 44$\pm$10  \\ 
S-RS$^{\rm (e)}$  &  5.7 $\pm$ 0.1    &  0.9 $\pm$ 0.1    & 0.77$^{+0.02} _{-0.01}$ &  573.5 (479) & 5000$\pm$1000 & 0.4$\pm$0.2 & 2.9$\pm$1.4  & 44$\pm$11 \\ 
W    &   6.2 $\pm$ 0.1    &  1.1 $\pm$ 0.1    &  1.47$^{+0.01} _{-0.02}$  & 641.5 (480) &   13000$\pm$1000 & 1.1$\pm$0.2  & 15.2$\pm$3.3  & 40$\pm$8  \\ 
%%%%%
\hline
%\multicolumn{9}{c}{Cassiopeia A}  \\
Cassiopeia A  \\ \cline{1-1}
N1 & 1.16$^{+0.02} _{-0.01}$ & 0.54$^{+0.03} _{-0.02}$ & 27.1$^{+0.3} _{-0.4}$ & 542 (398) & 4351 $\pm$ 322 & 0.60 $\pm$ 0.04 & 3.6 $\pm$ 0.6 & 36 $\pm$ 3 \\ 
N2 & 1.06 $\pm$ 0.02 & 0.44 $\pm$ 0.03 & 10.4$^{+0.1} _{-0.3}$ & 411 (338) & 4351 $\pm$ 322 & 0.60 $\pm$ 0.04 & 4.4 $\pm$ 0.7 & 39 $\pm$ 3 \\ 
S & 1.50$^{+0.03} _{-0.04}$ & 0.49 $\pm$ 0.03 & 19.3$^{+0.2} _{-0.4}$ & 512 (369) & 5479 $\pm$ 483 & 0.75 $\pm$ 0.07 & 6.3 $\pm$ 1.2 & 37$^{+4} _{-3}$ \\ 
NW & 1.33 $\pm$ 0.03 & 1.59$^{+0.26} _{-0.18}$ & 18.4$^{+0.1} _{-1.0}$ & 375 (376) & 4512 $\pm$ 483 & 0.62 $\pm$ 0.07 & 1.3 $\pm$ 0.3 & 25 $\pm$ 3 \\ 
SE & 1.98 $\pm$ 0.03 & 1.72 $\pm$ 0.09 & 10.0$^{+0.1} _{-0.2}$ & 1249 (661) & 5157 $\pm$ 644 & 0.71 $\pm$ 0.09 & 1.6 $\pm$ 0.4 & 24 $\pm$ 3  \\ 
E & 1.12 $\pm$ 0.04 & 0.89$^{+0.21} _{-0.15}$ & 4.3$^{+0.1} _{-0.4}$ & 310 (267) & 5157 $\pm$ 322 & 0.65 $\pm$ 0.04 & 3.1$^{+0.8} _{-0.6}$ & 30 $\pm$ 3 \\ 
NE & 1.45 $\pm$ 0.02 & 1.55 $\pm$ 0.09 & 16.2 $\pm$ 0.2 & 896 (689) & 4835 $\pm$ 483 & 0.67 $\pm$ 0.07 & 1.5 $\pm$ 0.3 & 25 $\pm$ 3 \\ 
%%%%%
\hline
%\multicolumn{9}{c}{Kepler}  \\
Kepler  \\ \cline{1-1}
1 & 0.38$^{+0.02} _{-0.01}$ & 1.54$^{+1.73} _{-0.58}$ & 1.08$^{+0.04} _{-0.54}$ & 334 (170) & 2067 $\pm$ 247 & 0.49 $\pm$ 0.09 & 0.3$^{+0.3} _{-0.1}$ & 22$^{+9} _{-4}$ \\ 
2 & 0.50 $\pm$ 0.02 & 0.39$^{+0.24} _{-0.12}$ & 0.40$^{+0.02} _{-0.28}$ & 273 (150) & 1972 $\pm$ 284 & 0.47 $\pm$ 0.10 & 1.0$^{+0.7} _{-0.4}$ & 35$^{+9} _{-6}$ \\ 
3 & 0.41 $\pm$ 0.02 & 0.76$^{+0.19} _{-0.14}$ & 1.51$^{+0.04} _{-0.21}$ & 217 (178) & 2825 $\pm$ 322 & 0.63 $\pm$ 0.10 & 1.1$^{+0.4} _{-0.3}$ & 28$^{+4} _{-3}$ \\ 
4 & 0.61$^{+0.05} _{-0.02}$ & 0.64$^{+0.15} _{-0.09}$ & 1.26$^{+0.03} _{-0.14}$ & 143 (149) & 3622 $\pm$ 284 & 0.68 $\pm$ 0.08 & 2.1$^{+0.6} _{-0.4}$ & 29 $\pm$ 3 \\ 
5 & 0.64$^{+0.05} _{-0.04}$ & 0.49$^{+0.08} _{-0.06}$ & 1.51$^{+0.04} _{-0.15}$ & 148 (164) & 3906 $\pm$ 303 & 0.70 $\pm$ 0.07 & 3.2$^{+0.7} _{-0.6}$ & 32 $\pm$ 3 \\ 
6 & 0.60 $\pm$ 0.03 & 1.65$^{+0.72} _{-0.41}$ & 1.63$^{+0.02} _{-0.34}$ & 170 (179) & 5727 $\pm$ 910 & 0.98 $\pm$ 0.17 & 2.0$^{+1.1} _{-0.8}$ & 21 $\pm$ 4 \\ 
7 & 0.51 $\pm$ 0.03 & 0.53$^{+0.12} _{-0.09}$ & 0.92$^{+0.03} _{-0.12}$ & 177 (142) & 3375 $\pm$ 569 & 0.63 $\pm$ 0.13 & 2.2$^{+0.9} _{-0.8}$ & 31$^{+6} _{-5}$ \\ 
8 & 0.51$^{+0.08} _{-0.06}$ & 0.85$^{+0.29} _{-0.19}$ & 1.19$^{+0.04} _{-0.19}$ & 176 (147) & 4589 $\pm$ 474 & 0.82 $\pm$ 0.09 & 2.5$^{+1.0} _{-0.8}$ & 27$^{+4} _{-3}$ \\ 
9 & 0.46$^{+0.03} _{-0.02}$ & 0.80$^{+0.24} _{-0.18}$ & 1.13$^{+0.03} _{-0.22}$ & 227 (154) & 3906 $\pm$ 398 & 0.71 $\pm$ 0.09 & 1.9$^{+0.7} _{-0.6}$ & 27$^{+4} _{-3}$ \\ 
10 & 0.36 $\pm$ 0.04 & 1.94$^{+4.32} _{-0.88}$ & 0.56$^{+0.01} _{-0.55}$ & 215 (126) & 3072 $\pm$ 417 & 0.55 $\pm$ 0.10 & 0.5$^{+1.1} _{-0.3}$ & 20$^{+15} _{-4}$ \\ 
11 & 0.49$^{+0.05} _{-0.04}$ & 1.14$^{+1.62} _{-0.48}$ & 0.34$^{+0.005} _{-0.281}$ & 96 (92) & 4020 $\pm$ 759 & 0.72 $\pm$ 0.15 & 1.4$^{+2.1} _{-0.8}$ & 24$^{+12} _{-6}$ \\ 
12 & 0.65$^{+0.09} _{-0.10}$ & 1.06$^{+0.72} _{-0.33}$ & 0.59$^{+0.01} _{-0.31}$ & 178 (119) & 3413 $\pm$ 512 & 0.62 $\pm$ 0.11 & 1.1$^{+0.8} _{-0.5}$ & 25$^{+7} _{-4}$ \\ 
13 & 0.54 $\pm$ 0.02 & 1.53$^{+2.61} _{-0.64}$ & 1.27$^{+0.03} _{-1.18}$ & 456 (177) & 1441 $\pm$ 607 & 0.34 $\pm$ 0.16 & 0.1$^{+0.3} _{-0.1}$ & 22$^{+15} _{-9}$ \\ 
14 & 0.12 $\pm$ 0.04 & 0.42$^{+0.16} _{-0.10}$ & 1.00$^{+0.03} _{-0.28}$ & 314 (171) & 2162 $\pm$ 266 & 0.50 $\pm$ 0.09 & 1.1$^{+0.5} _{-0.4}$ & 34$^{+6} _{-5}$ \\ 
%%%%%
\hline
%\multicolumn{9}{c}{Tycho}  \\
Tycho  \\ \cline{1-1}
1 & 0.89 $\pm$ 0.04 & 0.24 $\pm$ 0.02 & 5.5$^{+0.2} _{-0.3}$ & 484 (283) & 2380 $\pm$ 238 & 0.37 $\pm$ 0.04 & 2.5 $\pm$ 0.5 & 40 $\pm$ 4 \\ 
2 & 0.77 $\pm$ 0.04 & 0.18 $\pm$ 0.01 & 9.8$^{+0.2} _{-0.3}$ & 401 (301) & 3000 $\pm$ 500 & 0.51 $\pm$ 0.08 & 5.0 $\pm$ 1.7 & 43 $\pm$ 7 \\ 
3 & 0.59 $\pm$ 0.03 & 0.33 $\pm$ 0.01 & 23.8 $\pm$ 0.4 & 518 (369) & 3200 $\pm$ 320 & 0.55 $\pm$ 0.05 & 3.2 $\pm$ 0.6 & 35 $\pm$ 3 \\ 
4 & 0.66 $\pm$ 0.01 & 0.38 $\pm$ 0.01 & 29.8$^{+0.3} _{-0.4}$ & 561 (399) & 3580 $\pm$ 358 & 0.56 $\pm$ 0.06 & 3.4 $\pm$ 0.7 & 34 $\pm$ 3 \\ 
5 & 0.91 $\pm$ 0.04 & 0.40 $\pm$ 0.02 & 30.5$^{+0.3} _{-0.6}$ & 717 (404) & 3700 $\pm$ 370 & 0.58 $\pm$ 0.06 & 3.5 $\pm$ 0.7 & 33 $\pm$ 3 \\ 
6 & 0.76$^{+0.05} _{-0.04}$ & 0.42$^{+0.03} _{-0.02}$ & 26.3$^{+0.3} _{-0.6}$ & 640 (396) & 3850 $\pm$ 385 & 0.60 $\pm$ 0.06 & 3.6 $\pm$ 0.8 & 33 $\pm$ 3 \\ 
7 & 0.77 $\pm$ 0.01 & 0.36 $\pm$ 0.01 & 26.4$^{+0.3} _{-0.5}$ & 605 (396) & 3920 $\pm$ 648 & 0.61 $\pm$ 0.10 & 4.3 $\pm$ 1.4 & 34 $\pm$ 5 \\ 
8 & 0.78$^{+0.02} _{-0.01}$ & 0.39 $\pm$ 0.02 & 20.4$^{+0.3} _{-0.5}$ & 419 (363) & 3980 $\pm$ 418 & 0.62 $\pm$ 0.07 & 4.1 $\pm$ 0.9 & 33 $\pm$ 3 \\ 
9 & 0.81 $\pm$ 0.02 & 0.55$^{+0.04} _{-0.03}$ & 16.9$^{+0.2} _{-0.4}$ & 486 (366) & 4060 $\pm$ 634 & 0.63 $\pm$ 0.10 & 3.1 $\pm$ 1.0 & 30 $\pm$ 4 \\ 
10 & 1.16 $\pm$ 0.06 & 0.35 $\pm$ 0.03 & 11.8$^{+0.2} _{-0.3}$ & 651 (335) & 3780 $\pm$ 378 & 0.59 $\pm$ 0.06 & 4.2 $\pm$ 0.9 & 35 $\pm$ 3 \\ 
11 & 0.74 $\pm$ 0.03 & 0.41$^{+0.03} _{-0.02}$ & 17.7$^{+0.2} _{-0.4}$ & 427 (371) & 3240 $\pm$ 468 & 0.54 $\pm$ 0.07 & 2.6 $\pm$ 0.8 & 33 $\pm$ 5 \\ 
12 & 0.64$^{+0.05} _{-0.04}$ & 0.42 $\pm$ 0.03 & 14.9$^{+0.2} _{-0.5}$ & 536 (360) & 3480 $\pm$ 538 & 0.59 $\pm$ 0.08 & 3.0 $\pm$ 0.9 & 33 $\pm$ 5 \\ 
13 & 0.74 $\pm$ 0.03 & 0.40$^{+0.04} _{-0.03}$ & 9.2$^{+0.2} _{-0.5}$ & 452 (308) & 3330 $\pm$ 333 & 0.57 $\pm$ 0.05 & 2.8 $\pm$ 0.6 & 33 $\pm$ 3 \\ 
14 & 1.09 $\pm$ 0.04 & 0.40 $\pm$ 0.04 & 12.8$^{+0.2} _{-0.4}$ & 601 (351) & 3510 $\pm$ 351 & 0.55 $\pm$ 0.05 & 3.1 $\pm$ 0.7 & 33 $\pm$ 3 \\ 
15 & 0.55 $\pm$ 0.03 & 0.44 $\pm$ 0.03 & 12.0$^{+0.2} _{-0.5}$ & 458 (344) & 3110 $\pm$ 311 & 0.49 $\pm$ 0.05 & 2.2 $\pm$ 0.5 & 32 $\pm$ 3 \\ 
16 & 0.73 $\pm$ 0.03 & 0.41$^{+0.03} _{-0.02}$ & 11.9$^{+0.2} _{-0.4}$ & 601 (329) & 2210 $\pm$ 378 & 0.38 $\pm$ 0.06 & 1.2 $\pm$ 0.4 & 33 $\pm$ 5 \\ 
17 & 0.65 $\pm$ 0.02 & 0.57 $\pm$ 0.02 & 56.6$^{+0.4} _{-0.6}$ & 810 (428) & 2000 $\pm$ 400 & 0.36 $\pm$ 0.06 & 0.7 $\pm$ 0.3 & 30 $\pm$ 6 \\ 
18 & 1.10$^{+0.14} _{-0.06}$ & 0.28 $\pm$ 0.02 & 12.5$^{+0.2} _{-0.4}$ & 664 (341) & 2360 $\pm$ 236 & 0.39 $\pm$ 0.04 & 2.1 $\pm$ 0.4 & 37 $\pm$ 4 \\ 
19 & 0.80 $\pm$ 0.02 & 0.41 $\pm$ 0.02 & 17.5$^{+0.2} _{-0.4}$ & 583 (371) & 3310 $\pm$ 331 & 0.52 $\pm$ 0.05 & 2.7 $\pm$ 0.6 & 33 $\pm$ 3 \\ 
20 & 0.58 $\pm$ 0.02 & 0.38 $\pm$ 0.02 & 17.1$^{+0.2} _{-0.3}$ & 530 (372) & 3660 $\pm$ 366 & 0.57 $\pm$ 0.06 & 3.6 $\pm$ 0.7 & 34 $\pm$ 3 \\ 
%%%%%
\hline
%\multicolumn{9}{c}{G330.2$+$1.0}  \\
G330.2$+$1.0  \\ \cline{1-1}
whole    &   2.5 (fixed)   &  1.17 $^{+0.32} _{-0.22}$   & 2.52$^{+0.39} _{-0.03}$   & 244.6 (269) &     7000$\pm$2000 & 0.9$\pm$0.3 & 4.3$^{+2.7} _{-2.6}$  & 17$\pm$7 \\ 
%whole    &   2.4 (fixed)   &  0.65 $^{+0.07} _{-0.06}$   &    & 356.2 (166) &     7000$\pm$2000 & 0.9$\pm$0.3 & 6.0$\pm$4.1  & 18 \\ 
%NW1 & 2.4 (fixed) &  1.02$\pm$0.51 &  & & 9100$\pm$2600 & 0.9$\pm$0.3 &  8.3$\pm$8.9  & 17 \\ 
%NW2 & 2.4 (fixed) &  1.31$\pm$0.30 &  & & 8400$\pm$2900 & 0.9$\pm$0.3 &  5.5$\pm$5.0  & 15 \\ 
%SW1 & 2.4 (fixed) &  1.82$\pm$0.37 &   & & 3700$\pm$1400 & 0.9$\pm$0.3 &  0.8$\pm$0.7  & 14 \\ 
%%%%%
\hline
%\multicolumn{9}{c}{SN 1006}  \\
SN 1006 \\ \cline{1-1}
1 & 0.13 $\pm$ 0.02 & 0.15 $\pm$ 0.01 & 6.3$^{+0.3} _{-0.6}$ & 292 (279) & 6500 $\pm$ 500 & 0.75 $\pm$ 0.06 & 28.6 $\pm$ 4.7 & 26 $\pm$ 2 \\ 
2 & 0.11 $\pm$ 0.01 & 0.37 $\pm$ 0.01 & 75 $\pm$ 1 & 609 (505) & 5900 $\pm$ 500 & 0.68 $\pm$ 0.06 & 9.6 $\pm$ 1.6 & 20 $\pm$ 2 \\ 
3 & 0.10 $\pm$ 0.01 & 0.50 $\pm$ 0.02 & 88 $\pm$ 1 & 566 (525) & 5500 $\pm$ 800 & 0.64 $\pm$ 0.09 & 6.2 $\pm$ 1.8 & 18 $\pm$ 2 \\ 
4 & 0.09 $\pm$ 0.01 & 0.37$^{+0.02} _{-0.01}$ & 65$^{+1} _{-2}$ & 555 (431) & 5000 $\pm$ 500 & 0.64 $\pm$ 0.06 & 6.9 $\pm$ 1.4 & 20 $\pm$ 2 \\ 
5 & 0.12 $\pm$ 0.01 & 0.31 $\pm$ 0.02 & 17$^{+0} _{-1}$ & 290 (277) & 4900 $\pm$ 500 & 0.62 $\pm$ 0.06 & 8.0 $\pm$ 1.7 & 21 $\pm$ 2 \\ 
6 & 0.09 $\pm$ 0.01 & 0.20 $\pm$ 0.01 & 21 $\pm$ 1 & 403 (318) & 5800 $\pm$ 500 & 0.74 $\pm$ 0.06 & 17.2 $\pm$ 3.1 & 24 $\pm$ 2 \\ 
7 & 0.11 & 0.13$^{+0.03} _{-0.04}$ & 1.3$^{+0.3} _{-0.7}$ & 325 (171) & 7000 $\pm$ 1200 & 0.89 $\pm$ 0.15 & 37.4$^{+16.1} _{-16.4}$ & 27 $\pm$ 5 \\ 
8 & 0.11 & 0.56$^{+0.56} _{-0.23}$ & 5.6$^{+0.3} _{-4.2}$ & 233 (179) & 7200 $\pm$ 1000 & 0.92 $\pm$ 0.13 & 9.5$^{+9.9} _{-4.7}$ & 17$^{+6} _{-3}$ \\ 
9 & 0.14$^{+0.05} _{-0.04}$ & 0.09 $\pm$ 0.01 & 2.3$^{+0.3} _{-0.4}$ & 295 (202) & 5000 $\pm$ 800 & 0.64 $\pm$ 0.10 & 27.4$^{+9.2} _{-9.1}$ & 31 $\pm$ 5 \\ 
10 & 0.15 $\pm$ 0.02 & 0.20 $\pm$ 0.01 & 27 $\pm$ 1 & 327 (319) & 5000 $\pm$ 800 & 0.64 $\pm$ 0.10 & 12.5 $\pm$ 4.1 & 24 $\pm$ 4 \\ 
11 & 0.14 $\pm$ 0.01 & 0.30 $\pm$ 0.01 & 60 $\pm$ 1 & 574 (414) & 5500 $\pm$ 500 & 0.70 $\pm$ 0.06 & 10.4$^{+2.0} _{-1.9}$ & 21 $\pm$ 2 \\ 
12 & 0.11 $\pm$ 0.01 & 0.31 $\pm$ 0.01 & 48 $\pm$ 1 & 498 (457) & 5900 $\pm$ 800 & 0.75 $\pm$ 0.10 & 11.4 $\pm$ 3.1 & 21 $\pm$ 3 \\ 
13 & 0.10$^{+0.02} _{-0.01}$ & 0.32 $\pm$ 0.02 & 37 $\pm$ 1 & 439 (352) & 7500 $\pm$ 400 & 0.96 $\pm$ 0.05 & 18.0 $\pm$ 2.2 & 21 $\pm$ 1 \\ 
14 & 0.12 $\pm$ 0.02 & 0.23 $\pm$ 0.02 & 10.3$^{+0.4} _{-0.9}$ & 216 (228) & 5500 $\pm$ 500 & 0.70 $\pm$ 0.06 & 13.6 $\pm$ 2.7 & 23 $\pm$ 2 \\ 
15 & 0.01$^{+0.04} _{-0.01}$ & 0.17$^{+0.01} _{-0.02}$ & 2.3$^{+0.2} _{-0.5}$ & 223 (197) & 7000 $\pm$ 1000 & 0.89 $\pm$ 0.13 & 28.8$^{+8.5} _{-9.1}$ & 25$^{+3} _{-4}$ \\ 
16 & 0.23$^{+0.05} _{-0.03}$ & 0.03$^{+0.02} _{-0.01}$ & 0.04$^{+0.01} _{-0.03}$ & 283 (184) & 3000 $\pm$ 1200 & 0.44 $\pm$ 0.18 & 31.4$^{+30.7} _{-28.2}$ & 46$^{+19} _{-18}$ \\ 
17 & 0.12 & 0.03$^{+0.02} _{-0.01}$ & 0.53$^{+0.06} _{-0.49}$ & 173 (122) & 3000 $\pm$ 1200 & 0.38 $\pm$ 0.15 & 27.8$^{+27.3} _{-23.2}$ & 44$^{+18} _{-17}$ \\ 
%% 1713 
\hline
%\multicolumn{9}{c}{\rxj-NW}  \\
\rxj\ \\ \cline{1-1}
NW      &   0.75 $\pm$ 0.01    & 1.14$\pm$0.06  &  11.5$^{+0.3} _{-0.2}$  & 687.9 (587) &  3900$\pm$300 & 0.7$\pm$0.1  & 1.4$\pm$0.3  & 9.7$\pm$0.9 \\ 
%whole\check\ Tanaka08  & & 0.94$\pm$0.03  & & & 3900$\pm$300 & 0.7$\pm$0.1 & 1.7$\pm$0.5  & 9 \\ 
%%%%%
\hline
%\multicolumn{9}{c}{RCW 86 }  \\
RCW 86 \\ \cline{1-1}
NE    &   0.33 (fixed)   &  0.32 $\pm$0.01   &  3.22$^{+0.03} _{-0.05}$  & 527 (407) &   2500$\pm$700 & 0.3$\pm$0.1  & 2.0$\pm$1.1  & 14$\pm$4 \\ 
%NE    &   0.33 (fixed)   &  0.38 $^{+0.02} _{-0.01}$   &  3.17$^{+0.04} _{-0.03}$  & 396.4 (340) &   2500$\pm$700 & 0.3$\pm$0.1  & 2.1$\pm$1.3  & 14$^{+4}_{-3}$ \\ 
%%%%%
\hline
%\multicolumn{9}{c}{Vela Jr.}  \\
Vela Jr.  \\ \cline{1-1}
NW    &   0.67 (fixed)   &  0.54 $\pm$ 0.04    &  3.9$\pm$0.1  & 289.3 (258) &     2000$\pm$600 & 0.5$\pm$0.2 & 0.7$\pm$0.5  & 8$^{+3}_{-4}$ \\
%whole   &   0.67 (fixed)   &  0.31 $\pm$ 0.01    &  143$^{+2} _{-1}$  & 509.9 (265) &     2000$\pm$600 & 0.5$\pm$0.2 & 1.3$\pm$0.8  & 9 \\  
%%%%%
\hline
%\multicolumn{9}{c}{HESS J1731$-$347}   \\
HESS J1731$-$347 \\ \cline{1-1}
NE    &   1 (fixed)   &  0.97 $^{+0.46} _{-0.26}$   &  3.6$^{+1.2} _{-0.1}$  & 102.7 (86) &  2500$\pm$1000 & 0.5$\pm$0.1 & 0.65$^{+0.61}_{-0.55}$  & 6$^{+3}_{-12}$ \\ 
%%%%%
\hline
%\multicolumn{9}{c}{SN 1987A}  \\
SN 1987A \\ \cline{1-1}
%jointfit_7_chandraAll_1_Thermal-free.xcm
%whole & 0.24 (\check) & 0.69$^{+0.14} _{-0.11}$ &  2.11$^{+0.17} _{-0.02}$  & 76.3 (64)  &  6700$\pm$800 & 0.9$\pm$0.1 & 6.6$\pm$2.6  & 165$^{+16}_{-17}$ \\ 
%whole-RS$^{\rm (e)}$ & 0.24 (fixed) & 0.69$^{+0.14} _{-0.11}$ &  2.11$^{+0.17} _{-0.02}$  & 76.3 (64) &  4800$\pm$300 & 0.9$\pm$0.1  & 3.4$\pm$1.0  & 165$^{+11}_{-13}$ \\ 
whole & 0.18$\pm$0.02  & 0.70$^{+0.36} _{-0.08}$ &  2.49$^{+0.07} _{-0.02}$  & 610.7 (497)  &  6700$\pm$800 & 0.9$\pm$0.1 & 6.5$\pm$2.6  & 165$^{+16}_{-17}$ \\ 
whole-RS$^{\rm (e)}$ & 0.18$\pm$0.02  & 0.70$^{+0.36} _{-0.08}$ &  2.49$^{+0.07} _{-0.02}$  & 610.7 (497) &  4800$\pm$300 & 0.9$\pm$0.1  & 3.4$\pm$1.0  & 165$^{+11}_{-13}$ \\ 
%%%%%%%%%%%%%%%%
%% end
\enddata
\tablecomments{
%\tnote{\dag} X-ray observation data used for the spectral fitting. C: \chandra. N: \nustar. S: \suzaku. \\ %X: \xmm. 
(a) Flux is calculated for the 4--6 keV band in Cassiopeia A, Kepler, Tycho, and SN 1006 in units of $10^{-14}$~\flux, while is for the 2--10 keV band in the other SNRs in units of $10^{-12}$~\flux. \\
(b) References (see also \tabref{tab:dataset_SNRs}). G1.9$+$0.3: \cite{Borkowski2017}. Cassiopeia A: \cite{Patnaude2009}. Kepler: \cite{Vink2008_kepler,Katsuda2008_kepler}. Tycho: \cite{Katsuda2010_tycho,Williams2013}. G330.2$+$1.0: \cite{Borkowski2018}. SN 1006: \cite{Winkler2014}. \rxj: \cite{Tsuji2016,Acero2017}. RCW 86: \cite{Yamaguchi2016}. Vela Jr.: \cite{Katsuda2008_velajr,Allen2015}. HESS J1731$-$347: \cite{HESS2011_hessj1731}. SN 1987A: \cite{Frank2016}. \\
(c) Expansion parameter. \\
(d) Lower limit of magnetic field required for cooling-limited assumption (\eqref{eq:Blow}). \\
(e) RS: Assuming the reverse shock or reflection shock. \\
}
\end{deluxetable*}

%% file: make_FigTab/vshe0_diagram.tex
%%%%%%%%%%%%%%%%%%%%%%%%%%%%
%%%%%%%%%%%%%%%%%%%%%%%%%%%%

%%%%%%%%%%%%%%%%%%%%%%%%%%%%%%%%%%%%%%
\begin{figure*}[ht!]
%\gridline{
%\includegraphics[bb=0 0 539 386,width=0.33\textwidth]{\pathwork/Vshe0/Vshe0_case2_G1.9_20191127.pdf}{}{G1.9$+$0.3} }   
\gridline{
        \fig{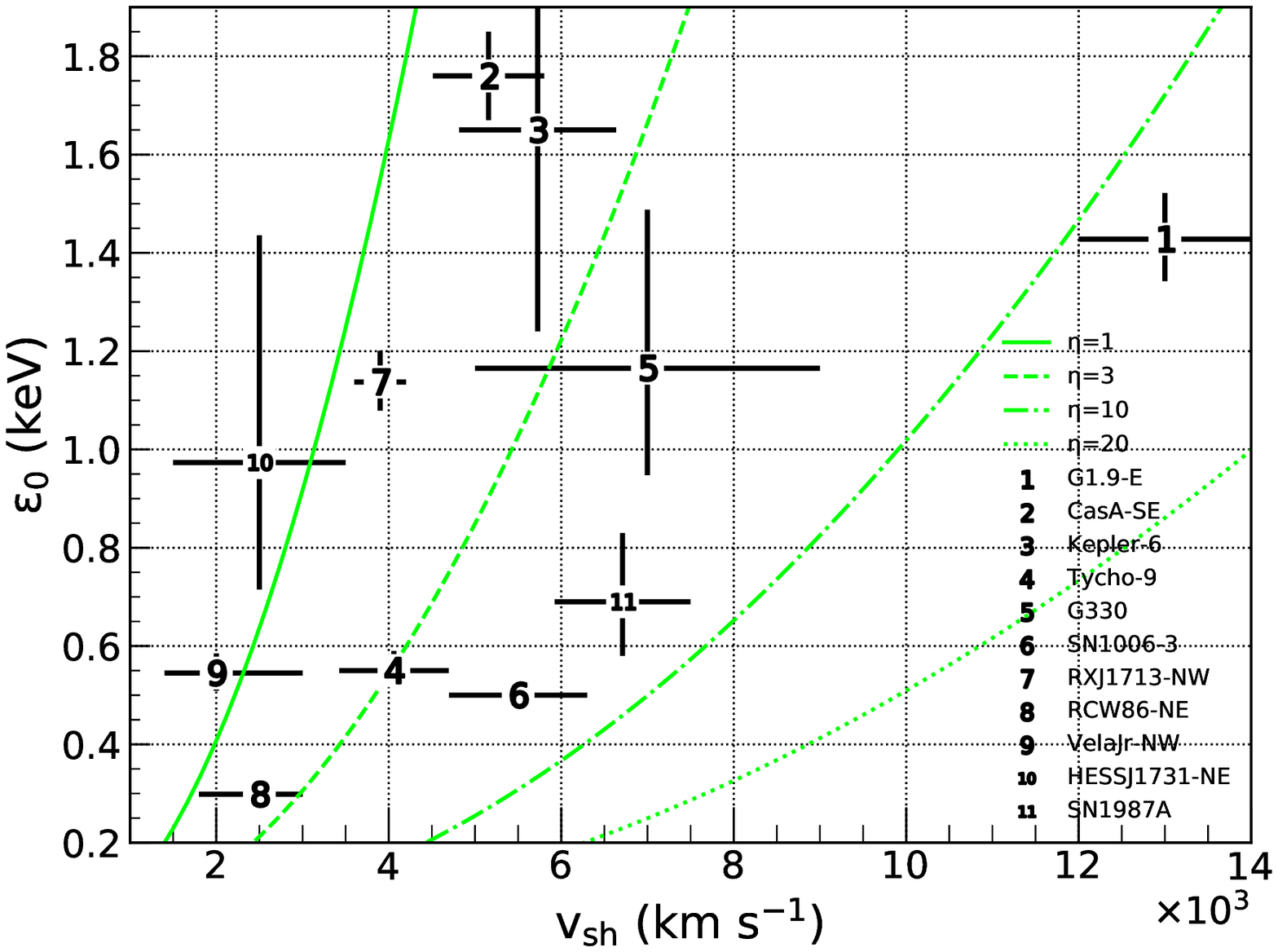}{0.33\textwidth}{All SNRs (the maximum-$\varepsilon_0$ regions)}
        \fig{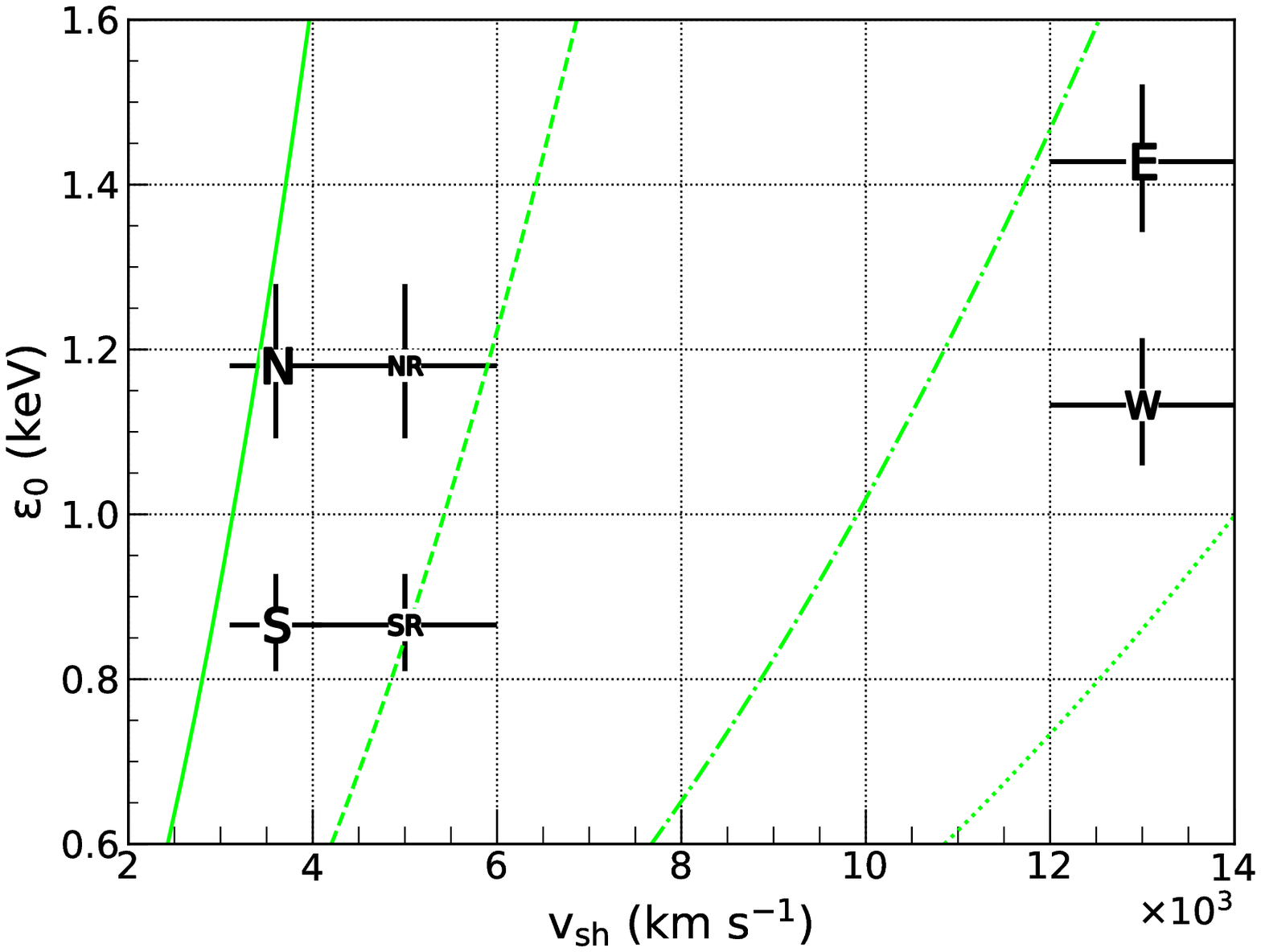}{0.33\textwidth}{G1.9$+$0.3}
        \fig{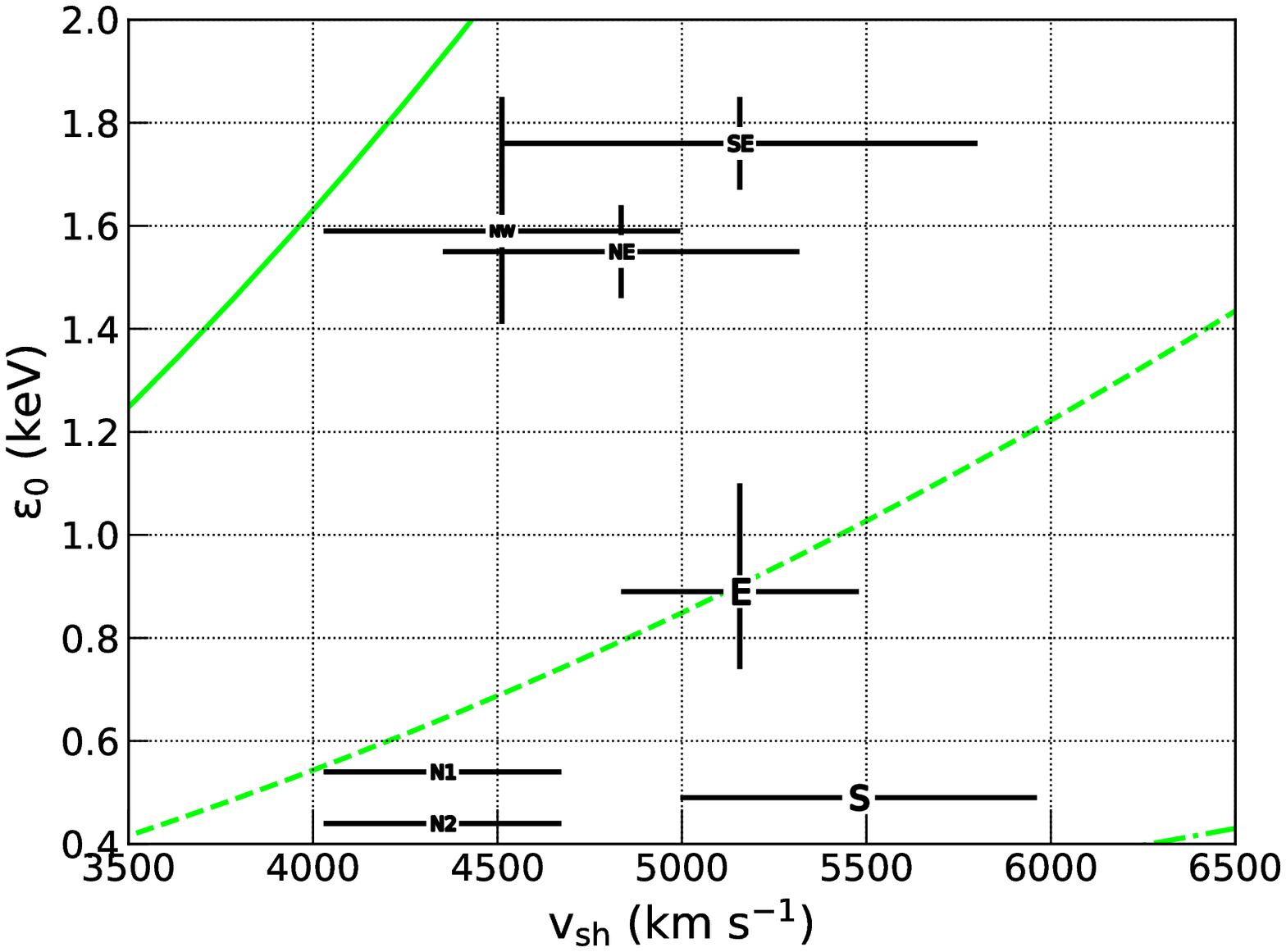}{0.33\textwidth}{Cassiopeia A} 
          }
\gridline{
        \fig{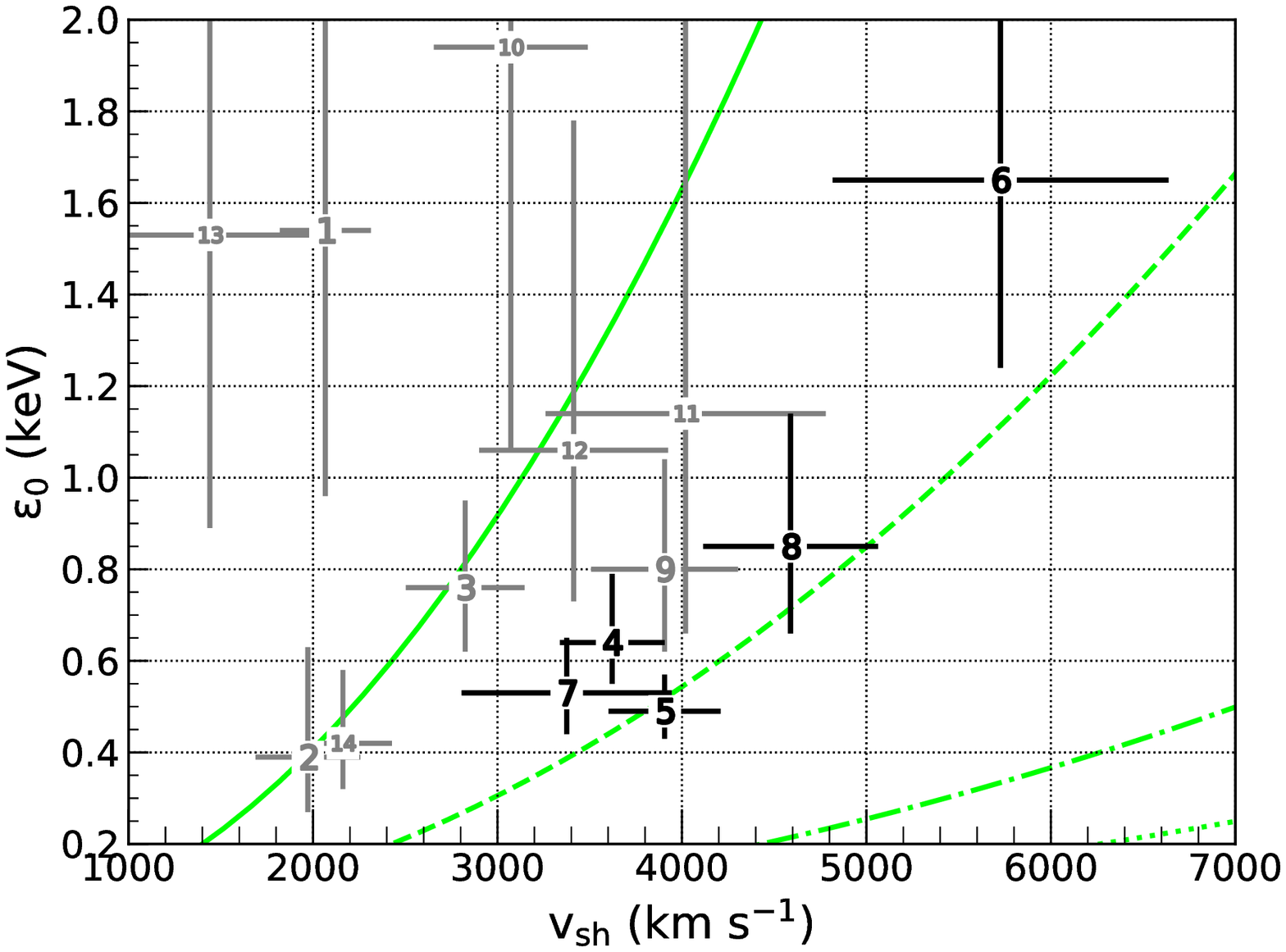}{0.33\textwidth}{Kepler}
        \fig{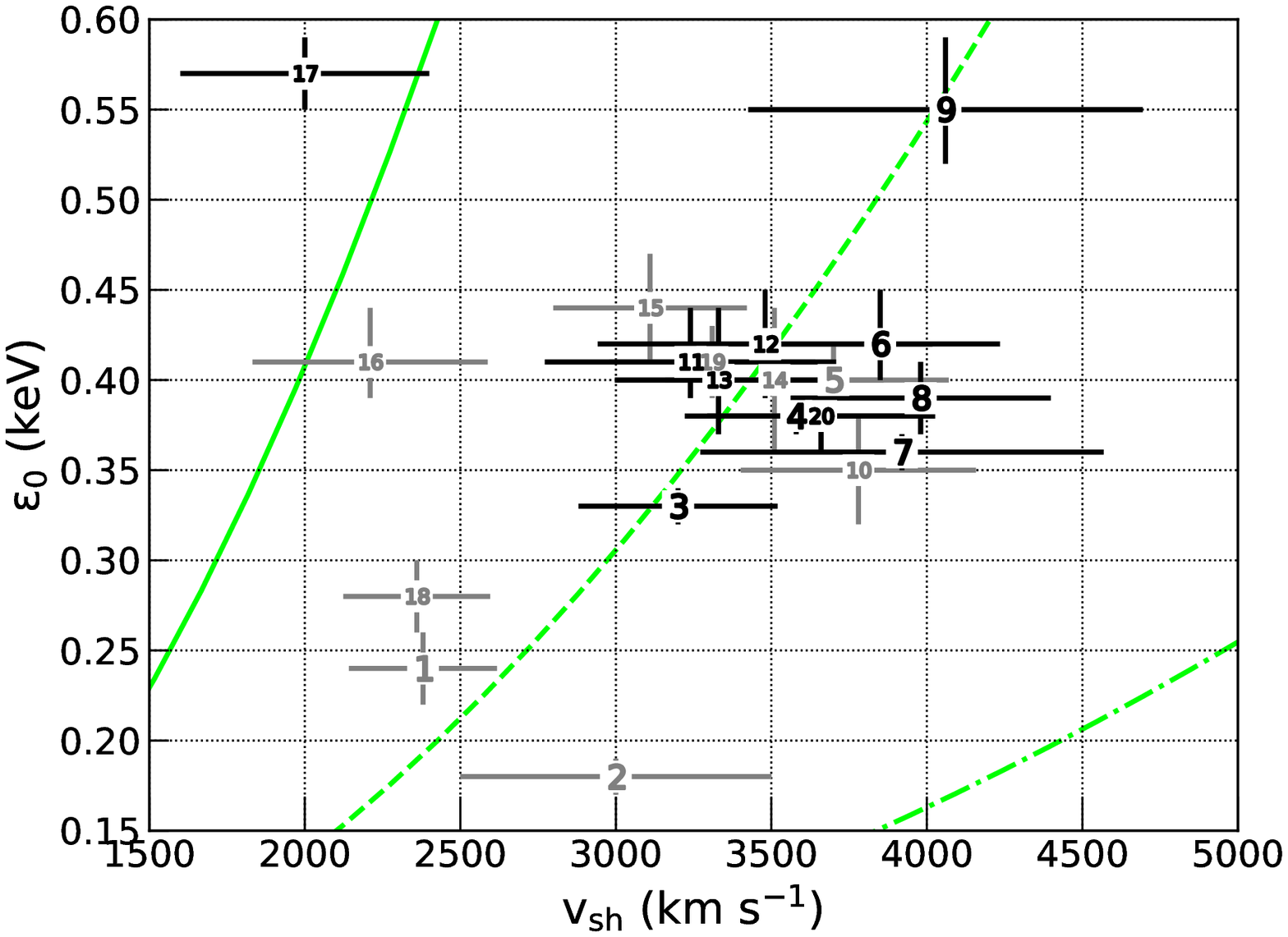}{0.33\textwidth}{Tycho}
        \fig{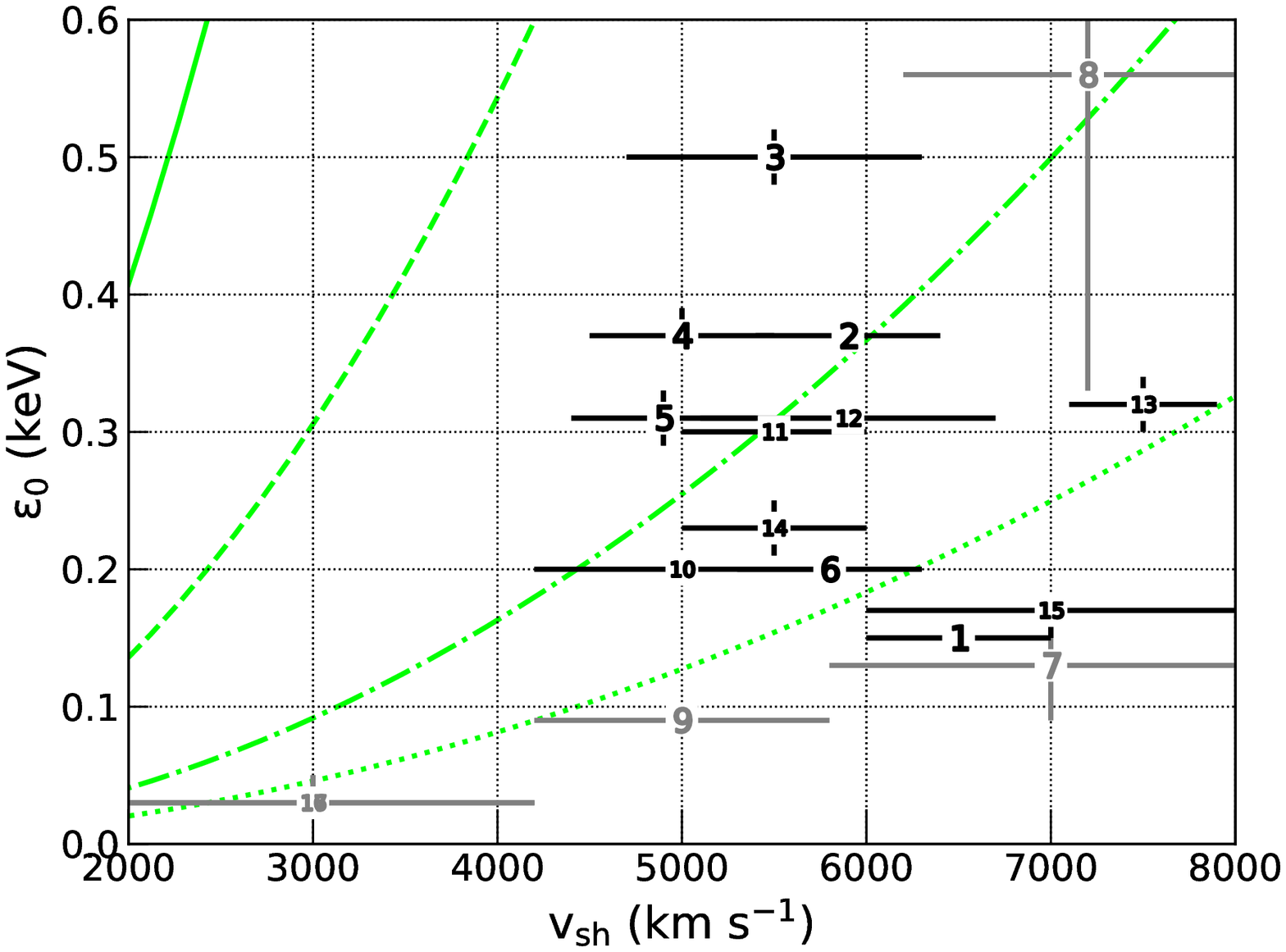}{0.33\textwidth}{SN 1006}
        }
\caption{
$\varepsilon_0$--$v_{\rm sh}$ diagram.
Green solid, dashed, dash-dotted, and dotted lines indicate $\eta$ of 1, 3, 10, and 20, respectively.
Top left: Each plot is taken from the region having the largest $\varepsilon_0$ (or the maximum $v_{\rm sh}$) in each SNR, where the region is highlighted in \figref{fig:Bohm_image_all}.
For Kepler, Tycho, and SN 1006, results of the synchrotron-dominated regions are shown in black, while those of the thermal-dominated regions are shown in grey (see \secref{sec:synch_spectra}).
In G1.9$+$0.3, NR and SR indicate the results assuming that the emission originates from the reverse shock in the north and south regions, respectively.
}
\label{fig:vshe0}
\end{figure*}

%%%%%%%%%%%%%%%%%%%%%%%%%%%%
%%%%%%%%%%%%%%%%%%%%%%%%%%%%

%% file: SNRs_nonthermal_individual_vshe0.tex
\subsubsection{G1.9$+$0.3}
G1.9$+$0.3, known as the youngest SNR in our galaxy, showed that the cutoff energy parameters were slightly variable inside the remnant:
$\varepsilon_0$ was obtained to be 1.1--1.2 keV in the entire remnant, the western and northern regions,
whereas it was slightly higher in the eastern rim ($\varepsilon_0 \approx 1.4$ keV) and lower in the southern part ($\varepsilon_0 \approx 0.9$ keV).
It should be noted that these differences in the measured cutoff energy parameters may have been underestimated because of the overlapped subregions and the limited angular resolution of \nustar.
In the case of forward shock, the Bohm factor was estimated to be 12--15 in the bright rim of E and W with the faster shock velocity of $\sim$13000 \kms. This was roughly consistent with $\eta \sim 20$ obtained in \cite{Aharonian2017}.
However, $\eta$ was 1--2 in the fainter area of N and S, with the slower shock speed of $\sim$4000 \kms.
The $\eta$ parameter averaged over the entire remnant (the N, E, S, and W regions) was $7.5 \pm 0.8$.
If we assume the outward proper motions in the N and S rims, which were measured in \cite{Borkowski2017}, correspond to the reverse shock,
the upstream speed in the rest frame of the reverse shock is given by
\begin{eqnarray}
u_1 = \frac{R_{\rm ref}}{t_{\rm age}} - v_{\rm obs} 
\label{eq:u12}
\end{eqnarray}
%which is same as \eqref{eq:u1} in \secref{sec:nustar_rxj1713}.
\citep{Sato2018,Tsuji2019}.
Indeed the reverse shock might play a major role in the northern region where the radio synchrotron emission is strong, and the proper motion of the shock is reduced, as proposed in \citep{Brose2019}.
Using \eqref{eq:u12}, $u_1$ in N and S is approximately calculated to be 5000 \kms, resulting in $\eta\sim $2--3.
We apparently see a significant difference in the acceleration efficiency across the remnant:
the Bohm factor in the bright E--W rim is approximately one order of magnitude higher than that in the N--S part in the case of either the forward shock or reverse shock.
%Any comment about the difference\check.

\subsubsection{Cassiopeia A}
The seven sectors (N1--2, NE, E, SE, S, and NW) are roughly compatible with the regions which \cite{Patnaude2009} measured proper motions and showed that the speeds are almost constant of $\sim$5000 \kms, assuming a distance of 3.4 kpc.
%Because the widths of all subregions are 15\arcsec\ and the lengths range from 26\arcsec (S1) to 63\arcsec (NE1), we used only the \chandra\ data for spectral fitting.
% (fitting) The \chandra\ spectra were fitted in 3.3--8 keV except for 6.4--6.8 keV (Fe lime emission) using the absorbed ZA07 model with a fixed column density ($N_H$) of 0.8$\times 10^{22}~\columnd$. The results were consistent when fitting with the ZA07 plus thermal (VNEI) model in 0.5--8 keV.
The cutoff energy parameters indicate significant varieties, showing the highest value in the SE rim and  the lowest value in the N2 region.
Because of the constant shock speeds and different cutoff energy parameters, the $\eta$ values also varied from 1.5 in SE and 6.3 in S.
This trend was consistent with the previous result in \cite{Stage2006}.
In the case of the forward shock in Cassiopeia A, the variation of $\eta$ is attributed to the different values of $\varepsilon_0$, which is opposite to G1.9$+$0.3 showing the variable shock speed and roughly constant $\varepsilon_0$.
If we averaged the $\eta$ parameters obtained in the seven small regions, the averaged $\eta$ was derived to be $3.1 \pm 0.3$.
%Comment on the thermal component and the acceleration\check.

\subsubsection{Kepler's SNR}
We defined 14 subregions that are almost comparable to those defined by \cite{Katsuda2008_kepler}, who performed proper motion measurements.
Five of these subregions (labeled as 4--8) turned out to be synchrotron dominated and highlighted in \figref{fig:vshe0}.
The observed $v_{\rm sh}$--$\varepsilon_0$ relation of these five regions indicated a clear correlation of $\varepsilon_0$ and $v_{\rm sh}$ (i.e., the higher the shock speed is, the larger the cutoff energy parameter is).
In addition, the correlation is well described with the theoretical curve with $\eta$ of 2--3.
This clarifies that the acceleration efficiency is constant ($\eta = $2--3) and independent of the sites within the forward shock located in the outermost eastern and southern rims of Kepler.
The averaged $\eta$ over the 5 synchrotron dominated regions was $2.4 ^{+0.3}_{-0.4}$.
Our analysis, for the first time, found that the particle acceleration in Kepler at the evolutional age of $\sim$400 years significantly deviates from the maximum rate of $\eta =1$.

\subsubsection{Tycho's SNR}
\figref{fig:vshe0} presents the $v_{\rm sh}$--$\varepsilon_0$ scatter plot measured in Tycho,
in which 11 regions (subregions 3, 4, 6--9, 11--13, 17, and 20) are synchrotron dominated.
%Note that nine subregions (boxes 1, 2, 5, 10, 14, 15, 16, 18, and 19) are dominated by the thermal components, which result in larger reduced chi-squared values with $\chi^2_{\rm red} \geq 3$ (\tabref{tab:Bohm_result_all_in_one}), and these thermal regions are removed in \figref{fig:vshe0}. 
Tycho shows slightly variable but theoretical predicted plots with $\varepsilon_0 = 0.3$--0.6 keV and $v_{\rm sh} = 3000$--4000 \kms\ \citep{Katsuda2010_tycho,Williams2013}.
This corresponds to nearly constant acceleration efficiency with $\eta$ of 3--4 (and the average value of $3.1 \pm 0.3$) in the forward shock of Tycho.
One exceptional region is subregion 17 located in the eastern rim.
The $\eta$ value should be $\sim$1 on the subregion 17, inferred from a slow shock speed of 2000 \kms\ and a higher cutoff energy parameter of 0.57 keV.
This nonthermal radiation in the E rim might be unique and behave differently from the other parts because a precursor has been detected \citep{Lee2010}, and a dense clump known as {\it knot g} is present (see, for example, \cite{Ghavamian2000}).
The dense density might affect the lower shock velocity of {\it knot g}.
Note that our result was roughly consistent with that of \cite{Lopez2015}, although the model and region definition were different (i.e., they also showed the greater roll-off energies in the higher shock speeds and the exception of {\it knot g}).

\subsubsection{SN~1006}
%%%%%%%%%%%%%%%%%%%%%%%%%%%%%%%%%%%%

\if0
SN~1006 is one of the best-studied and historical SNRs in our galaxy.
The high galactic latitude of 14.6\degr\ makes it an ideal laboratory to demonstrate supernova explosions or particle acceleration in an SNR shock.
%It has been well-studied in multi-wavelengths from radio to TeV gamma-ray.
The nonthermal (synchrotron) radiation of SN~1006 is concentrated in the northeastern and southwestern limbs, whereas the thermal emission is detected in the northwestern and southeastern limbs and in the interior region.

%% spectral fitting
The two limb regions, fil1--3 with an angular size of $5.0\arcmin \times 1.2\arcmin$, were analyzed using both \chandra\ and \nustar.
The subregions NE1--7 and SW1--7 had angular sizes of $2.0\arcmin \times 2.0\arcmin$, for which we used the results obtained by \cite{Li2018}.
The other subregions (NE0, E1--3, S1, SW8--9, and NW1--4) were extracted with \chandra\ and fitted using the absorbed ZA07 plus thermal (Vpshock) model.
\fi

%% Vsh - e0 in SN1006
The filaments in the NE and SW limbs were dominated by synchrotron emission.
Thus, the spectra of the subregions 2--4 (NE) and 11--13 (SW), extracted both with \chandra\ and \nustar, represent the typical properties of the synchrotron radiation from this remnant.
%The fitting results of these spectra using \chandra\ and \nustar\ were as follows (see also \figref{fig:Bohm_result_SN1006}). In fil1 (NE), the cutoff energy parameter was obtained as 0.40 $\pm$ 0.01 keV, resulting in $\eta=9.0\pm1.7$ with a shock speed of 5900 $\pm$ 500 \kms. In fil2 and fil3 (SW), the cutoff energy parameter was obtained as 0.32 $\pm$ 0.01 keV, resulting in $\eta \approx 10$ with a shock speed of 5500 $\pm$ 500 \kms.
They showed somewhat small cutoff energy parameters ($\varepsilon_0 = 0.3$--0.4 keV) and  relatively high shock speeds ($v_{\rm sh} = 5000$--8000 \kms), suggesting inefficient particle acceleration ($\eta \sim 10$) in SN~1006.
Note that the cutoff energy parameters obtained in this paper were roughly consistent with those of \cite{Li2018}.
%% Vsh - e0 (sub-regions) in SN1006
%The $v_{\rm sh}$--$\varepsilon_0$ plots of the subregions along the outermost rim of SN~1006 are shown in \figref{fig:vshe0}.
%To produce the plots of NE1--7 and SW1--7, we used the values in the literature (i.e., the shock velocities were taken from \cite{Winkler2014}, and the cutoff energy parameters were taken from \cite{Li2018} in which they derived $\varepsilon_0$ using \chandra\ and \nustar). The cutoff energy parameters in NE tended to appear larger than those in SW, although the spectral property of each subregion may have been mixed with that of the neighbors due to the limited angular resolution of \nustar, as noted in \citep{Li2018}.
The other subregions (1, 5--10, and 14--17) were analyzed using only \chandra, as \nustar\ covered only two limbs in the NE and SW parts.
The synchrotron dominated regions (1--6 and 10--15) showed a clear variation in $\varepsilon_0$ but a nearly constant value in $v_{\rm sh}$.
The averaged $\eta$ parameter over the synchrotron dominated spectra was $14 \pm 1$.

%{\bf
The validity of the loss-limited assumption is controversial in SN~1006.
\cite{Katsuda2010_sn1006} suggested that the accelerated electrons may not be loss-limited, while \cite{Miceli2013} found evidence supporting the loss limited scenario.
In this work, we obtained $B$ should be larger than 20--30 \uG\ for the ZA07 model to be applicable to SN~1006.
This condition can be sufficiently fulfilled, adopting $B \sim 300$~\uG\ that was estimated by the filament width \citep{Bamba2003,Volk2005}. 
%}

\begin{figure}
\begin{center}
    %% image
%\includegraphics[width=0.48\linewidth]{\pathwork/Azimuth_anything/Azimuth_All_case2_SN1006_20191127.eps}
%\includegraphics[width=0.48\linewidth]{\pathwork/Azimuth_anything/ThetaEta_case2_SN1006_20191127.eps}
%\includegraphics[width=0.48\linewidth]{\pathwork/paper/thetaEta_SN1006.eps}
%\includegraphics[width=0.48\linewidth]{make_FigTab/SN1006-output.csv_theta_Bn-Eta.eps}
\plotone{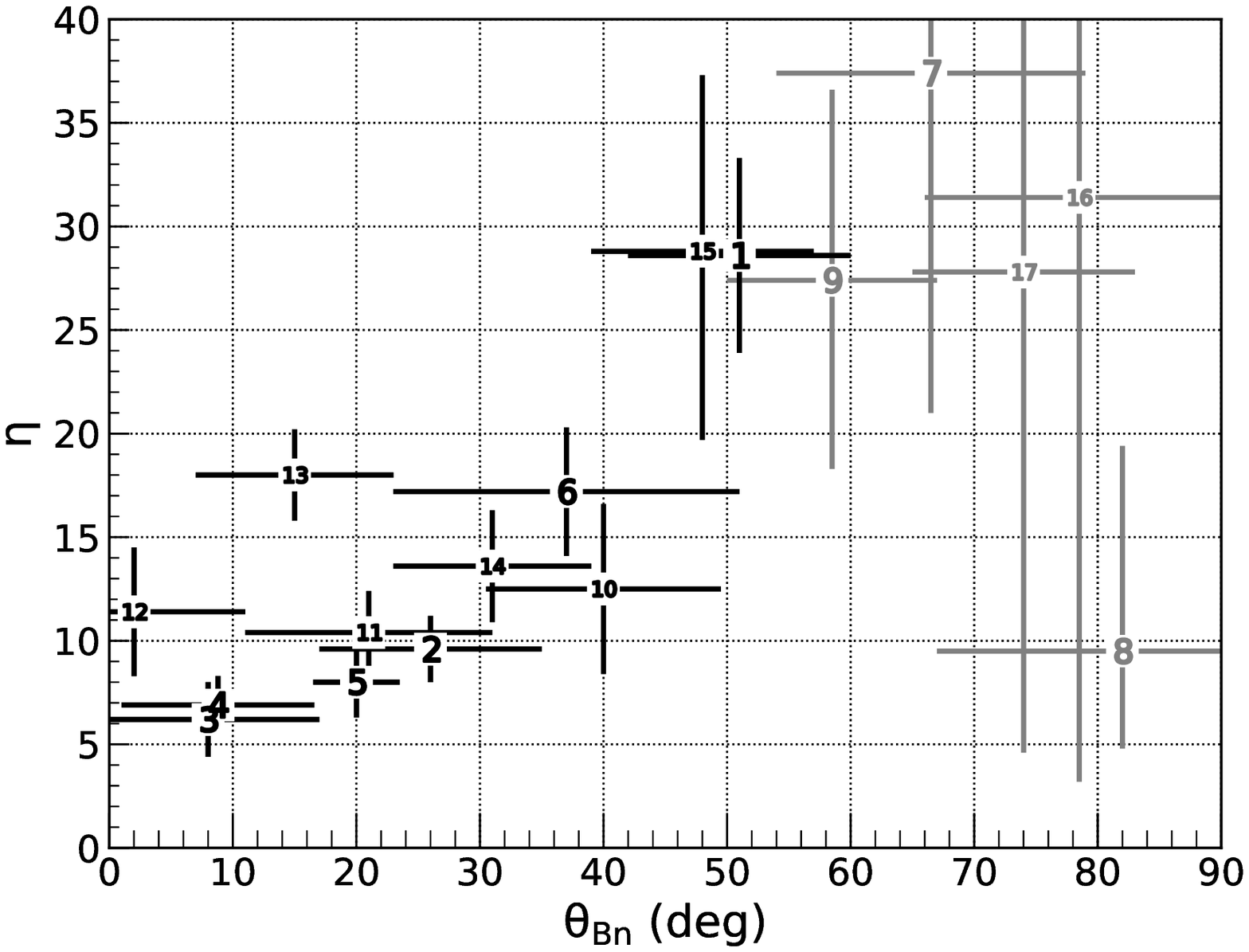}
\caption{
%Left: The azimuthal profiles of $\eta$, $\varepsilon_0$, and $v_{\rm sh}$ (counterclockwise from the north). The two limbs in the NE and SW are highlighted with gray bands.
The $\eta$ variation in SN~1006 as a function of shock obliquity, assuming the shock inclination is 60\degr\ from the north.
%(remove this figure and refer to \cite{Miceli2009} about the obliquity dependence\check)
The black and grey plots show the synchrotron dominated and thermal dominated regions, respectively.
}
\label{fig:Bohm_result_SN1006}
\end{center}
\end{figure}

%% azimuth
We also investigated a shock-obliquity dependency on particle acceleration in SN~1006.
SN~1006 is a unique remnant of which the ambient magnetic field was reported to be along the galactic plane that is approximately 60\degr\ counterclockwise inclined from the north \citep[see, e.g.,][]{Reynoso2013}.
This oriented magnetic field, combined with the Type Ia explosion in a high Galactic latitude of 14.6\degr, makes SN~1006 an ideal laboratory for studying the dependence of particle acceleration on magnetic field configurations (i.e., {\it parallel} or {\it perpendicular} shocks).
%We showed the azimuthal dependence of the cutoff energy parameters in \figref{fig:Bohm_result_SN1006}.
The azimuthal variations of roll-off frequencies (with the SRCUT model in XSPEC) and shock velocities were studied in \cite{Rothenflug2004,Miceli2009,Winkler2014}.
Our measurements of $\varepsilon_0$ indicated the larger cutoff energy parameters near the polar regions (NE and SW), which is consistent with the previous studies.
The observed variations of $\varepsilon_0$ and $v_{\rm sh}$ enabled us to reproduce a clear correlation between the shock obliquity $\theta_{\rm Bn}$ and acceleration efficiency $\eta$ in the synchrotron dominated regions, as illustrated in \figref{fig:Bohm_result_SN1006}. 
It should be noted that in NW and SE (subregions 7--9 and 16--17), where $\theta_{\rm Bn}$ is close to 90\degr,
the thermal emission dominates and the proper motion there is representative of X-ray emitting ejecta knots, not shock waves.
%To produce this scatter plot, we assumed the inclination of the magnetic field to be 60\degr\ from the north.
The shock-obliquity dependence on particle acceleration is discussed in \secref{sec:discussion}.

%{\bf 
\subsubsection{SN~1987A}
%%%%%%%%%%%%%%%%%%%%%%%%%%%%%%%%%%%%
Although the soft X-ray spectrum of SN~1987A below $\sim 7$ keV is well-fitted by a thermal model of two-temperature plasmas (i.e., Vpshock and Vequil in XSPEC), there exists an apparent excess in the \nustar\ spectrum above 10 keV, which cannot not described by the thermal model.
The origin of the hard X-ray emission with \nustar\ remains unclear \citep{Boggs2015,Reynolds2015}. %\footnote{https://ui.adsabs.harvard.edu/abs/2015AAS...22514022R/abstract}
\cite{Malyshev2019} recently reported GeV emission detected with \fermi\ likely from SN~1987A. 
Presuming that this GeV radiation originates from SN~1987A, there should be a population of nonthermal particles, and thus its synchrotron emission could be considered as a plausible interpretation of the hard X-ray excess.
% the hard X-ray emission can be synchrotron radiation from accelerated particles which are also responsible for the gamma-ray emission.
In this paper, we simply assumed that the hard component of \nustar\ is described by the nonthermal (synchrotron) radiation, resulting in $\varepsilon_0 \approx 0.7$ keV.

%%% shock speed
The expansion velocity of SN~1987A was reduced from $\sim$7000 \kms\ to $\sim$1000 \kms\ when impacting with the equatorial ring around 2003. 
We are interested in the shock velocity which is relevant for the acceleration of the X-ray emitting particles, thus adopted the former value (before the impact) for the shock speed of SN~1987A.
The following two factors should be addressed: 
(i) X-ray emitting particles cannot be accelerated if the shock speed is as slow as 1000 \kms. 
(ii) The cooling time of TeV electrons even in a strong magnetic field of $B \sim 300$ \uG\ exceeds significantly 10 years, $t_{\rm syn} = 140~{\rm yr}~(E/1~{\rm TeV})^{-1}~(B/300~{\rm \uG})^{-2}$. 
Therefore, in the framework on the adopted stationary approach model by \cite{ZA07}, we should use the shock velocity relevant to the moment of TeV particle acceleration (i.e., we chose $v_{\rm sh} \approx 7000$~\kms).
Combined with the cutoff energy parameter, the Bohm factor was estimated to be $\sim$7.
\if0
%% 2020-11-05
(1) X-ray emitting particles cannot be accelerated if the shock speed is as slow as 1000 km/s. (2) During 10 years (from the ER impact in 2003 to 2012--2014 when the NuSTAR data were taken) accelerated electrons in the TeV range do not loose their energy (i.e., energy loss time scale due to the synchrotron radiation is much larger than 10 years even with B=300 \uG) and still contain the information of acceleration before 2003 even though the ER impact might have stopped accelerating particles. (3) The ZA07 model assumes a steady state.
Combined with the cutoff energy parameter, the Bohm factor was obtained to be 7.
%% 2-10 keV speed; and radio
It also should be noted that the speed of the shock accelerating particles is ambiguous. The 2-10 keV morphology showed a speed of 3100 \kms\ after the impact, which is comparable with the radio expansion speed of $\approx$3900 \kms (ATCA 9 GHz; synchrotron radiation; \cite{Ng2013}). 
It has been suggested that the soft/cool and hard/warm X-ray emission respectively represent slow transmitted shocks in the dense clump and shocks (including reflected shocks) moving in the lower density ring \citep[see, e.g., ][]{Frank2016}. 
Adopting $v_{\rm sh}=3100$ \kms\ and $\varepsilon_0 = 0.7$ keV, $\eta$ is estimated to be 1.4.
\fi
The detailed discussion of particle acceleration in SN~1987A will be given in the future publication.
%}

%\if0
%%%%%%%%%%%%%%
%\fi

\subsubsection{Others}
%%%%%%%%%%%%%%%%%%%%%%%%%%%%%%%%%%%%
%Short comments on the other SNRs \check.
We could not investigate the spatially resolved $\varepsilon_0$--$v_{\rm sh}$ relation in the other SNRs (i.e., G330.2$+$1.0, \rxj, RCW 86, Vela Jr., HESS J1731$-$347, and SN~1987A) due to lacking observational data.
The proper motions were measured along the rim of G330.2$+$1.0 \citep{Borkowski2018}, but due to the large uncertainty of the shock speed and the cutoff energy parameter the spatially resolved $\varepsilon_0$--$v_{\rm sh}$ plot was not presented in this paper.
The $\varepsilon_0$--$v_{\rm sh}$ diagram of the NW rim in \rxj\ was presented in \cite{Tsuji2019}.
To date the shock speeds were obtained only in specific regions, the NE rim in RCW 86 and the NW rim in Vela Jr., since these two SNRs have relatively large angular sizes of $\sim$0.7\degr\ and $\sim$2\degr, respectively.

%% file: make_FigTab/EvolvingEta_table.tex
%%     
%%% ------------------------------------------------%%%
\begin{deluxetable}{l | cccc | cccc}
\tablecaption{Parameters of evolution of $\eta$ (\eqref{eq:Bohm_eta_age} and \eqref{eq:mEta})
\label{tab:eta} }
%\tablewidth{700pt}
\tabletypesize{\small}
\tablehead{
\colhead{} &  \colhead{$C_{\rm age}$ }&    \colhead{$\delta_{\rm age}$ }            & \colhead{$\chi^2$ (dof)}    & \colhead{$\rho ^{({\rm a} )}$} &  \colhead{$C_m$ }&    \colhead{$\delta_m$ }            & \colhead{$\chi^2$ (dof)}    & \colhead{$\rho ^{({\rm a} )}$} 
%\colhead{}                         & \colhead{(yr)}          &  \colhead{}    & \colhead{(kpc)} & \colhead{(\kms)}   & \colhead{} 
} 
\startdata
%%%%%%%%%%%%%%
All SNRs & 1.5$\pm$0.2   &     0.41$\pm$0.08     &    17 (8)  & $-$0.70 (2.4$\sigma$)     &  5.2$\pm$1.0     & 4.0$\pm$0.8 & 19 (8) & 0.60 (1.9$\sigma$)  \\
All SNRs with averaged $\eta$ & 2.2$\pm$0.2   &     0.37$\pm$0.06     &    89 (8)  & $-$0.65 (2.2$\sigma$)    &  3.3$\pm$0.4     & 0.67$\pm$0.34 & 118 (8) & 0.52 (1.6$\sigma$)  \\
%All SNRs with averaged $\eta$; w/o 1006 & 2.0$\pm$0.2   &     0.41$\pm$0.06     &    17 (8)  & $-$0.77 (2.8$\sigma$)    &  3.3$\pm$0.4     & 0.93$\pm$0.36 & 48 (8) & 0.61 (2.0$\sigma$)  \\
%% type I
\hline
Type Ia & 0.85$\pm$0.55   &     1.5$\pm$0.5     &    6.8 (2)  & $-$0.53 (0.9$\sigma$)     &  4.4$\pm$1.2     & 0.69$\pm$0.63 & 11 (2) & 0.50 (0.86$\sigma$)  \\
Type Ia with averaged $\eta$ & 4.4$\pm$0.6   &     -0.29$\pm$0.17     &    75 (2)  & $-$0.17 (0.27$\sigma$)     &  3.5$\pm$0.4     & 0.023$\pm$0.30 & 77 (2) & 0.32 (0.5$\sigma$)  \\
%Type Ia with averaged $\eta$; w/o 1006 & 1.2$\pm$0.3   &     1.0$\pm$0.2     &    7 (2)  & $-$0.48 (0.8$\sigma$)     &  3.4$\pm$0.4     & 0.26$\pm$0.34 & 17 (2) & 0.52 (0.9$\sigma$)  \\
%% type II
\hline
Type II & 1.4$\pm$0.3   &     0.41$\pm$0.09     &    2.5 (3)  & $-$0.86 (2.2$\sigma$)     &  9.6$\pm$3.2     & 6.2$\pm$1.3    & 1.6 (3) & 0.95 (2.9$\sigma$)  \\
Type II with averaged $\eta$ & 1.8$\pm$0.2  &     0.40$\pm$0.07     &    1.9 (3)  & $-$0.90 (2.4$\sigma$)     &  5.8$\pm$2.2     & 3.1$\pm$1.2& 19 (3) & 0.94 (2.8$\sigma$)  \\
%%%%%%%%%%%%%%%%
%% end
\enddata
\tablecomments{
$^{({\rm a})}$ Pearson's correlation coefficient.
}
\end{deluxetable}

%% file: make_FigTab/Bohm_dataset_Chandra_thispaper.tex
%%%%%%%%%%%%%%%%%%%%%%%%%%%%%%%%%%%%
%%%%%%%%%%%%%%%%%%%%%%%%%%%%%%%%%%%%

\startlongtable
\begin{deluxetable*}{ccccccc}
\tablecaption{Log of \chandra\ observations
\label{tab:Bohm_dataset_chandra}
}
%\tablewidth{700pt}
\tabletypesize{\scriptsize}
\tablehead{
\colhead{Name} & \colhead{Obs ID} & \colhead{Effective time} & \colhead{Date} & \colhead{RA}  & \colhead{Dec}  & \colhead{Roll} \\
 \colhead{}   & \colhead{} & \colhead{(ks)} & \colhead{(yyyy-mm-dd)} & \colhead{(deg)} & \colhead{(deg)} & \colhead{(deg)}  
} 
\startdata
%%%%%%%%%%%%%%
 G1.9$+$0.3 &  6708 &  23.9 &  2007-02-10 &  267.2 &  -27.2 &  91.8   \\
 G1.9$+$0.3 &  8521 &  25.7 &  2007-03-03 &  267.2 &  -27.2 &  91.8   \\
 G1.9$+$0.3 &  10111 &  68.3 &  2009-07-23 &  267.2 &  -27.2 &  270.0   \\
 G1.9$+$0.3 &  10112 &  50.8 &  2009-07-18 &  267.2 &  -27.2 &  283.2   \\
 G1.9$+$0.3 &  10928 &  35.4 &  2009-07-13 &  267.2 &  -27.2 &  270.2   \\
 G1.9$+$0.3 &  10930 &  82.1 &  2009-07-26 &  267.2 &  -27.2 &  270.0   \\
 G1.9$+$0.3 &  12689 &  155.6 &  2011-07-14 &  267.2 &  -27.2 &  277.3   \\
 G1.9$+$0.3 &  12690 &  48.2 &  2011-05-16 &  267.2 &  -27.2 &  79.2   \\
 G1.9$+$0.3 &  12691 &  184.0 &  2011-05-09 &  267.2 &  -27.2 &  79.2   \\
 G1.9$+$0.3 &  12692 &  162.6 &  2011-05-12 &  267.2 &  -27.2 &  79.2   \\
 G1.9$+$0.3 &  12693 &  127.5 &  2011-05-18 &  267.2 &  -27.2 &  79.2   \\
 G1.9$+$0.3 &  12694 &  159.3 &  2011-05-20 &  267.2 &  -27.2 &  79.2   \\
 G1.9$+$0.3 &  12695 &  39.5 &  2011-05-23 &  267.2 &  -27.2 &  79.2   \\
 G1.9$+$0.3 &  13407 &  48.4 &  2011-07-18 &  267.2 &  -27.2 &  277.3   \\
 G1.9$+$0.3 &  13509 &  55.3 &  2011-07-22 &  267.2 &  -27.2 &  277.3   \\
 G1.9$+$0.3 &  16947 &  38.8 &  2015-05-04 &  267.2 &  -27.2 &  86.7   \\
 G1.9$+$0.3 &  16948 &  39.6 &  2015-07-14 &  267.2 &  -27.2 &  271.6   \\
 G1.9$+$0.3 &  16949 &  9.1 &  2015-05-20 &  267.2 &  -27.2 &  75.2   \\
 G1.9$+$0.3 &  17651 &  111.6 &  2015-05-05 &  267.2 &  -27.2 &  86.7   \\
 G1.9$+$0.3 &  17652 &  26.2 &  2015-05-09 &  267.2 &  -27.2 &  86.7   \\
 G1.9$+$0.3 &  17663 &  56.5 &  2015-07-24 &  267.2 &  -27.2 &  271.6   \\
 G1.9$+$0.3 &  17699 &  19.8 &  2015-07-17 &  267.2 &  -27.2 &  271.6   \\
 G1.9$+$0.3 &  17700 &  14.9 &  2015-08-31 &  267.2 &  -27.2 &  260.2   \\
 G1.9$+$0.3 &  17702 &  36.9 &  2015-07-15 &  267.2 &  -27.2 &  271.6   \\
 G1.9$+$0.3 &  17705 &  9.9 &  2015-07-25 &  267.2 &  -27.2 &  271.6   \\
 G1.9$+$0.3 &  18354 &  29.7 &  2015-09-01 &  267.2 &  -27.2 &  260.2   \\
(total) & & 1659.6    \\
 \hline
 Cassiopeia A &  4634 &  148.6 &  2004-04-28 &  350.9 &  58.8 &  59.2 \\
 Cassiopeia A &  4635 &  135.0 &  2004-05-01 &  350.9 &  58.8 &  59.2 \\
 Cassiopeia A &  4636 &  143.5 &  2004-04-20 &  350.9 &  58.8 &  49.8 \\
 Cassiopeia A &  4637 &  163.5 &  2004-04-22 &  350.9 &  58.8 &  49.8 \\
 Cassiopeia A &  4638 &  164.5 &  2004-04-14 &  350.9 &  58.8 &  40.3 \\
 Cassiopeia A &  4639 &  79.0 &  2004-04-25 &  350.9 &  58.8 &  49.8 \\
 Cassiopeia A &  5196 &  49.5 &  2004-02-08 &  350.9 &  58.8 &  325.5 \\
 Cassiopeia A &  5319 &  42.3 &  2004-04-18 &  350.9 &  58.8 &  49.8 \\
 Cassiopeia A &  5320 &  54.4 &  2004-05-05 &  350.9 &  58.9 &  65.1 \\
 (total) & & 980.3    \\
  \hline
 Kepler &  116 &  48.8 &  2000-06-30 &  262.7 &  -21.5 &  261.1 \\
 Kepler &  4650 &  46.2 &  2004-10-26 &  262.7 &  -21.5 &  268.8 \\
 Kepler &  6714 &  157.8 &  2006-04-27 &  262.7 &  -21.4 &  89.0 \\
 Kepler &  6715 &  159.1 &  2006-08-03 &  262.7 &  -21.5 &  265.7 \\
 Kepler &  6716 &  158.0 &  2006-05-05 &  262.7 &  -21.4 &  89.5 \\
 Kepler &  6717 &  106.8 &  2006-07-13 &  262.7 &  -21.5 &  264.2 \\
 Kepler &  6718 &  107.8 &  2006-07-21 &  262.7 &  -21.5 &  264.8 \\
 Kepler &  7366 &  51.5 &  2006-07-16 &  262.7 &  -21.5 &  264.2 \\
 Kepler &  16004 &  102.7 &  2014-05-13 &  262.7 &  -21.5 &  92.7 \\
 Kepler &  16614 &  36.4 &  2014-05-16 &  262.7 &  -21.5 &  92.7 \\
 (total) & & 975.1    \\
\hline
 Tycho  &  7639 &  108.9 &  2007-04-23 &  6.3 &  64.1 &  29.2 \\
 Tycho  &  8551 &  33.3 &  2007-04-26 &  6.3 &  64.1 &  29.2 \\
 Tycho  &  10093 &  118.4 &  2009-04-13 &  6.3 &  64.1 &  29.2 \\
 Tycho  &  10094 &  90.0 &  2009-04-18 &  6.3 &  64.1 &  29.2 \\
 Tycho  &  10095 &  173.4 &  2009-04-23 &  6.3 &  64.1 &  29.2 \\
 Tycho  &  10096 &  105.7 &  2009-04-27 &  6.3 &  64.1 &  29.2 \\
 Tycho  &  10097 &  107.4 &  2009-04-11 &  6.3 &  64.1 &  26.3 \\
(total) & & 737.1 \\
\hline
 G330.2$+$1.0 &  6687 &  50.0 &  2006-05-21 &  240.2 &  -51.6 &  3.3 \\
 G330.2$+$1.0 &  19163 &  74.1 &  2017-05-02 &  240.2 &  -51.6 &  30.2 \\
 G330.2$+$1.0 &  20068 &  74.1 &  2017-05-05 &  240.2 &  -51.5 &  30.2 \\
 (total) & & 198.2    \\
 \hline
 SN 1006 N &  13743 &  92.6 &  2012-04-28 &  225.8 &  -41.7 &  19.9 \\
 SN 1006 NE &  9107 &  68.9 &  2008-06-24 &  226.0 &  -41.9 &  280.4 \\
 SN 1006 NE &  732 &  68.1 &  2000-07-10 &  226.0 &  -41.9 &  280.2 \\
 SN 1006 SW &  13739 &  100.1 &  2012-05-04 &  225.6 &  -42.1 &  9.1 \\
 SN 1006 NW &  1959 &  89.0 &  2001-04-26 &  225.6 &  -41.8 &  30.2 \\
 SN 1006 NW &  13737 &  87.1 &  2012-04-20 &  225.6 &  -41.8 &  31.7 \\
 SN 1006 W &  13738 &  73.5 &  2012-04-23 &  225.4 &  -42.0 &  25.3 \\
 SN 1006 W &  14424 &  25.4 &  2012-04-27 &  225.4 &  -42.0 &  25.3 \\
 SN 1006 SE &  13741 &  98.5 &  2012-04-25 &  226.0 &  -42.0 &  24.6 \\
 SN 1006 S &  13742 &  79.0 &  2012-06-15 &  225.8 &  -42.1 &  289.1 \\
 (total) & & 782.2    \\
 \hline
\rxj\ NW &  736 &  29.6 &  2000-07-25 &  258.0 &  -39.6 &  282.5  \\
\rxj\ NW &  6370 &  29.8 &  2006-05-03 &  257.9 &  -39.6 &  64.8   \\
\rxj\ NW &  10090 &  28.4 &  2009-01-30 &  257.9 &  -39.5 &  98.6   \\
\rxj\ NW &  10091 &  29.6 &  2009-05-16 &  257.9 &  -39.5 &  53.8   \\
\rxj\ NW &  10092 &  29.2 &  2009-09-10 &  257.9 &  -39.6 &  266.1 \\
\rxj\ NW &  12671 &  89.9 &  2011-07-01 &  257.9 &  -39.6 &  304.5 \\
 (total) & & 236.5    \\
\hline
RCW 86 &  1993 &  92.0 &  2001-02-01 &  220.2 &  -62.7 &  80.2   \\
 RCW 86 NE &  4611 &  71.7 &  2004-06-15 &  221.3 &  -62.4 &  295.2  \\
 RCW 86 NE &  7642 &  69.2 &  2007-06-20 &  221.3 &  -62.3 &  299.0  \\
 RCW 86  &  10699 &  2.0 &  2009-06-14 &  220.5 &  -62.6 &  304.4  \\
 RCW86 &  13748 &  36.1 &  2013-02-14 &  220.1 &  -62.7 &  70.7   \\
 RCW 86 &  14890 &  26.7 &  2013-02-03 &  220.4 &  -62.2 &  75.2   \\
 RCW 86 &  15608 &  29.2 &  2013-02-05 &  220.4 &  -62.2 &  75.2   \\
 RCW 86 &  15609 &  37.6 &  2013-02-10 &  220.4 &  -62.2 &  75.2   \\
 RCW86 &  15610 &  23.1 &  2013-02-17 &  220.1 &  -62.7 &  70.7   \\
 RCW86 &  15611 &  25.9 &  2013-02-12 &  220.1 &  -62.7 &  70.7   \\
 RCW 86 NE &  16952 &  67.2 &  2015-06-25 &  221.3 &  -62.4 &  293.6  \\
(total) & & 480.7    \\
 \hline
 Vela Jr. NW &  3846 &  39.5 &  2003-01-05 &  132.3 &  -45.6 &  30.2  \\
 Vela Jr. NW &  4414 &  34.5 &  2003-01-06 &  132.3 &  -45.6 &  30.2  \\
% Vela Jr. South &  7638 &  54.5 &  2007-10-16 &  133.3 &  -47.3 &  98.2 \\
 Vela Jr. NORTH &  9123 &  39.7 &  2008-08-31 &  132.3 &  -45.7 &  146.2 \\
% Vela Jr. &  18640 &  39.5 &  2016-08-30 &  133.9 &  -46.7 &  150.9 \\
 (total) & & 113.7    \\
 \hline
  HESS J1731$-$347 &  9139 &  29.2 &  2008-04-28 &  263.0 &  -34.7 &  81.2 \\
 \hline 
 SN 1987A &  14697 &  67.6 &  2013-03-21 &  83.9 &  -69.3 &  264.2  \\    
 SN 1987A &  14698 &  68.5 &  2013-09-28 &  83.9 &  -69.3 &  79.4  \\    
 SN 1987A &  15809 &  70.5 &  2014-03-19 &  83.9 &  -69.3 &  266.1  \\    
 SN 1987A &  15810 &  48.3 &  2014-09-20 &  83.9 &  -69.3 &  84.2  \\    
 SN 1987A &  17415 &  19.4 &  2014-09-17 &  83.9 &  -69.3 &  84.2  \\  
  (total) & & 274.2   \\ 
\if0
\hline
 SN 1987A &  122 &  8.6 &  2000-01-17 &  83.9 &  -69.3 &  327.7 \\
 SN 1987A &  1967 &  98.8 &  2000-12-07 &  83.9 &  -69.3 &  8.6 \\
 SN 1987A &  1044 &  17.8 &  2001-04-25 &  83.8 &  -69.3 &  228.9 \\
 SN 1987A &  2831 &  49.4 &  2001-12-12 &  83.9 &  -69.3 &  3.9 \\
 SN 1987A &  2832 &  44.3 &  2002-05-15 &  83.8 &  -69.3 &  210.3 \\
 SN 1987A &  3829 &  49.0 &  2002-12-31 &  83.9 &  -69.3 &  345.3 \\
 SN 1987A &  3830 &  45.3 &  2003-07-08 &  83.8 &  -69.3 &  158.9 \\
 SN 1987A &  4614 &  46.5 &  2004-01-02 &  83.9 &  -69.3 &  343.0 \\
 SN 1987A &  4615 &  48.8 &  2004-07-22 &  83.8 &  -69.3 &  144.9 \\
 SN 1987A &  5579 &  31.9 &  2005-01-09 &  83.9 &  -69.3 &  335.2 \\
 SN 1987A &  5580 &  23.7 &  2005-07-11 &  83.9 &  -69.3 &  153.2 \\
 SN 1987A &  6668 &  42.3 &  2006-01-28 &  83.9 &  -69.3 &  316.4 \\
 SN 1987A &  6669 &  36.4 &  2006-07-27 &  83.9 &  -69.3 &  139.9 \\
 SN 1987A &  7636 &  33.5 &  2007-01-19 &  83.9 &  -69.3 &  325.1 \\
 SN 1987A &  7637 &  25.7 &  2007-07-13 &  83.9 &  -69.3 &  153.7 \\
 SN 1987A &  9142 &  6.6 &  2008-01-09 &  83.9 &  -69.3 &  335.2 \\
 SN 1987A &  9143 &  8.6 &  2008-07-04 &  83.9 &  -69.3 &  161.8 \\
 SN 1987A &  10130 &  6.0 &  2009-01-05 &  83.9 &  -69.3 &  339.4 \\
(total) & & 623.2    \\
\fi
%% end
\enddata
%\tablecomments{ }
\end{deluxetable*}

%% file: make_FigTab/Bohm_dataset_NuSTAR_thispaper.tex
%%%%%%%%%%%%%%%%%%%%%%%%%%%%
%%%%%%%%%%%%%%%%%%%%%%%%%%%%

\startlongtable
\begin{deluxetable*}{ccccccc}
\tablecaption{Log of \nustar\ observations
\label{tab:Bohm_dataset_nustar}
}
%\tablewidth{700pt}
\tabletypesize{\scriptsize}
\tablehead{
\colhead{Name} & \colhead{Obs ID} & \colhead{Effective time} & \colhead{Date} & \colhead{RA}  & \colhead{Dec}  & \colhead{Roll} \\
 \colhead{}   & \colhead{} & \colhead{(ks)} & \colhead{(yyyy-mm-dd)} & \colhead{(deg)} & \colhead{(deg)} & \colhead{(deg)}  
} 
\startdata
%%%%%%%%%%%%%%%%%%%%%%%%%%%%
  G1.9$+$0.3 &  40001015003 &  85.4 &  2013-07-08 &  267.2 &    $-$27.2 &  327.3 \\    
 G1.9$+$0.3 &  40001015005 &  121.6 &  2013-07-14 &  267.2 &    $-$27.2 &  327.3 \\    
 G1.9$+$0.3 &  40001015007 &  144.7 &  2013-07-27 &  267.2 &    $-$27.2 &  327.3 \\   
 (total)  &    & 351.7 \\
  \hline
 Cassiopeia A &  40021002002 &  270.9 &  2012-11-23 &  350.8 &  58.8 &  338.3 \\    
 Cassiopeia A &  40021002006 &  135.6 &  2013-03-02 &  350.9 &  58.8 &  248.7 \\    
 Cassiopeia A &  40021002008 &  189.3 &  2013-03-05 &  350.9 &  58.8 &  248.7 \\    
 Cassiopeia A &  40021003003 &  197.8 &  2013-05-28 &  350.9 &  58.8 &  151.2 \\    
 Cassiopeia A &  40021001002 &  170.1 &  2012-08-27 &  350.8 &  58.8 &  75.7 \\    
 Cassiopeia A &  40021001004 &  25.7 &  2012-10-07 &  350.7 &  58.8 &  33.0 \\    
 Cassiopeia A &  40021001005 &  184.5 &  2012-10-07 &  350.8 &  58.8 &  33.0 \\    
 Cassiopeia A &  40021002010 &  12.4 &  2013-03-09 &  350.9 &  58.8 &  248.7 \\    
 Cassiopeia A &  40021003002 &  12.4 &  2013-05-28 &  350.9 &  58.8 &  151.2 \\    
 Cassiopeia A &  40021011002 &  235.1 &  2013-10-30 &  350.9 &  58.8 &  6.8 \\    
 Cassiopeia A &  40021012002 &  205.8 &  2013-11-27 &  350.8 &  58.8 &  335.2 \\    
 Cassiopeia A &  40021015002 &  74.4 &  2013-12-21 &  350.9 &  58.8 &  312.3 \\    
 Cassiopeia A &  40021015003 &  136.9 &  2013-12-23 &  350.9 &  58.8 &  312.2 \\  
 (total)  &    & 1850.9 \\
 \hline
\if0
  Tycho &  40020001002 &  338.7 &  2014-04-12 &  6.4 &  64.1 &  222.9 \\    
 Tycho &  40020001004 &  262.1 &  2014-07-18 &  6.4 &  64.1 &  124.2 \\    
 Tycho &  40020011002 &  146.8 &  2014-05-31 &  6.4 &  64.1 &  169.1 \\
 (total)  &    & 672.1 \\
 \hline 
 Kepler &  40001020002 &  212.8 &  2014-10-11 &  262.7 &    $-$21.5 &  336.2 \\    
 Kepler &  90201021002 &  106.6 &  2017-02-07 &  262.7 &    $-$21.5 &  158.2 \\    
 Kepler &  90201021004 &  41.4 &  2017-04-22 &  262.7 &    $-$21.4 &  156.1 \\    
 Kepler &  90201021006 &  47.0 &  2017-10-08 &  262.6 &    $-$21.5 &  336.4 \\    
 Kepler &  90201021008 &  35.0 &  2018-06-04 &  262.7 &    $-$21.4 &  147.8 \\    
 Kepler &  90201021010 &  31.3 &  2018-06-17 &  262.6 &    $-$21.6 &  6.3 \\    
 Kepler &  10501005002 &  86.1 &  2019-03-17 &  262.7 &    $-$21.4 &  157.2 \\    
 (total)  &    & 560.2  \\
 \hline
\fi
 SN1006 NE &  40110001002 &  198.5 &  2016-03-02 &  225.9 &    $-$41.8 &  180.0 \\    
 SN1006 SW &  40110002002 &  204.8 &  2016-03-08 &  225.5 &    $-$42.0 &  180.0 \\ 
 \hline
 \rxj\ NW & 40111001002  &   43 &  2015-09-27 &  257.86 & $-$39.52    & 343.3  \\   
\rxj\ NW & 40111002002  &    49 & 2016-03-30   &  257.93 & $-$39.58   & 165.6   \\
 (total)  &    & 92.0 \\
\hline
 Vela Jr. NW &  40101011002 &  69.0 &  2015-07-07 &  132.2 &$-$45.7 &  40.4 \\
 Vela Jr. NW &  40101011004 &  102.1 &  2015-07-16 &  132.2 &    $-$45.7 &  44.4 \\     
 (total)  &    & 142.2 \\
\hline
 SN1987A &  40001014002 &  57.5 &  2012-09-07 &  84.0 &    $-$69.2 &  149.1 \\    
 SN1987A &  40001014003 &  113.3 &  2012-09-08 &  84.0 &    $-$69.2 &  149.1 \\    
 SN1987A &  40001014004 &  198.2 &  2012-09-11 &  83.9 &    $-$69.2 &  149.2 \\    
 SN1987A &  40001014006 &  45.2 &  2012-10-20 &  84.0 &    $-$69.3 &  190.5 \\    
 SN1987A &  40001014007 &  173.2 &  2012-10-21 &  83.9 &    $-$69.3 &  190.5 \\    
 SN1987A &  40001014010 &  160.0 &  2012-12-12 &  83.9 &    $-$69.3 &  242.9 \\    
 SN1987A &  40001014013 &  403.1 &  2013-06-29 &  83.8 &    $-$69.2 &  80.1 \\    
 SN1987A &  40001014015 &  83.5 &  2014-04-21 &  83.8 &    $-$69.3 &  13.1 \\    
 SN1987A &  40001014016 &  379.4 &  2014-04-22 &  83.8 &    $-$69.3 &  13.1 \\    
 SN1987A &  40001014018 &  170.5 &  2014-06-15 &  83.8 &    $-$69.2 &  65.0 \\    
 SN1987A &  40001014020 &  237.4 &  2014-06-19 &  83.8 &    $-$69.2 &  70.2 \\    
 SN1987A &  40001014023 &  397.7 &  2014-08-01 &  83.9 &    $-$69.2 &  111.4 \\    
 (total)  &    & 2419.0 \\
 %%%%%%%%%%%%%%%%%%%%%%%%%%%%
 %% end
\enddata
%\tablecomments{ }
\end{deluxetable*}

%%%%%%%%%%%%%%%%%%%%%%%%%%%%
%%%%%%%%%%%%%%%%%%%%%%%%%%%%

%% file: make_FigTab/Bohm_result_tab_thermal.tex
%%%%%%%%%%%%%%%%%%%%%%%%%%%%
%%%%%%%%%%%%%%%%%%%%%%%%%%%%

%\begin{longtable}
\startlongtable
\begin{deluxetable*}{cccccccccccc}
\tablecaption{Thermal parameters.
\label{tab:Bohm_result_tab_thermal}
}
%\tablewidth{700pt}
\tabletypesize{\tiny}
%\tabletypesize{\scriptsize}
\tablehead{
\colhead{Region} & 
%\colhead{$N_H$}   & 
%\colhead{$\varepsilon_0$}   & 
\colhead{$kT$} & \colhead{O} &  \colhead{Ne} & \colhead{Mg} & \colhead{Si} & \colhead{S} & \colhead{Ar} & \colhead{Ca} & \colhead{Fe} & \colhead{$nt$} & \colhead{Norm} 
%\colhead{$\chi^2$} & 
%\colhead{d.o.f}  & 
\\
\colhead{} & \colhead{(keV)} & \colhead{(O$_\odot$)} &  \colhead{(Ne$_\odot$)} & \colhead{(Mg$_\odot$)} & \colhead{(Si$_\odot$)} & \colhead{(S$_\odot$)} & \colhead{(Ar$_\odot$)} & \colhead{(Ca$_\odot$)} & \colhead{(Fe$_\odot$)} & \colhead{ ($10^{10}$~s~$\cc$)} & \colhead{($10^{-5}~{\rm cm}^{-5}$)}   
} 
\startdata
%%%%%%%%%%%%%%
Cassiopeia A  \\ \cline{1-1}
N1 & 1.70 & 1.0 & 1.0 & 1.0 $\pm$ 0.6 & 11 $\pm$ 2 & 10 $\pm$ 2 & 12$^{+7} _{-6}$ & 12$^{+16} _{-12}$ & 1.0 & 6.9 & 5.0 $\pm$ 0.9 \\ 
N2 & 1.70 & 1.0 & 1.0 & 1.8$^{+0.5} _{-0.4}$ & 7.9$^{+1.1} _{-0.9}$ & 6.9$^{+1.2} _{-1.0}$ & 9.5$^{+4.2} _{-3.9}$ & 8.0$^{+10.2} _{-8.0}$ & 1.0 & 6.9 & 5.0 $\pm$ 0.6 \\ 
S & 0.69 & 1.0 & 1.0 & 1.3 $\pm$ 0.1 & 1.1 $\pm$ 0.1 & 1.4 $\pm$ 0.2 & 3.0$^{+1.3} _{-1.2}$ & 1.0 & 1.0 & 22.6 & 118$^{+14} _{-13}$ \\ 
NW & 1.61 & 1.0 & 1.0 & 0.1$^{+2.3} _{-0.1}$ & 20$^{+36} _{-8}$ & 12$^{+21} _{-5}$ & 42$^{+82} _{-23}$ & 60$^{+129} _{-49}$ & 1.0 & 9.6 & 1.2$^{+0.9} _{-0.8}$ \\ 
SE & 1.77 & 1.0 & 1.0 & 1.4 & 6.0 & 3.3 & 1.0 & 1.0 & 1.0 & 10.4 & 1.4 $\pm$ 0.4   \\ 
E & 1.84 & 1.0 & 1.0 & 0.5 $\pm$ 0.5 & 4.4$^{+1.0} _{-0.8}$ & 2.5$^{+0.9} _{-0.8}$ & 1.0 & 1.0 & 1.0 & 14.3 & 3.6$^{+0.7} _{-0.6}$ \\ 
NE & 1.76 & 1.0 & 1.0 & 0.8 & 11.0 & 8.8 & 10.2 & 23.1 & 1.0 & 14.0 & 1.3 $\pm$ 0.2 \\ 
%%%%%
\hline
%\multicolumn{9}{c}{Kepler}  \\
Kepler  \\ \cline{1-1}
1 & 0.80 & 0.42 & 1.0 & 1.1 & 11 $\pm$ 1 & 19 $\pm$ 1 & 1.0 & 1.0 & 1.0 & 5.9 & 5.8 $\pm$ 0.3 \\ 
2 & 0.58 & 0.21 & 1.0 & 0.4 & 15 $\pm$ 1 & 27 $\pm$ 3 & 1.0 & 1.0 & 1.0 & 9.5 & 12 $\pm$ 1 \\ 
3 & 0.60 & 0.98 & 1.0 & 0.4 & 21 $\pm$ 2 & 26 $\pm$ 4 & 1.0 & 1.0 & 1.0 & 82.1 & 3.9$^{+0.4} _{-0.3}$ \\ 
4 & 0.34 & 1.00 & 1.0 & 1.0 & 1 & 1.0 & 1.0 & 1.0 & 1.0 & 100.0 & <0.35 \\ 
5 & 0.50 & 1.00 & 1.0 & 1.0 & 1 & 1.0 & 1.0 & 1.0 & 1.0 & 60.0 & 0.28$^{+0.43} _{-0.28}$ \\ 
6 & 0.53 & 1.00 & 1.0 & 1.0 & 44 & 28 & 130.8 & 1.0 & 1.0 & 69.8 & 0.15 $\pm$ 0.06 \\ 
7 & 0.52 & 1.00 & 1.0 & 1.0 & 39 & 29 & 1.0 & 1.0 & 1.0 & 60.0 & 0.29 $\pm$ 0.07 \\ 
8 & 0.45 & 2.21$^{+3.37} _{-2.21}$ & 1.0 & 2.7$^{+4.2} _{-2.7}$ & 67$^{+77} _{-29}$ & 105$^{+130} _{-54}$ & 1.0 & 1.0 & 1.0 & 42.1 & 0.47$^{+0.36} _{-0.25}$ \\ 
9 & 0.57 & 0.18 & 1.0 & 0.4 $\pm$ 0.3 & 18 $\pm$ 2 & 22 $\pm$ 4 & 1.0 & 1.0 & 1.0 & 20.5 & 3.4$^{+0.4} _{-0.3}$ \\ 
10 & 0.57 & 0.61 & 1.0 & 0.4 & 39$^{+6} _{-5}$ & 42$^{+9} _{-7}$ & 15$^{+28} _{-15}$ & 1.0 & 1.0 & 87.1 & 1.8 $\pm$ 0.3 \\ 
11 & 0.51 & 0.18 & 1.0 & 0.4 & 34$^{+6} _{-5}$ & 43$^{+11} _{-10}$ & 1.0 & 1.0 & 1.0 & 12.7 & 1.2 $\pm$ 0.2 \\ 
12 & 0.50 & 0.72 & 1.0 & 0.7 & 21$^{+7} _{-5}$ & 34$^{+15} _{-10}$ & 62$^{+87} _{-62}$ & 1.0 & 1.0 & 9.7 & 2.5$^{+0.8} _{-0.7}$ \\ 
13 & 0.69 & 0.59$^{+0.08} _{-0.07}$ & 1.0 & 1.0 $\pm$ 0.1 & 7.5$^{+0.5} _{-0.4}$ & 15 $\pm$ 1 & 34 $\pm$ 5 & 1.0 & 1.2 $\pm$ 0.1 & 8.2 & 26 $\pm$ 1 \\ 
14 & 2.11 & 0.24$^{+0.14} _{-0.11}$ & 1.0 & 2.8 $\pm$ 0.3 & 17$^{+2} _{-1}$ & 17 $\pm$ 2 & 28 $\pm$ 5 & 1.0 & 1.0 & 3.0 & 2.5$^{+0.3} _{-0.2}$ \\ 
%%%%%
\hline
%\multicolumn{9}{c}{Tycho}  \\
Tycho  \\ \cline{1-1}
1 & 0.78 & 1.0 & 1.0 & 1.0 & 33$^{+4} _{-3}$ & 45$^{+6} _{-5}$ & 129$^{+20} _{-18}$ & 316$^{+67} _{-65}$ & 1.0 & 6.8 & 20 $\pm$ 2 \\ 
2 & 2.17 & 1.9$^{+0.3} _{-0.2}$ & 1.0 & 0.8 & 12 $\pm$ 1 & 11 $\pm$ 1 & 13 & 18.8 & 1.0 & 2.3 & 30 $\pm$ 3 \\ 
3 & 3.38 & 2.3 $\pm$ 0.5 & 1.0 & 1.0 & 10$^{+3} _{-2}$ & 7.2$^{+2.3} _{-1.7}$ & 1 & 1.0 & 1.0 & 2.3 & 5.6$^{+1.4} _{-1.2}$ \\ 
4 & 5.46 & 10.9 & 1.0 & 1.0 & 107 & 68 & 74 & 1.0 & 1.0 & 2.6 & 0.94 $\pm$ 0.03 \\ 
5 & 1.30 & 2.7 $\pm$ 0.4 & 1.0 & 1.0 & 32 $\pm$ 4 & 39$^{+6} _{-5}$ & 55$^{+10} _{-9}$ & 129$^{+28} _{-25}$ & 1.0 & 3.5 & 30$^{+5} _{-4}$ \\ 
6 & 1.81 & 6.1$^{+2.5} _{-1.2}$ & 1.0 & 1.8$^{+1.4} _{-0.7}$ & 57$^{+50} _{-20}$ & 61$^{+54} _{-22}$ & 1 & 1.0 & 1.0 & 2.9 & 3.5$^{+2.0} _{-1.7}$ \\ 
7 & 2.64 & 4.0 & 1.0 & 3.5 & 31 & 23 & 1 & 1.0 & 1.0 & 3.0 & 3.6 $\pm$ 0.1 \\ 
8 & 1.79 & 1.0 & 1.0 & 4.2 & 107 & 102 & 121 & 276.1 & 1.0 & 4.3 & 0.93 $\pm$ 0.03 \\ 
9 & 2.52 & 5.5 & 1.0 & 5.1 & 68$^{+21} _{-13}$ & 54$^{+18} _{-11}$ & 65 & 1.0 & 1.0 & 3.2 & 1.2 $\pm$ 0.3 \\ 
10 & 1.22 & 7.1 $\pm$ 0.7 & 1.0 & 1.2 $\pm$ 0.1 & 20$^{+3} _{-2}$ & 22$^{+4} _{-3}$ & 27$^{+5} _{-4}$ & 60$^{+17} _{-14}$ & 1.0 & 3.9 & 43$^{+7} _{-6}$ \\ 
11 & 1.25 & 5.1$^{+1.2} _{-1.0}$ & 1.0 & 1.3$^{+0.7} _{-0.6}$ & 40$^{+12} _{-8}$ & 45$^{+15} _{-10}$ & 87 & 1.0 & 1.0 & 3.6 & 3.6 $\pm$ 0.9 \\ 
12 & 1.73 & 3.2 $\pm$ 0.6 & 1.0 & 1.4$^{+0.6} _{-0.5}$ & 32$^{+13} _{-8}$ & 32$^{+14} _{-9}$ & 1 & 1.0 & 1.0 & 2.8 & 3.9$^{+1.4} _{-1.1}$ \\ 
13 & 1.64 & 3.4$^{+0.6} _{-0.5}$ & 1.0 & 1.9 $\pm$ 0.3 & 24 $\pm$ 2 & 22 & 1 & 1.0 & 1.0 & 2.6 & 5.7 $\pm$ 0.4 \\ 
14 & 1.95 & 6.4 $\pm$ 0.5 & 1.0 & 1.1 $\pm$ 0.1 & 17$^{+2} _{-1}$ & 13 $\pm$ 1 & 10 & 13.5 & 1.0 & 2.3 & 32 $\pm$ 3 \\ 
15 & 0.73$^{+0.05} _{-0.02}$ & 1.0 & 1.0 & 1.0 & 30$^{+6} _{-4}$ & 37$^{+7} _{-6}$ & 36$^{+11} _{-5}$ & 1.0 & 1.0 & 29$^{+13} _{-12}$ & 22 $\pm$ 5 \\ 
16 & 0.61 & 0.8 $\pm$ 0.1 & 1.0 & 1.9$^{+0.2} _{-0.1}$ & 18 $\pm$ 2 & 40 $\pm$ 4 & 124 $\pm$ 18 & 1.0 & 1.0 & 7.1 & 37 $\pm$ 3 \\ 
17 & 1.15$^{+0.23} _{-0.11}$ & 1.1$^{+0.2} _{-0.1}$ & 1.0 & 1.0 & 23 $\pm$ 2 & 26$^{+3} _{-4}$ & 46$^{+12} _{-13}$ & 1.0 & 1.0 & 3.0$^{+0.5} _{-0.6}$ & 32$^{+4} _{-5}$ \\ 
18 & 0.77 & 1.0 & 1.0 & 1.3$^{+0.4} _{-0.3}$ & 46$^{+18} _{-15}$ & 55$^{+22} _{-18}$ & 96$^{+40} _{-34}$ & 329$^{+141} _{-115}$ & 1.0 & 5.8 & 53$^{+29} _{-15}$ \\ 
19 & 1.15 & 1.7 & 1.0 & 1.4 & 38$^{+4} _{-3}$ & 40$^{+4} _{-3}$ & 61 $\pm$ 9 & 196$^{+30} _{-29}$ & 1.0 & 3.8 & 18 $\pm$ 2 \\ 
20 & 1.76 & 1.6 & 1.0 & 2.0 & 56$^{+32} _{-15}$ & 49$^{+29} _{-14}$ & 31 & 45.6 & 1.0 & 3.6 & 1.9 $\pm$ 0.7 \\ 
%%%%%
\hline
%\multicolumn{9}{c}{SN 1006}  \\
SN 1006 \\ \cline{1-1}
1 & 0.41 & 5.0 & 4.7 & 12.0 & 29 & 1.0 & 1.0 & 1.0 & 1.0 & 1.6 & 1.1$^{+0.4} _{-0.3}$ \\ 
2 & 0.10$^{+0.06} _{-0.10}$ & 5.0 & 4.7 & 12.0 & 29 & 1.0 & 1.0 & 1.0 & 1.0 & 1.6 & 63$^{+166} _{-53}$ \\ 
3 & 0.38$^{+0.08} _{-0.09}$ & 4.4 & 1.5 & 15.0 & 50 & 1.0 & 1.0 & 1.0 & 1.0 & 1.3 & 1.5$^{+0.5} _{-0.4}$ \\ 
4 & 0.41$^{+0.04} _{-0.05}$ & 4.4 & 1.5 & 15.0 & 50 & 1.0 & 1.0 & 1.0 & 1.0 & 1.3 & 2.9$^{+0.5} _{-0.4}$ \\ 
5 & 0.29$^{+0.05} _{-0.06}$ & 4.4 & 1.5 & 15.0 & 50 & 1.0 & 1.0 & 1.0 & 1.0 & 1.3 & 2.8$^{+1.2} _{-0.7}$ \\ 
6 & 0.37 $\pm$ 0.02 & 4.4 & 1.5 & 15.0 & 50 & 1.0 & 1.0 & 1.0 & 1.0 & 1.3 & 6.1 $\pm$ 0.9 \\ 
7 & 2.1$^{+0.9} _{-0.5}$ & 9.0$^{+1.4} _{-0.9}$ & 1.5 & 11$^{+2} _{-1}$ & 44$^{+11} _{-6}$ & 1.0 & 1.0 & 1.0 & 1.0 & 0.35$^{+0.05} _{-0.03}$ & 3.3$^{+0.4} _{-0.5}$ \\ 
8 & 1.4$^{+0.8} _{-0.4}$ & 6.7$^{+0.9} _{-0.7}$ & 1.5 & 8.8$^{+1.4} _{-1.2}$ & 48$^{+15} _{-12}$ & 1.0 & 1.0 & 1.0 & 1.0 & 0.47$^{+0.11} _{-0.07}$ & 10 $\pm$ 2 \\ 
9 & 0.39$^{+0.02} _{-0.03}$ & 4.4 & 1.5 & 15.0 & 241$^{+63} _{-45}$ & 1.0 & 1.0 & 1.0 & 1.0 & 1.3 & 13$^{+5} _{-4}$ \\ 
10 & 0.40$^{+0.04} _{-0.03}$ & 4.4 & 1.5 & 15.0 & 50 & 1.0 & 1.0 & 1.0 & 1.0 & 1.3 & 10$^{+3} _{-2}$ \\ 
11 & 0.39 $\pm$ 0.06 & 4.4 & 1.5 & 15.0 & 50 & 1.0 & 1.0 & 1.0 & 1.0 & 1.3 & 7.4$^{+2.4} _{-2.0}$ \\ 
12 & 0.38 $\pm$ 0.06 & 4.4 & 1.5 & 15.0 & 50 & 1.0 & 1.0 & 1.0 & 1.0 & 1.3 & 4.5$^{+1.4} _{-1.2}$ \\ 
13 & 0.12$^{+0.16} _{-0.12}$ & 5.0 & 4.7 & 12.0 & 29 & 1.0 & 1.0 & 1.0 & 1.0 & 1.6 & 25$^{+225} _{-22}$ \\ 
14 & 0.09$^{+0.19} _{-0.09}$ & 5.0 & 4.7 & 12.0 & 29 & 1.0 & 1.0 & 1.0 & 1.0 & 1.6 & 27$^{+62} _{-26}$ \\ 
15 & 0.14$^{+0.15} _{-0.14}$ & 5.0 & 4.7 & 12.0 & 29 & 1.0 & 1.0 & 1.0 & 1.0 & 1.6 & 2.6$^{+50.4} _{-2.2}$ \\ 
16 & 0.27$^{+0.02} _{-0.03}$ & 6.3$^{+1.3} _{-0.7}$ & 4.3$^{+0.9} _{-1.0}$ & 12.0 & 342$^{+108} _{-88}$ & 1.0 & 1.0 & 1.0 & 1.0 & 1.6 & 41$^{+33} _{-12}$ \\ 
17 & 5.5$^{+-5.5} _{-3.2}$ & 20$^{+24} _{-9}$ & 13$^{+10} _{-4}$ & 9.0$^{+5.1} _{-3.1}$ & 29 & 1.0 & 1.0 & 1.0 & 1.0 & 0.12$^{+0.24} _{-0.02}$ & 1.6$^{+1.5} _{-0.7}$ \\ 
%%%%%
\hline
%\multicolumn{9}{c}{RCW 86 }  \\
RCW 86 \\ \cline{1-1}
NE  & 0.44 & 1 & 1.9 & 1.9 & 1.9 & 1 & 1 & 1.2 & 1 & 2.7 & 17 $\pm$1  \\
%NE  & 0.44 & 1 & 1.9 & 1.9 & 1.9 & 1 & 1 & 1.2 & 1 & 2.7 & 15 $\pm$1  \\
\hline 
%\multicolumn{9}{c}{SN 1987A}  \\
SN 1987A \\ \cline{1-1}
%jointfit_7_chandraAll_1_Thermal-free.xcm
%whole & 0.56/2.4$\pm$0.3\chech & 0.28 &  0.71 & 0.72 & 0.8 & 1.2 & 1 & 1 & 0.6 & 25 & 85$\pm$7/85$\pm$7\check \\ 
whole & 0.68$\pm$0.03/2.5$^{+0.3}_{-0.2}$ & 0.34 & 3.0 & 1.0 & 1.4 & 2.5 & 2.7 & 1 & 0.61 & 8.5$^{+2.3}_{-1.7}$ & 290$\pm$12/50$\pm$4 \\ 
%%%%%%%%%%%%%%%%
%% end
\enddata
\tablecomments{
The parameter values without uncertainties are fixed. 
For the thermal model, VNEI is used for Cassiopeia A, Kepler, and Tycho, while Vpshock is used for SN 1006 and RCW 86 (see also the text). 
In SN 1987A, the first and second terms of $kT$ and Norm indicate the best-fit parameters of the Vequil and Vpshock models, respectively.
%He/He$_\odot$=13 and N/N$_\odot$=3.2. %, and gaussian at 6.6 keV is added.
}
\end{deluxetable*}